\documentclass[mathpazo]{cicp}
\usepackage{mathrsfs}
\usepackage{amsmath}
\usepackage{stmaryrd}
\usepackage{bbding}
\usepackage{dcolumn}
\usepackage{graphicx}
\usepackage{amsfonts}
\usepackage{amssymb}
\usepackage{psfrag}
\usepackage{wrapfig}
\usepackage{subfigure}
\usepackage{makeidx}
\usepackage{bm}
\usepackage{epsf}
\usepackage{color}
\usepackage{epsfig}
\usepackage{setspace}
\usepackage{graphicx}
\usepackage{epstopdf}
\usepackage{psfrag}
\usepackage{subfigure}
\usepackage{multirow}
\usepackage{enumerate}
\usepackage{verbatim}
\newtheorem{rmk}{Remark}

\usepackage{makecell}
\usepackage{float}
\usepackage{esint}
\usepackage{comment}
\newcommand{\mathsym}[1]{{}}
\newcommand{\unicode}[1]{{}}
\newcommand{\tabincell}[2]{\begin{tabular}{@{}#1@{}}#2\end{tabular}}
\begin{document}

\title{Performance Enhancement for High-order Gas-kinetic Scheme Based on WENO-adaptive-order Reconstruction }

 \author[Ji X et.~al.]{Xing Ji\affil{1}, and Kun Xu\affil{1,2}\comma\corrauth}
 \address{\affilnum{1}\ Department of Mathematics, Hong Kong University of Science and Technology, Clear Water Bay, Kowloon, Hong Kong SAR. \\
           \affilnum{2}\ Department of Mechanical and Aerospace Engineering, Hong Kong University of Science and Technology, Clear Water Bay, Kowloon, Hong Kong SAR.}
 \emails{{\tt xjiad@connect.ust.hk} (X.~Ji), {\tt makxu@ust.hk} (K.~Xu)}

	\begin{abstract}
	High-order gas-kinetic scheme (HGKS) has been well-developed in the past years.
	Abundant numerical tests including hypersonic flow, turbulence, and  aeroacoustic problems,  have been used to validate its accuracy, efficiency, and robustness.
	However, there are still rooms for its further improvement.
	Firstly, the reconstruction in the previous scheme mainly achieves a third-order accuracy for the initial non-equilibrium states due to the use of standard WENO reconstruction for cell interface values only, where the slopes inside each cell are not provided. In the previous approach,
the slopes have to be reconstructed from the cell interface values and cell averages again, and the same order of accuracy for slopes as the original WENO scheme cannot be guaranteed.
At the same time, the equilibrium state in space and time in HGKS has to be reconstructed separately.
    Secondly, it is complicated to get reconstructed data at Gaussian points from the WENO-type method in high dimensions.
    For HGKS, besides the point-wise values at the Gaussian points it also requires the slopes in both normal and tangential directions
    of a cell interface.
    Thirdly, there exists visible spurious overshoot/undershoot at weak discontinuities from the previous HGKS with the standard WENO reconstruction.
   In order to overcome these difficulties, in this paper we use an improved reconstruction for HGKS.
   The WENO with adaptive order (WENO-AO) \cite{balsara2016efficient} method is  implemented for reconstruction.
   Equipped with WENO-AO reconstruction, the performance enhancement of HGKS is fully explored.
   WENO-AO not only provides the interface values, but also the slopes. In other words, a whole polynomial inside each cell is provided in
   WENO-AO reconstruction. The available polynomial may not benefit the high-order schemes based on the Riemann solver, where only
   points-wise values at the cell interface are needed. But, it can be fully utilized in the HGKS.
      As a result, the HGKS becomes simpler than the previous one with the direct implementation of cell interface values and their slopes from WENO-AO. The additional reconstruction of equilibrium state at the beginning of each time step can be avoided as well by dynamically merging the reconstructed non-equilibrium slopes.
   The new HGKS essentially releases or totally removes the above existing problems in previous HGKS.
   The accuracy of the scheme from 1D to 3D from the new HGKS can recover the theoretical order of accuracy of the WENO reconstruction.
   In the two- and three-dimensional simulations, the new HGKS shows better robustness and efficiency than the previous scheme in all test cases.

\end{abstract}



\maketitle

\section{Introduction}

The gas-kinetic scheme (GKS) targets on the Euler and
Navier-Stokes solutions under the finite volume framework \cite{GKS-2001}.
Its interface flux is based on a time evolution
solution of the kinetic model equation, such as the Bhatnagar-Gross-Krook
(BGK) model \cite{BGK}.
High order gas kinetic scheme (HGKS) has
been developed systematically in the past decade \cite{3rdGKS-Li}.
In comparison with traditional Riemann solver based high-order CFD
methods \cite{toro-2013-rm-book,ttoro}, the distinguishable points of HGKS include the followings:
(i) The time evolving gas distribution function at a cell interface provides a multiple
scale flow physics from the kinetic particle transport to the
hydrodynamic wave propagation, which unifies the evolution from the
upwind flux vector splitting to the central difference Lax-Wendroff type discretization.
(ii) Both inviscid and viscous fluxes are
obtained from the moments of a single time-dependent gas distribution function.
(iii) The flux in GKS has the multi-dimensional properties
\cite{implicitGKS}, where both normal and tangential derivatives of
flow variables around a cell interface contribute the time
evolution solution of the gas distribution function.
(iv) The time evolving gas distribution function at the cell interface not only provides the
flux function, but also the time evolution of macroscopic flow variables. The updated interface flow variables at the beginning of next time step
can be directly used to construct higher-order compact schemes \cite{pan2015unstructuredcompact,pan2016unstructuredcompact,zhao2019compact}.
(v) Different from the Runge-Kutta (RK) time discretization for achieving high-order temporal accuracy,
the multi-stage multi-derivative (MSMD) provides a higher-order time evolution solution with less middle stages due to the
existence of the time-derivative of the interface flux function in HGKS.
Inspired initially by the higher-order generalized Riemann problem \cite{li2016twostage},
a two-stage fourth-order GKS is proposed \cite{Pan2016twostage}.
Recently a family of HGKS have been constructed with only two or three stages for a fifth-order time accurate solution \cite{ji2018family}.
Based on the same fifth-order WENO reconstruction, the performance of
HGKS shows great advantages in terms of efficiency, accuracy, and
robustness compared with traditional higher-order schemes with Riemann solver and Runge-Kutta time-stepping techniques.
Especially, HGKS can capture flow
structures, such as shear instabilities, significantly better than
the schemes based on the Riemann solver due to the multi-dimensional property in GKS flux function.
Among the existing HGKS,
the two-stage fourth-order method \cite{Pan2016twostage} seems to be
an optimal choice in practical computation, which is both efficient and accurate,
and is as robust as a second-order scheme.
It has been applied to compressible multi-component flow \cite{pan2017two-multicomponent},
direction simulation of compressible homogeneous turbulent flow \cite{pan2018two}, and hypersonic non-equilibrium multi-temperature flow \cite{cao2018physical}.
Besides, HGKS has been successfully extended in the DG \cite{luo2010bgk,ren2015multi,ren2016multi} and CPR \cite{zhang2018third} frameworks.

However, there are still rooms for the further improvement of HGKS.
Firstly, the reconstruction procedure proposed in \cite{3rdGKS-Luo} is still adopted in most of the existing HGKS \cite{Pan2016twostage,gks-benchmark,pan2017two-multicomponent,ji2018family,pan2018two,ji2018compact}.
Here the WENO-JS \cite{weno} and WENO-Z \cite{wenoz} reconstructions are directly implemented for the construction of the interface values for the non-equilibrium states.
Then, a simple third-order reconstruction is adopted to obtain the derivatives of flow variables at both sides of the interface
 by using the WENO-based reconstructed cell interface values and cell averages.
Rigorously it achieves only a third-order spatial accuracy and is consistent with the originally designed third-order scheme \cite{3rdGKS-Luo}.
Certainly, in smooth test cases higher order accuracy can be achieved, because
the equilibrium state instead of the above reconstructed non-equilibrium one contributes mostly in the final flux transport.
When the flow is discontinuous, the order of accuracy cannot be properly defined.
However, in special cases, such as low Reynolds number flow computation with both smooth flow and strong shocks, the above
third-order reconstruction does suffer the decrease of order of accuracy.
Actually, the function from the large stencils used in the fifth-order reconstruction has not been fully utilized in the above approach.
Secondly, some spurious overshoots/undershoots have been observed in some test cases.
They typically appear around the corner of weak discontinuities.
Thirdly, for the higher-order tangential reconstruction at a cell interface,
the optimal weights for WENO-JS/Z reconstruction might become non-positive at the targeted Gaussian points.
For example, it is negative for the central point if three Gaussian points are used at a cell interface.
Theoretically, it is a general problem for many other higher order methods as well under the finite volume framework.
A way to resolve this problem is to use the splitting technique \cite{shi2002technique}.
But, it increases the complexity of the algorithm and the robustness of the scheme decreases with the existence of strong shocks.
Overall, the HGKS has a high requirement on the initial reconstruction because  the derivatives of flow variables
at each Gaussian point are needed as well.
The above third-order reconstruction for non-equilibrium state becomes a common choice in previous HGKS \cite{ji2018family}.

Instead of concentrating on the reconstruction of interface values, there exists another class of WENO methods
to reconstruct a complete polynomial inside each cell based on all stencils \cite{levy2000compact,dumbser2007arbitrary,zhu2016new,balsara2016efficient}.
One of the outstanding strategies is named as the WENO with adaptive order (WENO-AO) method \cite{balsara2016efficient}.
Using the same stencils from original WENO scheme, the WENO-AO could reconstruct a polynomial with fifth-order accuracy in smooth region, and  automatically approach to the smoothest quadratic sub-stencil in discontinuous region.
The WENO-AO is more suitable for HGKS to get the initial reconstruction under finite volume framework on Cartesian mesh.
The benefits include the followings.
(i) The linear weights at the locations of all Gaussian-points become positive.
They have fixed values in \cite{balsara2016efficient} and work properly in all test cases in present paper.
There is no need to include more free parameter.
(ii) The non-equilibrium states, including point-wise values and slopes, can be reconstructed at once through a unique polynomial inside each cell,
and they keep the same spatial order of accuracy.
The HGKS benefits more from the WENO-AO reconstruction than the Riemann solver-based schemes, where only point-wise interface values are needed.
(iii) The previous HGKS needs extra reconstruction for the equilibrium state across the cell interface with high-order spatial accuracy in smooth region.
In the new approach, a unified way is adopted  to model the equilibrium state with the same order of accuracy \cite{GKS-lecture} from the non-equilibrium ones directly through particle colliding dynamics. 
The absence of additional reconstruction for the equilibrium state makes the scheme be simpler, especially for three dimensional flow computations.
(iv) The new scheme becomes more robust than the previous one due to up-winding mechanism in the construction of the equilibrium state
which has a upwinding biased weighting functions.
As a result, the scheme avoids oscillation around weak discontinuities due to the consistent reconstructions of both the equilibrium and the non-equilibrium states from a single initial WENO-AO reconstruction.
(v) The previous HGKS obtains accurate results in smooth region \cite{Pan2016twostage,pan2018two} mainly due to the linear reconstruction for the equilibrium state across the cell interface. The new reconstruction for the equilibrium state can recover the previous result in the smooth region and preserve the advantage of the original HGKS.

In this paper, the HGKS with WENO-AO reconstruction will be developed.
In section 2, a review of the conventional HGKS framework is presented. The contents include
the time marching strategy, the GKS flux function, and the original WENO reconstructions from 1-D to 3-D cases.
Then, the two-stage fourth-order GKS is introduced \cite{Pan2016twostage,ji2018family}.
In section 3, the new HGKS with WENO-AO reconstruction is presented and the comparison with the previous one is included.
Section 4 provides inviscid and viscous test cases from one-dimensional to three dimensional flows.
The accuracy, efficiency, and   robustness of the scheme are validated.
The last section is the conclusion.

\section{Review of high-order gas-kinetic scheme (HGKS)}
This paper focuses on the initial reconstruction for HGKS. The reconstruction techniques from 1-D to 3-D are presented in detail.
However, for the flux evaluation and temporal discretization in HGKS, in order to clearly present the idea only one dimensional formulation will be fully reviewed.
The multidimensional flux function in 2D and 3D can be found in \cite{xu2014direct}.

The conservation laws
\begin{equation*}
	\begin{split}
		\textbf{W}_t+ \nabla \cdot \textbf{F}(\textbf{W})=0,\textbf{W}(0,x)=\textbf{W}_0(x),x\in \Omega \subseteq \mathbb{R}
	\end{split}
\end{equation*}
can be written as
$$\textbf{W}_t=-\nabla \cdot {\textbf{F}}(\textbf{W}),$$
for the conservative variables  $\textbf{W}$ and the corresponding flux $\textbf{F}$.
With the spatial discretization $\textbf{W}^h$ and
appropriate evaluation $-\nabla \cdot {\textbf{F}}(\textbf{W})$,
the original PDEs become a system of  ordinary differential equation (ODE)
\begin{equation}\label{ode}
	\begin{split}
		\textbf{W}
		^h_t=\mathcal{L}(\textbf{W}
		^h),t=t_n,
	\end{split}
\end{equation}
where $\mathcal{L}(\textbf{W}^h)$ is the spatial operator of flux.

\subsection{Two-stage fourth-order temporal discretization}
The two-stage fourth-order time marching scheme can be used to solve the above initial value problem, which is given by
\begin{equation}\label{step-hyper-1}
\begin{aligned}
\textbf{W}
^*=\textbf{W}
^n+\frac{1}{2}\Delta t\mathcal
{L}(\textbf{W}
^n)+\frac{1}{8}\Delta t^2\frac{\partial}{\partial
	t}\mathcal{L}(\textbf{W}
^n),
\end{aligned}
\end{equation}
\begin{equation}\label{step-hyper-2}
\begin{aligned}
\textbf{W}
^{n+1}=\textbf{W}
^n+\Delta t\mathcal
{L}(\textbf{W}
^n)+\frac{1}{6}\Delta t^2\big(\frac{\partial}{\partial
	t}\mathcal{L}(\textbf{W}
^n)+2\frac{\partial}{\partial
	t}\mathcal{L}(\textbf{W}
^*)\big),
\end{aligned}
\end{equation}
where $ \partial \mathcal{L}(\textbf{W}) / \partial t $ is the time  derivative of spatial operator.
It was derived independently in \cite{li2016twostage} for hyperbolic conservation laws.
The above temporal discretization has been used in many higher-order GKS \cite{Pan2016twostage,gks-benchmark,pan2018two,ji2018compact,cao2018physical}.

\begin{rmk}
The well established numerical scheme for ODE can be used to solve the Eq.\eqref{ode} by several ways.
If we define
\begin{equation*}
\begin{split}
\textbf{W}^{(m)}_t(t^n)=\frac{d^m \textbf{W}
^n}{dt^m}=\frac{d^{m-1}\mathcal{L}(\textbf{W}
^n)}{dt^{m-1}}=\mathcal{L}^{(m-1)},
\end{split}
\end{equation*}
a $m$th-order time marching scheme can be constructed straightforwardly if the  time derivatives of $\mathcal{L}^{(m)}$ up to $(m-1)$th-order are provided.
However, for the nonlinear system only a few low order derivatives can be obtained, such as $L$ for the
approximate Remiann solver, $\mathcal{L}^{(1)}$ for the generalized Riemann problem (GRP) solver \cite{grp} and the 2nd-order GKS flux function,
and $\mathcal{L}^{(2)}$ for the 3rd-order GKS flux function \cite{3rdGKS-Li}.
The computational cost grows tremendously if higher-order derivatives are required,
such as the 4th-order GKS flux function with the possible evaluation of $\mathcal{L}^{(3)}$ \cite{liu-tang}.

Another approach, which is similar to Runge-Kutta (RK) method, is to introduce the middle stages and update the solution at $t^{n+1}$ with a linear combination of $\mathcal{L}$ and their derivatives in the multiple stages, which is named the multi-stage multi-derivative (MSMD) method.
If $\mathcal{L}$ is used only, the traditional RK method is recovered.
Many middle stages are required in RK method to achieve higher-order temporal accuracy.
For example, $6$ stages are the minimum requirements for a 5th-order RK method \cite{fifth-RK}.
Recent research reveals that the usage of RK method with the time-independent $\mathcal{L}$ alone may generate an inconsistent higher-order method \cite{ben2019consistency}.
With the inclusion of $\mathcal{L}^{(1)}$, the multi-stage two-derivative  method can be constructed, such as the above two-stage fourth-order method.
\end{rmk}


For one dimensional conservation laws, Eq.\eqref{ode} can be written
as the following semi-discrete finite volume form
\begin{align}\label{semidiscrete}
\frac{\text{d} \textbf{W}
_{i}^n}{\text{d} t}=-\frac{1}{\Delta x}
(\textbf{F}
_{i+1/2}^n-\textbf{F}
_{i-1/2}^n):=\mathcal{L}(\textbf{W}
_{i}^n),
\end{align}
where $\mathcal{L}(\textbf{W}_{i}
)$ is the numerical operator for spatial difference of the flux functions.
For a time-dependent flux function $\textbf{F}
_{i \pm 1/2}=\textbf{F}
_{i \pm 1/2}(t)$,
the numerical fluxes and their time derivatives, such as
$\mathcal{L}(\textbf{W}
_{i}^n)$ and $\partial \mathcal{L}(\textbf{W}_{i}^n) /\partial t$, can be evaluated as follows
\begin{align*}
\mathcal{L}(\textbf{W}
_{i}^n)&= - \frac{1}{\Delta x}(\textbf{F}
_{i+1/2}(\textbf{W}
^n,t_n)-\textbf{F}
_{i-1/2}(\textbf{W}
^n,t_n)) \\
\frac{\partial}{\partial t}\mathcal{L}(\textbf{W}
_{i}^n)&= - \frac{1}{\Delta x}(\partial_t \textbf{F}
_{i+1/2}(\textbf{W}
^n,t_n)-\partial_t \textbf{F}
_{i-1/2}(\textbf{W}
^n,t_n)) .
\end{align*}
According to Eq.\eqref{step-hyper-1}, $\textbf{W}
_{i}^*$ at time $t_*$ can be updated. With the similar procedure,
the numerical fluxes and their time derivatives at the intermediate stage can be constructed as well,
where $ \partial \mathcal{L}(\textbf{W}_{i}^*)/\partial t$ is given by
\begin{align*}
\frac{\partial}{\partial t}\mathcal{L}(\textbf{W}
_{i}^*)&=- \frac{1}{\Delta x}(\partial_t \textbf{F}
_{i+1/2}(\textbf{W}
^*,t_*)-\partial_t \textbf{F}
_{i-1/2}(\textbf{W}
^*,t_*)) .
\end{align*}
Then, $\textbf{W}_{i}^{n+1}$ can be updated through Eq.\eqref{step-hyper-2}.
In the following, the detailed gas-kinetic flux function $\textbf{F}_{i \pm 1/2}(t)$ of HGKS is presented.

\subsection{On the construction of time-dependent evolution solution at a cell interface}

The one-dimensional gas-kinetic BGK equation \cite{BGK} can be
written as
\begin{equation}\label{bgk}
f_t+u\cdot\nabla f=\frac{g-f}{\tau},
\end{equation}
where $f$ is the gas distribution function, $g$ is the corresponding
equilibrium state, and $\tau$ is the collision time.

The equilibrium state is a Maxwellian distribution
\begin{equation}\label{maxwell}
\begin{split}
g=\rho(\frac{\lambda}{\pi})^{\frac{K+1}{2}}e^{\lambda((u-U)^2+\xi^2)},
\end{split}
\end{equation}
where $\lambda =m_0/2k_BT $, and $m_0, k_B, T$ represent the molecular mass, the Boltzmann constant, and temperature, $K$ is the number of internal degrees of freedom, i.e. $K=(3-\gamma)/(\gamma-1)$ for one-dimensional flows, and $\gamma$ is the specific heat ratio.
The collision
term satisfies the following compatibility condition
\begin{equation}\label{compatibility}
\int \frac{g-f}{\tau}\pmb{\psi} \text{d}\Xi=0,
\end{equation}
where $\pmb{\psi}=(1,u,\displaystyle \frac{1}{2}(u^2+\xi^2))$,
$\text{d}\Xi=\text{d}u\text{d}\xi_1...\text{d}\xi_{K}$, the internal variable $\xi^2=\xi^2_1+\xi^2_2+...+\xi^2_K$. The connections between macroscopic mass $\rho$, momentum $\rho U$, and energy $\rho E$ with the distribution function $f$ are
\begin{equation}\label{f-to-convar}
\left(
\begin{array}{c}
\rho\\
\rho U\\
\rho E\\
\end{array}
\right)
=\int \pmb{\psi} fd\Xi.
\end{equation}

Based on the Chapman-Enskog expansion for BGK equation
\cite{CE-expansion}, the gas distribution function in the continuum
regime can be expanded as
\begin{align*}
f=g-\tau D_{u}g+\tau D_{u}(\tau
D_{u})g-\tau D_{u}[\tau D_{u}(\tau
D_{u})g]+...,
\end{align*}
where $D_{u}={\partial}/{\partial t}+u\cdot
\nabla$. By truncating on different orders of $\tau$, the
corresponding macroscopic equations can be derived. For the Euler
equations, the zeroth order truncation is taken, i.e. $f=g$. For the
Navier-Stokes equations, the first order truncated distribution function is
\begin{align*}
f=g-\tau (ug_x+g_t).
\end{align*}

Taking moments of the BGK equation Eq.\eqref{bgk} and integrating with
respect to space, the semi-discrete form \eqref{semidiscrete} for the update of macroscopic variables could be recovered.
The numerical fluxes $F_{i+1/2}(t)$ can be obtained
as follows
\begin{equation}\label{macro-flux}
\textbf{F}
_{i+1/2}(t)=\int\pmb{\psi} u f(x_{i+1/2},t,u,\xi)\text{d}\Xi,
\end{equation}
where $f(x_{i+1/2},t,u,\xi)$ is the gas distribution
function at the cell interface. In order to construct the numerical
fluxes, the integral solution of BGK equation Eq.\eqref{bgk} is used
\begin{equation}\label{integral1}
f(x_{i+1/2},t,u,\xi)=\frac{1}{\tau}\int_0^t g(x',t',u,\xi)e^{-(t-t')/\tau}dt'\\
+e^{-t/\tau}f_0(-ut,u,\xi),
\end{equation}
where $x_{i+1/2}=0$ is the location for flux evaluation,
and $x_{i+1/2}=x'+u(t-t')$ is the trajectory of
particle. Here $f_0$ is the initial gas distribution function and $g$ is the corresponding
equilibrium state.
The integral solution mimics a physical process from the particle free transport in $f_0$ for the kinetic scale physics
to the hydrodynamic flow evolution in the integral of $g$ term.
The flow behavior at cell interface depends on the ratio of time step
to the  local particle collision time $\Delta t/\tau$.

To evaluate a time evolution solution at a cell interface,
the following notations are introduced first
\begin{align*}
a \equiv (\partial g/\partial x)/g=g_x/g,
A \equiv (\partial g/\partial t)/g=g_t/g,
\end{align*}
where $g$ is the equilibrium state.  The variables $(a, A)$, denoted by $\omega$,
depend on particle velocity in the form of
\cite{GKS-2001}
\begin{align*}
\omega=\omega_{1}+\omega_{2}u+\omega_{3}\displaystyle
\frac{1}{2}(u^2+\xi^2),
\end{align*}
in 1D case.
For the kinetic part of the integral solution in Eq.\eqref{integral1},
the initial gas distribution function can be constructed as
\begin{equation*}
f_0=f_0^l(x,u)\mathbb{H} (x)+f_0^r(x,u)(1- \mathbb{H}(x)),
\end{equation*}
where $\mathbb{H}(x)$ is the Heaviside function. Here $f_0^l$ and $f_0^r$ are the
initial gas distribution functions on both sides of a cell
interface, which have one to one correspondence with the initially
reconstructed macroscopic variables. For the 2nd-order scheme, the
Taylor expansion for the gas distribution function in space around
$x=0$ is expressed as
\begin{align}\label{third-1}
f_0^k(x)=f_G^k(0)&+\frac{\partial f_G^k}{\partial
	x}x,
\end{align}
for $k=l,r$. According to the Chapman-Enskog expansion, $f_{G}^k$ has the form
\begin{align}\label{third-2}
f_{G}^k(0)=g^k(0)-\tau(a^{k}u+A^k)g^k(0),
\end{align}
where $g^l,g^r$ are the equilibrium states with the form in Eq.\eqref{maxwell} which can be fully determined from  the
reconstructed macroscopic variables $\textbf{W}
^l, \textbf{W}
^r$ at the left and right sides of a cell interface,
\begin{align}\label{get-glr}
\int\pmb{\psi} g^{l}\text{d}\Xi=\textbf{W}
^l,\int\pmb{\psi} g^{r}\text{d}\Xi=\textbf{W}
^r.
\end{align}
Substituting Eq.\eqref{third-1} and Eq.\eqref{third-2} into Eq.\eqref{integral1},
the kinetic part in the integral solution can be written as
\begin{equation}\label{dis1}
\begin{aligned}
e^{-t/\tau}f_0^k(-ut,u,\xi)
=e^{-t/\tau}g^k[1-\tau(a^{k}u+A^k)-ta^{k}u],
\end{aligned}
\end{equation}
where the coefficients $a^{k},...,A^k, k=l,r$ are defined according
to the expansion of $g^{k}$.
Note that higher-order derivatives about $g_k$ have been dropped because we target on the N-S solutions.
After determining the kinetic part
$f_0$, the equilibrium state $g$ in the integral solution
Eq.\eqref{integral1} can be expanded in space and time as well
\begin{align}\label{equli}
g= g^{c}+\frac{\partial  g^{c}}{\partial x}x+\frac{\partial  g^{c}}{\partial t}t,
\end{align}
where $ g^{c}$ is the Maxwellian equilibrium state located on the interface, which
can be determined through the compatibility condition
Eq.\eqref{compatibility},
\begin{align}\label{compatibility2}
\int\pmb{\psi} g^{c}\text{d}\Xi=\textbf{W}^c=\int_{u>0}\pmb{\psi}
g^{l}\text{d}\Xi+\int_{u<0}\pmb{\psi} g^{r}\text{d}\Xi,
\end{align}
where $\textbf{W}^c$ are the macroscopic flow variables for the determination of the
equilibrium state $ g^{c}$. Substituting Eq.\eqref{equli} into Eq.\eqref{integral1}, the hydrodynamic part in  the integral solution
can be written as
\begin{equation}\label{dis2}
\begin{aligned}
\frac{1}{\tau}\int_0^t
g&(x',t',u,\xi)e^{-(t-t')/\tau}dt'
=C_1 g^{c}+C_2 a^{c} u g^{c} +C_3 A^{c} g^{c} ,
\end{aligned}
\end{equation}
where the coefficients
$a^{c},A^{c}$ are
defined from the expansion of the equilibrium state $ g^{c}$. The
coefficients $C_i, i=1,2,3$ in Eq.\eqref{dis2}
are given by
\begin{align*}
C_1=1-&e^{-t/\tau}, C_2=(t+\tau)e^{-t/\tau}-\tau, C_3=t-\tau+\tau e^{-t/\tau}.
\end{align*}
The coefficients in Eq.\eqref{dis1} and Eq.\eqref{dis2}
can be determined by the spatial derivatives of macroscopic flow
variables and the compatibility condition as follows
\begin{align}\label{co}
\langle a\rangle =\frac{\partial \textbf{W} }{\partial x}=\textbf{W}_x,
\langle A+au\rangle=0,
\end{align}
where $$ \langle (...) \rangle  = \int \pmb{\psi} (...) g d \Xi .$$
Finally, the second-order time dependent gas  distribution function at a cell interface is \cite{GKS-2001}
\begin{align}\label{flux}
f(x_{i+1/2},t,u,\xi)=&(1-e^{-t/\tau}) g^{c}+((t+\tau)e^{-t/\tau}-\tau) ug_x^{c}\nonumber\\
+&(t-\tau+\tau e^{-t/\tau}) g_t^{c}\nonumber\\
+&e^{-t/\tau}g^l[1-(\tau+t)ug^l_x-\tau g^l_t)]H(u) \nonumber\\
+&e^{-t/\tau}g^r[1-(\tau+t)ug^r_x-\tau g^r_t)](1-H(u))\nonumber\\
=&(1-e^{-t/\tau}) g^{c}+((t+\tau)e^{-t/\tau}-\tau)a^{c}u g^{c}\nonumber\\
+&(t-\tau+\tau e^{-t/\tau})A^{c}  g^{c}\nonumber\\
+&e^{-t/\tau}g^l[1-(\tau+t)a^{l}u-\tau A^l)]H(u)\nonumber\\
+&e^{-t/\tau}g^r[1-(\tau+t)a^{r}u-\tau A^r)] (1-H(u)).
\end{align}
The details for the evaluation of all terms in the above equation are given in Appendix C.

\subsection{On the computation of flux}

Eq.\eqref{flux} provies a time-dependent gas distribution function, which can be used to evaluate the fluxes for the
macroscopic flow variables through Eq.(\ref{macro-flux}).
In order to obtain $\textbf{F}_{i \pm 1/2}(\textbf{W})$ and $\partial_t\textbf{F}_{i \pm 1/2}(\textbf{W})$ at both $t_n$ and $t_*=t_n + \Delta t/2$,
the flux function can be approximated as a linear function of time within a time interval.

Let's define the following notation,
\begin{align*}
\mathbb{F}_{i+1/2}(\textbf{W}^n,\delta)=\int_{t_n}^{t_n+\delta} \textbf{F}_{i+1/2}(\textbf{W}^n,t)\text{d}t.
\end{align*}
At $t_n=0$,
the flux in the time interval $[t_n, t_n+\Delta t]$ is expanded as
the following linear form
\begin{align*}
\textbf{F}_{i+1/2}(\textbf{W}^n,t)=\textbf{F}_{i+1/2}^n+ t \partial_t \textbf{F}_{i+1/2}^n  .
\end{align*}
The coefficients $\textbf{F}_{i+1/2}^n$ and $\partial_t \textbf{F}_{i+1/2}^n$ can be
fully determined as follows
\begin{align*}
\textbf{F}_{i+1/2}(\textbf{W}^n,t_n)\Delta t&+\frac{1}{2}\partial_t
\textbf{F}_{i+1/2}(\textbf{W}^n,t_n)\Delta t^2 =\mathbb{F}_{i+1/2}(\textbf{W}^n,\Delta t) , \\
\frac{1}{2}\textbf{F}_{i+1/2}(\textbf{W}^n,t_n)\Delta t&+\frac{1}{8}\partial_t
\textbf{F}_{i+1/2}(\textbf{W}^n,t_n)\Delta t^2 =\mathbb{F}_{i+1/2}(\textbf{W}^n,\Delta t/2).
\end{align*}
By solving the linear system, we have
\begin{equation}\label{second}
\begin{aligned}
\textbf{F}_{i+1/2}(\textbf{W}^n,t_n)&=(4\mathbb{F}_{i+1/2}(\textbf{W}^n,\Delta t/2)-\mathbb{F}_{i+1/2}(\textbf{W}^n,\Delta t))/\Delta t,\\
\partial_t \textbf{F}_{i+1/2}(\textbf{W}^n,t_n)&=4(\mathbb{F}_{i+1/2}(\textbf{W}^n,\Delta t)-2\mathbb{F}_{i+1/2}(\textbf{W}^n,\Delta t/2))/\Delta
t^2.
\end{aligned}
\end{equation}

\begin{rmk}	
	For inviscid smooth flow with $\tau=0$, the time evolution solution in Eq.\eqref{flux} reduces to
	\begin{align}\label{euler-smooth-flux}
	f(x_{i+1/2},t,u,\xi)= g^{c}+g^{c}_{t}t= g^{c}+\overline{A} g^{c}t.
	\end{align}
	The coefficients in Eq.(\ref{second}) can be
	simplified as
	\begin{align*}
	\textbf{F}_{i+1/2}(\textbf{W}^n,t_n)&=\int\pmb{\psi} u  g^{c}(n)d\Xi,\\
	\partial_t \textbf{F}_{i+1/2}(\textbf{W}^n,t_n)&=\int\pmb{\psi} u  g^{c}_{t}(n)d\Xi,
	\end{align*}
    which are equivalent to the formulation of $\textbf{F}$ and $\textbf{F}_t$ through the macroscopic Euler equations.
\end{rmk}

\begin{rmk}	
For smooth viscous flow, the full time dependent solution could be simplified as \cite{GKS-2001},
\begin{align}\label{NS-smooth-flux}
f(x_{i+1/2},t,u,\xi)= g^{c}-\tau (a^c u +A^c)+A^c g^{c}t
\end{align}	
under the assumptions of $g^l=g^r=g^c$,$g^l_x=g^r_x=g^c_x$.
The above gas-kinetic solver for smooth flow has less numerical dissipations than the full GKS solver in Eq.\eqref{flux} with the
inclusion of possible discontinuities.
The scheme has been used for the purely smooth flow simulations \cite{xu2003lattice}.
The above solver has only first-order time accuracy for the dissipative terms in the NS equations \cite{Pan2016twostage}.

\end{rmk}

Up to now, the 1-D HGKS is fully reviewed once we determined the $g^{l,r,c}$ and $g_x^{l,r,c}$ i.e. $\textbf{W}^{l,r,c}$ and $\textbf{W}_x^{l,r,c}$ through reconstruction.
For 2-D and 3-D HGKS, the formulation of spatial operators and fluxes could be found in \cite{Pan2016twostage,pan2018two}.

\subsection{Previous GKS with fifth-order WENO reconstruction} \label{origin-reconstruction}

In the following, we first review the conventional reconstruction procedure in GKS for 1-D case \cite{3rdGKS-Luo,Pan2016twostage}, then
extend it to 2-D and 3-D cases \cite{Pan2016twostage,pan2018two}.

\subsubsection{Reconstruction of non-equilibrium states $g^{l,r}$ ($W^{l,r}$) by WENO(Z)} \label{1-d-step-1}

The key idea of WENO is to construct the desired values $Q$ on targeted locations
by the linear combination of the sub-stencil values through the optimal weights.
$Q$ could be either conservative variable, characteristic variable, or primitive variable.

To reconstruct the left interface value $Q_{i+1/2}^l$ at the cell interface
$x_{i+1/2}$, three sub-stencils are selected
\begin{align*}
S_0=\{I_{i-2},I_{i-1}, I_i\}, ~~S_1=\{I_{i-1},I_i, I_{i+1}\}, ~~S_2=\{ I_i,
I_{i+1},I_{i+2}\}.
\end{align*}
The quadratic  polynomials
$w^{r3}_{k}(x)$ corresponding to the sub-stencils $S_k, k=0,1,2$ are
constructed by requiring
\begin{align*}
\frac{1}{\Delta x}\int_{I_{i-j-k-1}}p^{r3}_{k}(x)dx=\overline{Q}_{i-j-k-1},~j=-1,0,1,
\end{align*}
where $\overline{Q}$ represents the cell-averaged quantity.
Each of them can achieve a third-order spatial accuracy $r=3$ in smooth case.
For the reconstructed polynomials, the point value at the cell
interface $x_{i+1/2}$ is given in terms of the cell averages as follows
\begin{align*}
p_{0}^{r3}(x_{i+1/2})&=\frac{1}{3}\overline{Q}_{i-2}-\frac{7}{6}\overline{Q}_{i-1}+\frac{11}{6}\overline{Q}_{i},\\
p_{1}^{r3}(x_{i+1/2})&=-\frac{1}{6}\overline{Q}_{i-1}+\frac{5}{6}\overline{Q}_i+\frac{1}{3}\overline{Q}_{i+1},\\
p_{2}^{r3}(x_{i+1/2})&=\frac{1}{3}\overline{Q}_{i}+\frac{5}{6}\overline{Q}_{i+1}-\frac{1}{6}\overline{Q}_{i+2}.
\end{align*}

On the large stencil $\mathbb{S}_3 = \{S_0, S_1, S_2\}$,
a fourth-order polynomial $p_3^{r5}(x)$ can be constructed according to the following conditions
\begin{align*}
\frac{1}{\Delta x}\int_{I_{i+j}}Q(x)dx=\overline{Q}_{i+j}, ~j=-2,-1, 0,1,2,
\end{align*}
and the point value at the cell interface $x_{i+1/2}$ can be written as
\begin{align*}
p_{3}^{r5}(x_{i+1/2})=\frac{1}{60}(47\overline{Q}_{i}-13\overline{Q}_{i-1}+2\overline{Q}_{i-2}+27\overline{Q}_{i+1}-3\overline{Q}_{i+2}).
\end{align*}

The linear weights $\gamma_k, k=0,1,2,$ can be found such that
\begin{align*}
p_{3}^{r5}(x_{i+1/2})=\sum_{k=0}^{2}\gamma_{k}p_{k}^{r3}(x_{i+1/2}),
\end{align*}
where $\displaystyle\gamma_{0}=\frac{1}{10},\gamma_{1}=\frac{3}{5}, \gamma_{2}=\frac{3}{10}$.
These three weights are called optimal weights, which are unique.
It lifts the reconstructed low order value from the small stencils to a higher-order one from the large stencil.

To deal with discontinuities, the non-normalized WENO-Z type nonlinear weight \cite{wenoz}  is
introduced as follows
\begin{align*}
\omega_k=\gamma_k(1+\frac{\delta}{\beta_{k}+\epsilon}),
\end{align*}
where the global smooth indicator $\delta$ is designed as
\begin{align*}
\delta=|\beta_0-\beta_2|.
\end{align*}

The normalized
weights $\overline{\omega_k}$ is defined as follows
\begin{align*}
\overline{\omega}_k=\frac{\omega_k}{\sum_0^2 \omega_l},
\end{align*}
where $\epsilon$ is a small parameter. The $\beta_k$ are the smoothness indicators which are defined as \cite{weno}
\begin{equation}\label{smooth-indicator}
\beta_k=\sum_{q=1}^{q_k}\Delta x^{2q-1}\int_{x_{i-1/2}}^{x_i+1/2}\big(\frac{\text{d}^q}{\text{d}x^q}p_k(x)\big)^2dx=O(\Delta x^2),
\end{equation}
where $q_k$ is the order of $p_k(x)$. For $p_k^{r3},  k=0,1,2$, $q_k=2$; for $p_3^{r5}$, $q_3=4$.
$\epsilon=10^{-8}$ is taken in current work. The formulae for the smooth indicators are given explicitly in Appendix A.

Thus, the reconstructed left interface value $Q_{i+1/2}^l$ can be written as
\begin{align*}
Q_{i+1/2}^l=\sum_{k=0}^{2}\overline{\omega}_{k}p_{k}^{r3}(x_{i+1/2}).
\end{align*}
Finally, $\textbf{Q}$ should be changed  to the corresponding conservative variables $\textbf{W}$.
The above reconstruction has the following properties.
\begin{itemize}
	\item \textbf{Benefits}:
	1) The WENO reconstruction can be easily adopted in 1-D HGKS;
	2) The optimal weights are unique, which means no free parameter is introduced;
	3) It is efficient since only the low-order smoothness indicators are needed in computation.
	\item \textbf{Deficiencies}:
	1) Only interface values are reconstructed while GKS also requires the derivatives at the interfaces;
	2) The optimal weights may become negative on the different locations. For example at $x=x_i$, the optimal weights are $\displaystyle\gamma_{0}=-\frac{8}{90},\gamma_{1}=\frac{49}{40}, \gamma_{2}=-\frac{8}{90}$.
	\item \textbf{Improvement}: The above optimal weights only provide the reconstructed data at certain fixed location.
    It may not give optimal performance for HGKS due to the additional requirement of slopes in gas-kinetic evolution model.
\end{itemize}

\subsubsection{Reconstruction of non-equilibrium states $g_x^{l,r}$ ($W_x^{l,r}$)}\label{1-d-step-2}

Once the discontinuities appear, not only the $W^{l,r}$ but also $W_x^{l,r}$ shall be reconstructed through suitable limiting process. Theoretically, we could also use a unique linear combination of the derivatives of the above small stencils to obtain the derivatives of the above large stencil at the desired locations.
However, the linear weights need to be re-derived and it is not guaranteed to have all positive coefficients.
The non-linear weights need to be additionally computed.

In the original one-step third-order GKS \cite{3rdGKS-Luo}, the $W_x^{l,r}$ are obtained by constructing a second order polynomial by requiring
\begin{align*}
\frac{1}{\Delta x}\int_{I_{i}}p(x)dV=\overline{W}_{i},p(x_{i-1/2})=W_{i-1/2}^r,p(x_{i+1/2})=W_{i+1/2}^l,
\end{align*}
and the solutions are
\begin{align}\label{non-equ-slope-reconstruction-old}
	&p(x)=a_0+a_1 (x-x_i) + a_2 (x-x_i)^2, \nonumber \\
	&a_0 = \frac{1}{4} (-W_{i-1/2}^r - W_{i+1/2}^l + 6 W_{i}), \nonumber\\
	& a_1 =\frac{W_{i+1/2}^l-W_{i-1/2}^r}{\Delta x}, \nonumber\\
	& a_2 = \frac{3 (W_{i-1/2}^r + W_{i+1/2}^l - 2  W_{i})}{\Delta x^2}
\end{align}
with
\begin{align*}
(W_{x}^r)_{i-1/2}= -\frac{2 (2 W_{i-1/2}^r + W_{i+1/2}^l - 3 W_{i})}{\Delta x},
(W_{x}^l)_{i+1/2}= \frac{2 (W_{i-1/2}^r + 2 W_{i+1/2}^l - 3 W_{i})}{\Delta x}.	
\end{align*}
Lately, all fourth- and higher-order gas kinetic schemes, including compact schemes \cite{Pan2016twostage,gks-benchmark,ji2018family,ji2018compact,zhao2019compact}, follow the above recipe to reconstruct $W_x^{l,r}$.
Considering the fact that the non-equilibrium parts mainly take effects once there is discontinuity,
accurate results can be still obtained in most smooth test cases due to the main contributions from the equilibrium state presented below.

\begin{itemize}	 	
	\item \textbf{Benefits}: It is simple and practically robust with the weighted $ W^{l,r}_{i \pm 1/2}$. Little additional computational cost is needed after the reconstruction of $ W^{l,r}_{i \pm 1/2}$.
	\item \textbf{Deficiencies}: 1) Only third-order accuracy is achieved for the slopes on the targeted locations;
	2) The values of $ W_{i - 1/2}^r$ and $ W_{i + 1/2}^l$ may fall into different sides of a strong shock. In such a case,
 the linear construction in the cell $i$ by connecting the $ W_{i - 1/2}^r$ and $ W_{i + 1/2}^l$  may not be appropriate.
	\item \textbf{Solutions}: A simple and efficient WENO procedure for the reconstruction of both $W^{l,r}$ and $W_x^{l,r}$ with the
same accuracy is needed.
\end{itemize}

\subsubsection{Reconstruction of equilibrium state $g^{c}$ ($W^c$)}\label{1-d-step-3}

With the reconstructed $W_{i+1/2}^l$ and $W_{i+1/2}^r$ at both sides
of a cell interface $x_{i+1/2}$, the macroscopic variables
$W_{i+1/2}^{c}$ and  the corresponding equilibrium state $ g^{c}$ can
be determined according to Eq.\eqref{compatibility2}.

\begin{itemize}
	\item \textbf{Advantages}:
	1)
	The weighting function is coming from the instant collision among the particles that are going to across the cell interface.
	It is physically consistent with the mechanism to get the equilibrium state, and it also includes an upwind mechanics naturally.
	As a result,  the scheme is more robust than the use of arithmetic average in the construction of the equilibrium state;
	2) When $g^l=g^r$, we have $ g^{c}=g^l=g^r$. The above weighted average can keep the $ g^{c}$ the same order of accuracy of $g^l,g^r$.
\end{itemize}

\subsubsection{Reconstruction of slopes of equilibrium state $g_x^c$ ($W^c_x$)}\label{1-d-step-4}
To fully  determine the slopes of the equilibrium state across the cell
interface, the conservative variables across the cell interface is expanded as
\begin{align*}
w^{c}(x)=W_{i+1/2}^{c}+S_1(x-x_{i+1/2})+\frac{1}{2}S_2(x-x_{i+1/2})^2+\frac{1}{6}S_3(x-x_{i+1/2})^3+\frac{1}{24}S_4(x-x_{i+1/2})^4.
\end{align*}
With the following conditions,
\begin{align*}
\int_{I_{i+k}} w^{c}(x)=W_{i+k}, k=-1,...,2,
\end{align*}
the derivatives are determined  by
\begin{align}\label{g0slope}
\begin{aligned}
\displaystyle (W_x^{c})_{i+1/2}=S_1=\big[-\frac{1}{12}(\overline{W}_{i+2}-\overline{W}_{i-1})
+\frac{5}{4}(\overline{W}_{i+1}-\overline{W}_{i})\big]/\Delta x.
\end{aligned}
\end{align}

\begin{itemize}
	\item \textbf{Benefits}:
	1) The smooth reconstruction is consistent with the concept of equilibrium part.  Meanwhile it has the highest order of accuracy with the same stencil.
	\item \textbf{Deficiencies}:
	1) When discontinuities appear, the linear reconstruction of $g_x^0$ may not be appropriate and effect the robustness of the scheme;
	2) A separate module is used for the reconstruction of the equilibrium state and additional modules are needed for the reconstruction in the tangential direction in 2D and 3D cases.
	It increases the complexity of the algorithm.
	\item \textbf{Solutions}:
	In most cases, the contribution from the equilibrium state gets to a minimum contribution due to the enlarged particle collision time $\tau_n$ in the shock region.
	The weak undershoot/overshoot in the previous GKS can be effectively reduced from a newly developed reconstruction in Section \ref{new-reconstruction}.
\end{itemize}	
The reconstruction for the initial non-equilibrium and equilibrium
states are reviewed. In the following, the reconstructions in the 2D and 3D cases will be presented.

\subsubsection{Two dimensional reconstruction} \label{2-d-reconstuction}
The direction by direction reconstruction strategy is usually applied on
rectangular meshes \cite{accuracy-FVM}.
For a fourth-order scheme, two
Gaussian points on each interface are needed for numerical flux
integration.
Our target is to reconstruct
\begin{align*}
W^{l}, ~W^{l}_x, ~W^{l}_{y}, ~~W^{r}, ~W^{r}_x, ~W^{r}_{y},  ~~W^{c}, ~W^{c}_x, ~W^{c}_{y},
\end{align*}
at each Gaussian point $(x_{i+1/2},y_{j_m})$, $m=1,2$.
The reconstruction procedure for the Gaussian point $(i+1/2,j_m), m=0,1$ is
summarized as follows.
The conserved flow variables $\textbf{W}$ should be transferred into the corresponding variables $\textbf{Q}$ for reconstruction if necessary.
\begin{enumerate}[Step 1.]
	\item
	According to one dimensional WENO reconstruction in Sub-subsection \ref{1-d-step-1}, the line averaged reconstructed values $(Q^{l})_{i+1/2,j}, (Q^{r})_{i+1/2,j}$ can be constructed by using the cell averaged values
	$(\overline{Q})_{i+l,j}, l=-2,...2$, and $(\overline{Q})_{i+l+1,j}, l=-2,...,2$.
	
	Then the line averaged spatial derivatives
	$(W^{l}_x)_{i+1/2,j}, (W^{r}_x)_{i+1/2,j}$
	can be constructed with the method in Sub-subsection \ref{1-d-step-2}.
	\item
	Next the line averaged values $(W^{c})_{i+1/2,j}$ are obtained by the compatibility condition in Sub-subsection \ref{1-d-step-3}. 	
	The face averaged derivatives $(W^c_{x})_{i+1/2,j}$ are evaluated by the linear reconstruction in Sub-subsection \ref{1-d-step-4}.
	\item
	Again with the one-dimensional WENO reconstruction in Sub-subsection \ref{1-d-step-1} along the tangential direction,
    the point-wise values (the index $i+1/2$ is omitted)
	$(Q^{l})_{j \pm 1/2}$,
	$ (Q^{r})_{j \pm 1/2}$
	can be constructed by using the line averaged values
	\begin{align*}
	(Q^{l})_{j+l}, (Q^{r})_{j+l},l=-2,...,2.
	\end{align*}
	Then, the point-wise values and spatial derivatives
	\begin{align*}
	&(W^{l})_{j _m}, (W^{r})_{j_m},
	(W^{l}_y)_{j_m}, (W^{r}_y)_{j_m},
	\end{align*}
	with $y=y_{j_m},m=0,1$ can be determined with the method in Sub-subsection \ref{1-d-step-2}.
	
	Again with the one-dimensional WENO reconstruction in Sub-subsection \ref{1-d-step-1} along the tangential direction,
   the point-wise derivatives
	$(Q^{l}_x)_{j \pm 1/2}, (Q^{r}_x)_{j \pm 1/2}$
	can be constructed by using the line averaged derivatives
	$(Q^{l}_x)_{j+l},(Q^{r}_x)_{j+l},l=-2,...,2..$
	And the desired point-wise derivatives
	$(W^{l}_x)_{j_m}$, $ (W^{r}_x)_{j_m},$
	with $y=y_{j_m},m=0,1$ can be determined with the method in Sub-subsection \ref{1-d-step-2}.
	
	\item
	A linear fourth-order polynomial can be constructed by using the line averaged values
	$(W^c)_{j-l}, l=-2,...,2$,
	and the  expected values and derivatives
	$(W^c)_{j \pm 1/2}$, $(W^c_y)_{j \pm1/2}$
	at $y=y_{j_m},m=0,1$ can be obtained.
	
	Similarly a linear fourth-order polynomial by using the line averaged derivatives
	$(W^c_x)_{j-l}, l=-2,...,2$, and
	the expected values and derivatives $(W^c_x)_{j \pm1/2}$ at $y=y_{j_m},m=0,1$, are obtained.

\end{enumerate}

\subsubsection{Three dimensional reconstruction}

For the three dimensional computation, our target is to construct
\begin{align*}
W^{l}, ~W^{l}_x, ~W^{l}_{y},~W^{l}_{z}, ~~W^{r}, ~W^{r}_x, ~W^{r}_{y},~W^{r}_{z},  ~~W^{c}, ~W^{c}_x, ~W^{c}_{y},~W^{c}_{z},
\end{align*}
at each Gaussian point  $(x_{i+1/2},y_{j_m},z_{k_n})$, $m, n= 1,...,2$.
The detailed procedure is given as follows

\begin{enumerate}[Step 1.]
	\item
	According to one dimensional WENO reconstruction in Sub-subsection \ref{1-d-step-1}, the face averaged reconstructed values $(Q^{l})_{i+1/2,j,k}, (Q^{r})_{i+1/2,j,k}$ can be constructed by using the cell averaged values
	$(\overline{Q})_{i+l,j,k}, l=-2,...2$, and $(\overline{Q})_{i+l+1,j,k}, l=-2,...,2$.
		
	Then the face averaged spatial derivatives
	$(W^{l}_x)_{i+1/2,j,k}, (W^{r}_x)_{i+1/2,j,k}$
	can be constructed with the method in Sub-subsection \ref{1-d-step-2}.
	\item
	Next the face averaged values $(W^c)_{i+1/2,j,k}$ are obtained by the compatibility condition in Sub-subsection \ref{1-d-step-3}.
	
	The face averaged derivatives $(W^c_x)_{i+1/2,j,k}$ are determined by the linear reconstruction in Sub-subsection \ref{1-d-step-4}.
	\item \label{3-d-step-3}
	Again with the one-dimensional WENO reconstruction in Sub-subsection \ref{1-d-step-1} along the horizontal direction, the line averaged values (the index $i+1/2$ is omitted)
	$
	(Q^{l})_{j \pm 1/2,k}, (Q^{r})_{j \pm 1/2,k}
	$
	can be constructed by using the face averaged values
	\begin{align*}
	(Q^{l})_{j+l,k}, (Q^{r})_{j+l,k},l=-2,...,2.
	\end{align*}
	The averaged values and spatial derivatives
	\begin{align*}
	&(W^{l})_{j _m,k}, (W^{r})_{j _m,k},
	(W^{l}_y)_{j_m,k}, (W^{r}_y)_{j_m,k},
	\end{align*}
	with $y=y_{j_m},m=0,1$ can be determined with the method in Sub-subsection \ref{1-d-step-2}.
	
	Again with the one-dimensional WENO reconstruction in Sub-subsection \ref{1-d-step-1} along the horizontal direction, the line averaged derivatives
	$(Q^{l}_x)_{j \pm 1/2,k}, (Q^{r}_x)_{j \pm 1/2,k}$
	can be constructed by using the face averaged derivatives
	$(Q_x^{l})_{j+l,k},(Q_x^{r})_{j+l,k},l=-2,...,2.$
	The line averaged  derivatives
	$(W^{l}_x)_{j_m,k}$, $ (W^{r}_x)_{j_m,k},$
	with $y=y_{j_m},m=0,1$ are determined with the method in Sub-subsection \ref{1-d-step-2}.
	
	\item \label{3-d-step-4}
	 A linear fourth-order polynomial can be constructed by using the face averaged values
	$(W^c)_{j-l ,k}, l=-2,...,2$,
	and the line averaged values and derivatives
	$(W^c)_{j \pm 1/2,k}$, $(W^c_y)_{j \pm1/2,k}$
	at $y=y_{j_m},m=0,1$ are obtained.
	
	Similarly a linear fourth-order polynomial by using the face averaged derivatives
	$(W^c_x)_{j-l ,k}, l=-2,...,2$,
	 the line averaged values, and derivatives $(W^c_x)_{j \pm1/2,k}$ at $y=y_{j_m},m=0,1$, are obtained.

	\item With one-dimensional WENO reconstruction in the vertical direction,  the point-wise values and derivatives
	\begin{align*}
	(W^{l})_{j_m,k_n}, (W^{r})_{j_m,k_n},(W^{l}_z)_{j_m,k_n}, (W^{r}_z)_{j_m,k_n}
	\end{align*}
	are obtained by using the line averaged values
	\begin{align*}
	(Q^{l})_{j_m,k+l}, (Q^{r})_{j_m,k+l},l=-2,...,2
	\end{align*}
	with the exactly same method in Step \ref{3-d-step-3},
	same as the determination of
	\begin{align*}
(W^{l}_x)_{j_m,k_n}, &(W^{r}_x)_{j_m,k_n},\\
(W^{l}_y)_{j_m,k_n}, &(W^{r}_y)_{j_m,k_n}.
\end{align*}
	Similarly, the point-wise values and derivatives
	\begin{align*}
	(W^c)_{j_m,k_n},(W^c_z)_{j_m,k_n},
	\end{align*}
		are obtained by using the line averaged values
	\begin{align*}
	(W^c)_{j_m,k+l}, (W^c)_{j_m,k+l},l=-2,...,2
	\end{align*}
	with the exactly same method in Step \ref{3-d-step-4}.	
	And spatial derivatives
	\begin{align*}
	(W^c_x)_{j_m,k_n},(W^c_y)_{j_m,k_n}
	\end{align*}
	can be obtained in the same way.
\end{enumerate}

\begin{rmk} \label{3-d-module-old}
	In summary we need the following six modules for the reconstruction from a programmer's perspective,
	\begin{itemize}
		\item  reconstruction of non-equilibrium states for face-averaged value;
		\item  reconstruction of non-equilibrium states for line-averaged value;
		\item  reconstruction of non-equilibrium states for point value;
		\item  reconstruction of equilibrium state for face-averaged value;
		\item  reconstruction of equilibrium state for line-averaged value;
		\item  reconstruction of equilibrium state for point-value.
	\end{itemize}
\end{rmk}

\section{Fifth-order WENO-AO reconstruction for GKS performance enhancement}\label{new-reconstruction}

\subsection{Reconstruction of non-equilibrium sates $g^{l,r},g_x^{l,r}$ ($W^{l,r},W_x^{l,r}$) from one single polynomial} \label{new-1-d-step-1}
Instead reconstructing the point-wise values and their slopes separately in the previous GKS,
we can reconstruct the whole polynomial within each cell through a new WENO procedure.
Then, all required variables $g^l,g^r,g_x^l,g_x^r,g_{xx}^l,g_{xx}^r...$ can be obtained at once.
This  reconstruction method is named as WENO-AO by Balsara \cite{balsara2016efficient}.
To be compatible with the fourth-order temporal accuracy in current HGKS, the fifth-order WENO5-AO is adopted,
and the detailed formulation is the following.

	We start from rewriting $p_3^{r5}(x)$ as
\begin{align}\label{rewrite-r5}
p_{3}^{r5}(x)&=\gamma_3(\frac{1}{\gamma_3}p_3^{r5}(x)-\sum_0^2 \frac{\gamma_k}{\gamma_3}p_k^{r3}(x))+\sum_0^2 {\gamma_k}p_k^{r 3}(x), r_1 \neq 0,
\end{align}
where $r_{k}, l=0,1,2,3$ are defined as linear weights. Clearly Eq.\eqref{rewrite-r5} holds true for any choice of $r_{k}, l=0,1,2,3$.
Balsara et al. \cite{balsara2016efficient} chose them as
\begin{align*}
\gamma_3=\gamma_{Hi};~~ \gamma_0=(1-\gamma_{Hi})(1-\gamma_{Lo})/2; ~~\gamma_1=(1-\gamma_{Hi}); ~~\gamma_2=\gamma_0,
\end{align*}
which satisfy $r_{l}>0, l=0,1,2,3$ and $\sum_0^3 \gamma_{k} = 1$,
and suggest $\gamma_{Hi} \in [0.85,0.95]$ and $\gamma_{l_o} \in [0.85,0.95]$.
Here we choose $\gamma_{Hi} =0.85 $ and $\gamma_{l_o} =0.85 $ in the numerical tests if no specification values are provided.

To avoid the loss of order of accuracy at inflection points, the WENO-Z type \cite{wenoz} non-linear weights are used as
\begin{align}\label{weno-ao-non-linear-un-normalized-weights}
\omega=\gamma_k(1+ \frac{\delta^2}{(\beta_{k}+\epsilon)^2} ),
\end{align}
where the global smooth indicator $\delta$ is defined as
\begin{align}\label{weno-ao-global-smooth-indicator}
\tau=\frac{1}{3}(|\beta_3^{r5}-\beta_0^{r3}|+|\beta_3^{r5}-\beta_1^{r3}|+|\beta_3^{r5}-\beta_2^{r3}|) = O(\Delta x^4).
\end{align}
The normalized weights are given by
\begin{align*}
\overline{\omega_k}=\frac{\omega_k}{\sum_0^3 \omega_q}.
\end{align*}
Then the final form of the reconstructed polynomial is
\begin{align}\label{weno-ao-re-polynomial}
P^{AO(5,3)}(x)=\overline{\omega_3}(\frac{1}{\gamma_3}p_3^{r5}(x)-\sum_0^2 \frac{\gamma_k}{\gamma_3}p_k^{r3}(x))+\sum_0^2 {\overline{\omega_k}}p_k^{r 3}(x).
\end{align}
So all the desired quantities at cell interfaces can be fully determined as
\begin{align*}
&Q^r_{i-1/2}=P^{AO(5,3)}(x_{i-1/2}),~~Q^l_{i+1/2}=P^{AO(5,3)}(x_{i+1/2}), \nonumber \\
&(Q^r_{x})_{i-1/2}=P_x^{AO(5,3)}(x_{i-1/2}),~~(Q^l_{x})_{i+1/2}=P_x^{AO(5,3)}(x_{i+1/2}).
\end{align*}
\begin{rmk}
	Denote the exact flow distribution as $p(x)$,
	we rewrite \eqref{weno-ao-re-polynomial} as
	\begin{align}\label{weno-ao-accuracy-analysis-1}
	P^{AO(5,3)}(x)&=\frac{\overline{\omega_3}}{\gamma_3}p_3^{r5}(x)+\sum_0^2 (\overline{\omega_k} - \overline{\omega_3} \frac{\gamma_k}{\gamma_3})p_k^{r3}(x) \nonumber\\
	&=\frac{\overline{\omega_3}}{\gamma_3}(p(x)+O(\Delta x ^5))+\sum_0^2 (\overline{\omega_k} - \overline{\omega_3} \frac{\gamma_k}{\gamma_3}) (p(x)+O(\Delta x ^3))
	\nonumber \\
	&=p(x)[\frac{\overline{\omega_3}}{\gamma_3}+\sum_0^2 (\overline{\omega_k} - \overline{\omega_3} \frac{\gamma_k}{\gamma_3})]+
	\frac{\overline{\omega_3}}{\gamma_3}O(\Delta x ^5)+\sum_0^2 (\overline{\omega_k} - \overline{\omega_3} \frac{\gamma_k}{\gamma_3}) O(\Delta x ^3)
	\nonumber \\
	&=p(x)+\frac{\overline{\omega_3}}{\gamma_3}O(\Delta x ^5)+\sum_0^2 (\overline{\omega_k} - \overline{\omega_3} \frac{\gamma_k}{\gamma_3}) O(\Delta x^3)
	\end{align}
	with the constraints of $\sum_0^3 \gamma_{k} = 1$ and $\sum_0^3 \overline{\omega_k} = 1$.
	
	According to Eq.\eqref{smooth-indicator}, Eq.\eqref{weno-ao-non-linear-un-normalized-weights} and Eq.\eqref{weno-ao-global-smooth-indicator}, we have
	\begin{align}\label{weno-ao-accuracy-analysis-2}
	\overline{\omega_k}\sim \omega_k  = \gamma_k (1+O(\Delta x ^4)).
	\end{align}
	thus
	\begin{align}\label{weno-ao-accuracy-analysis-3}
	P^{AO(5,3)}(x)& = p(x)+(1+O(\Delta x ^4))O(\Delta h ^5) \nonumber \\
	&+ \sum_0^2 [\gamma_k (1+O(\Delta x ^4)) - \gamma_3 (1+O(\Delta x ^4)) \frac{\gamma_k}{\gamma_3}] O(\Delta x ^3) \nonumber \\
	&=p(x)+(\Delta x ^5).
	\end{align}
\end{rmk}

In comparison with traditional WENO-type method, the above reconstruction has the following properties.

\begin{itemize}
\item The new reconstruction is more expensive compared with the traditional WENO, mainly due to the requirement of the high-order smooth indicator.
 However, its benefit is not fully utilized when it is applied to schemes with Riemann solvers, where only point-wise values are needed.
 But, it becomes natural and efficient when it is used in GKS under the two-stage fourth-order formulation.
\item The reconstruction is flexible with the location of Gaussian points since the linear weights are independent of geometry.
In fact, similar idea is also adopted by Zhu et.al for designing new WENO scheme  on triangular meshes recently \cite{zhu2018new}.
\end{itemize}

\subsection{Reconstruction equilibrium states $ g^{c},g_x^{c}$ ($ W^{c},W_x^{c}$) locally}\label{new-1-d-step-2}
The reconstructions for the non-equilibrium states  have the uniform order and can be used to get the equilibrium state directly,
such as $ g^{c},g_x^{c}, g_{xx}^{c}..$ by a suitable average of $g^{l,r},g_x^{l,r}, g_{xx}^{l,r}..$.
The simplest way is to use the arithmetic average, but it is only applicable for smooth flow.
To be consistent with the construction of  $ g^{c}$, we make an analogy of the kinetic-based weighting for $g_x^{c},...$, which are given by
\begin{align}\label{new-equ-part}
	&\int\pmb{\psi} g^{c}\text{d}\Xi=\textbf{W}^c=\int_{u>0}\pmb{\psi}
	g^{l}\text{d}\Xi+\int_{u<0}\pmb{\psi} g^{r}\text{d}\Xi, \nonumber \\
	&\int\pmb{\psi} g^{c}_x\text{d}\Xi=\textbf{W}_x^c=\int_{u>0}\pmb{\psi}
	g_x^{l}\text{d}\Xi+\int_{u<0}\pmb{\psi} g_x^{r}\text{d}\Xi.
\end{align}
This method has been used in an early version of second-order GKS \cite{GKS-lecture}.
In this way, all components of the microscopic slopes across the interface have been obtained.

\subsection{Reconstruction procedure on higher dimension}
\subsubsection{Two dimensional reconstruction}
In 2D, the reconstructed values are
\begin{align*}
W^{l}, ~W^{l}_x, ~W^{l}_{y}, ~~W^{r}, ~W^{r}_x, ~W^{r}_{y},
\end{align*}
at each Gaussian point $(x_{i+1/2},y_{j_m})$, $m=1,2$.
The reconstruction procedure for the Gaussian point $(i+1/2,j_m), m=0,1$ is
summarized as follows. Here the time level $n$ is omitted.
\begin{enumerate}[Step 1.]
	\item
	According to the one dimensional WENO-AO reconstruction in Subsection \ref{new-1-d-step-1}, the line averaged reconstructed values and derivatives
	\begin{align*}
	(Q^{l})_{i+1/2,j}, (Q^{r})_{i+1/2,j}, (Q^{l}_x)_{i+1/2,j}, (Q^{r}_x)_{i+1/2,j}
	\end{align*}
	can be constructed by using the cell averaged values
	$(\overline{Q})_{i+l,j}, l=-2,...2$, and $(\overline{Q})_{i+l+1,j}, l=-2,...,2$.

	\item
	Again with the one-dimensional WENO-AO reconstruction in Subsection \ref{new-1-d-step-1} along the tangential direction, the values (the index $i+1/2$ is omitted)
	\begin{align*}
    (Q^{l})_{j _m}, (Q^{r})_{j_m},
    (Q^{l}_y)_{j_m}, (Q^{r}_y)_{j_m},
    \end{align*}
	can be constructed by using the line averaged values
	\begin{align*}
	(Q^{l})_{j+l}, (Q^{r})_{j+l},l=-2,...,2,
	\end{align*}
	with $y=y_{j_m},m=0,1$.	
	The details are given in Appendix B.
	In the same way,
	the desired point-wise derivatives
	$(Q^{l}_x)_{j_m}$, $ (Q^{r}_x)_{j_m},$
	with $y=y_{j_m},m=0,1$ can be determined with the method in Subsection \ref{new-1-d-step-1}
	by using the line averaged derivatives
	$(Q^{l}_x)_{j+l},(Q^{r}_x)_{j+l},l=-2,...,2$.
	\item All the quantities related to the equilibrium states are obtain by the unified weighting method in Subsection \ref{new-1-d-step-2}.
	
\end{enumerate}

\subsubsection{Three dimensional reconstruction}

For the three dimensional computation, the reconstruction procedure for the cell interface $x_{i+1/2,j,k}$ is given as an example.
Again our target is to construct
\begin{align*}
W^{l}, ~W^{l}_x, ~W^{l}_{y},~W^{l}_{z}, ~~W^{r}, ~W^{r}_x, ~W^{r}_{y},~W^{r}_{z},
\end{align*}
at each Gaussian point  $(x_{i+1/2},y_{j_m},z_{k_n})$, $m, n= 1,...,2$.
The detailed procedure is given as follows

\begin{enumerate}[Step 1.]
	\item
	According to one dimensional WENO-AO reconstruction in Subsection \ref{new-1-d-step-1}, the face averaged reconstructed values 	\begin{align*}
	(Q^{l})_{i+1/2,j,k}, (Q^{r})_{i+1/2,j,k}, (Q^{l}_x)_{i+1/2,j,k}, (Q^{r}_x)_{i+1/2,j,k}
	\end{align*}
	can be obtained  by using the cell averaged values
	$(\overline{Q})_{i+l,j,k}, l=-2,...2$, and $(\overline{Q})_{i+l+1,j,k}, l=-2,...,2$.
	
	Then the face averaged spatial derivatives
	$(Q^{l}_x)_{i+1/2,j,k}, (Q^{r}_x)_{i+1/2,j,k}$
	can be constructed with the method in Subsection \ref{new-1-d-step-1}.
	\item \label{new-3-d-step-2}
	With the one-dimensional WENO-AO reconstruction in Subsection \ref{new-1-d-step-1} along the horizontal direction, the line averaged values (the index $i+1/2$ is omitted)
	\begin{align*}
(Q^{l})_{j _m,k}, (Q^{r})_{j_m,k},
(Q^{l}_y)_{j_m,k}, (Q^{r}_y)_{j_m,k},
\end{align*}
	with $y=y_{j_m},m=0,1$ are determined  by using the face averaged values
	\begin{align*}
	(Q^{l})_{j+l,k}, (Q^{r})_{j+l,k},l=-2,...,2.
	\end{align*}
	In the same way, the line averaged derivatives
	$(Q^{l}_x)_{j \pm 1/2,k}$, $(Q^{r}_x)_{j \pm 1/2,k}$,
	$(Q^{l}_x)_{j_m,k}$, $ (Q^{r}_x)_{j_m,k}$,
	with $y=y_{j_m},m=0,1$ 	
	can be constructed by using the face averaged derivatives
	$(Q^{l}_x)_{j+l,k},(Q^{r}_x)_{j+l,k},l=-2,...,2.$.
		
	\item With one-dimensional WENO-AO reconstruction in the vertical direction,  the point values and derivatives
	\begin{align*}
	(Q^{l})_{j_m,k_n}, (Q^{r})_{j_m,k_n},(Q^{l}_z)_{j_m,k_n}, (Q^{r}_z)_{j_m,k_n}
	\end{align*}
	are obtained by using the line averaged values
	\begin{align*}
	(Q^{l})_{j_m,k+l}, (Q^{r})_{j_m,k+l},l=-2,...,2
	\end{align*}
	with the exactly same method in Step \ref{new-3-d-step-2},
	so are the quantities
	\begin{align*}
	(Q^{l}_x)_{j_m,k_n}, &(Q^{r}_x)_{j_m,k_n},\\
	(Q^{l}_y)_{j_m,k_n}, &(Q^{r}_y)_{j_m,k_n}.
	\end{align*}
\end{enumerate}

\begin{rmk}
	In summary, in the new scheme only the following three modules in the reconstruction are needed in comparison with the previous 3-D HGKS in Remark \ref{3-d-module-old},
	\begin{itemize}
	\item  reconstruction of non-equilibrium states for face-averaged value;
    \item  reconstruction of non-equilibrium states for line-averaged value;
    \item  reconstruction of non-equilibrium states for point-wise value.
	\end{itemize}

\end{rmk}

\subsection{Improvements}

The newly proposed reconstruction scheme has at least the following improvements in comparison with the previous one.
\begin{itemize}
	\item It becomes flexible to the reconstruction at the Gaussian points of the interface due to the free choice of the linear weights.
	The new reconstruction method, like the so-called multi-resolution WENO scheme \cite{zhu2018multiresolution}, can be adopted naturally in the current HGKS framework.
	\item Although the choice of the linear weights is not unique, the numerical solutions are not sensitive to it \cite{balsara2016efficient}.
	\item The reconstruction algorithm has been greatly simplified.
	\item The new reconstruction keeps the non-equilibrium states to have the same order of accuracy as the equilibrium one.
	\item The new scheme becomes more robust and less sensitive to the definition of numerical viscosity coefficient than the previous HGKS.
\end{itemize}

\begin{rmk}\label{linear-case}
For the smooth Euler solutions, the gas distribution function at a cell interface $i+1/2$ depends on $W^c_{i+1/2}$ and $(W^c_x)_{i+1/2}$ as shown in Eq.\eqref{euler-smooth-flux}.
With the linear weights, both the fifth-order WENO and WENO-AO methods get to the same polynomials reconstructed by the corresponding five cells
and yield the same point-wise values	
\begin{align*}
&(W^-_{WENO})_{i + 1/2}=(W^-_{WENO-AO})_{i + 1/2}=\frac{1}{60} (27 W_{i+1}+47 W_{i}-13 W_{i-1}+2 W_{i-2}-3 W_{i+2}),\\
&(W^+_{WENO})_{i + 1/2}=(W^+_{WENO-AO})_{i + 1/2}=\frac{1}{60} (47 W_{i+1}+27 W_{i}-3 W_{i-1}-13 W_{i+2}+2 W_{i+3}).
\end{align*}
And clearly they will have the same $(W^c)_{i+1/2}$.
As for $(W^c_x)_{i+1/2}$, the conventional reconstruction gives
\begin{align*}
({W}^c_{x})_{i + 1/2}=\frac{-15 W_{i}+W_{i-1}+15 W_{i+1}-W_{i+2}}{12 \Delta x},
\end{align*}
by the linear reconstruction in Sub-subsection \ref{1-d-step-4}.
For the new method, it gets to
\begin{align*}
&({W}^l_{x})_{i + 1/2}=(p_{i,3}^{r5})_x(x_{i+1/2}) = \frac{-15 W_{i}+W_{i-1}+15 W_{i+1}-W_{i+2}}{12 \Delta x},\\
&({W}^r_{x})_{i + 1/2}=(p_{i+1,3}^{r5})_x(x_{i+1/2}) = \frac{-15 W_{i}+W_{i-1}+15 W_{i+1}-W_{i+2}}{12 \Delta x}.
\end{align*}
Identical $({W}^c_{x})_{i + 1/2}$ has been obtained by the new method through Eq.\eqref{new-equ-part}.
Therefore, the new reconstruction procedure exactly recover the old one in the smooth case.
\end{rmk}

\section{Numerical Results}

In this section, numerical tests from 1-D to 3-D will be presented to validate the new reconstruction method.
For the
inviscid flow, the collision time $\tau$ is
\begin{align*}
\tau=c_1\Delta t+c_2\displaystyle|\frac{p_l-p_r}{p_l+p_r}|\Delta
t.
\end{align*}
Usually $c_1=0.05$ and $c_2=1$ are defined in the conventional HGKS. But, $c_1=0$ can be safely chosen for the new HGKS in most test cases.
For the viscous flow, the collision time is related to the viscosity coefficient,
\begin{align*}
\tau=\frac{\mu}{p}+c_2 \displaystyle|\frac{p_l-p_r}{p_l+p_r}|\Delta t,
\end{align*}
where $p_l$ and $p_r$ denote the pressure on the left and right
sides of the cell interface, $\mu$ is the dynamic viscous
coefficient, and $p$ is the pressure at the cell interface. In
smooth flow regions, it reduces to $\tau=\mu/p$.
The ratio of specific heats takes $\gamma=1.4$. The reason for including pressure
jump term in the particle collision time is to add artificial
dissipation in the discontinuous region to enlarge the shock thickness to the scale of
numerical cell size, and to keep the
non-equilibrium dynamics in the shock layer through the kinetic particle transport to mimic the real
physical mechanism inside the shock structure.
The time step is determined by
$$ \Delta t = C_{CFL} \mbox{Min} ( \frac{ \Delta x}{||\textbf{U}||+a_s}, \frac{ (\Delta x)^2}{4\nu}) ,$$
where $C_{CFL}$ is the CFL number, $||\textbf{U}||$ is the magnitude of velocities, $a_s$ is the sound speed, and $\nu = \mu /\rho$ is the kinematic viscosity coefficient.

The current WENO5-AO GKS is compared with the
WENO5-GKS in \cite{Pan2016twostage}. Both schemes take
the identical Gaussian points at each cell interface, and
two stage fourth order time marching strategy is used for the update of numerical solutions.
The reconstruction is based on characteristic variables for both
schemes.
Denote
\begin{align*}
F(W)=(\rho U_1, \rho U_1^2+p, \rho U_1U_2, \rho U_1U_3,
U_1(\rho E+p))
\end{align*}
in the local coordinate. The Jacobian
matrix $\partial F/\partial W$ can be diagonalized by the right
eigenmatrix $R$. For a specific cell interface, $R_*$ is the right
eigenmatrix of $\partial F/\partial W^*$, and $W^*$ are the averaged
conservative flow variables from both sides of the cell interface.
The characteristic variables for reconstruction are defined as
$Q=R_*^{-1}W$.
The WENO-Z type weights are chosen to avoid the accuracy decrease around the physical extrema.

\subsection{1-D test cases}

\subsubsection{Accuracy test in 1-D}

The advection of density perturbation is tested, and the initial
condition is given as follows
\begin{align*}
\rho(x)=1+0.2\sin(\pi x),\ \  U(x)=1,\ \ \  p(x)=1, x\in[0,2].
\end{align*}
With the periodic boundary condition, the analytic
solution is
\begin{align*}
\rho(x,t)=1+0.2\sin(\pi(x-t)),\ \ \  U(x,t)=1,\ \ \  p(x,t)=1.
\end{align*}
In the computation, a uniform mesh with $N$ points is used.
The collision time $\tau=0$ is set since the flow is smooth and inviscid.
The time step $\Delta t=0.2\Delta x$ is fixed.
Based on the above reconstruction and time-marching method, the current
scheme is expected to present a fifth-order spatial accuracy and a
fourth-order temporal accuracy as analyzed in \cite{ji2018family}.
The $L^1$, $L^2$ and $L^{\infty}$ errors and the corresponding orders at $t=2$ are given below.
Both WENO5-GKS and WENO5-AO-GKS are tested by replacing their non-linear weights by the linear ones.
It has been analyzed in Subsection \ref{linear-case} that the two schemes become identical in this case.
With the mesh refinement in Table \ref{accuracy-weno5-linear} and \ref{accuracy-weno5-ao-linear},
the expected orders of accuracy are obtained and the numerical errors are identical.
Next,  smooth indicators are used to obtain the non-linear weights.
Then WENO5-GKS follows the original reconstruction procedure in Subsection \ref{origin-reconstruction}
and WENO5-AO-GKS takes the new reconstruction procedure in Section \ref{new-reconstruction}. The expected orders have been obtained in Table \ref{accuracy-weno5-z} and Table \ref{accuracy-weno5-AO-weight1}.
In comparison of the results in Table \ref{accuracy-weno5-AO-weight1} and Table \ref{accuracy-weno5-AO-weight2},
the use of the linear weights for the WENO-AO reconstruction has almost no effect on the order of accuracy.
If the WENO5-AO reconstruction is replaced by WENO5 reconstruction in the new reconstruction procedure,
only third-order accuracy can be achieved as shown in Table \ref{accuracy-weno5-z-all-collision},
due to the low-order reconstruction for the non-equilibrium states described in \ref{1-d-step-2}.

\begin{table}[!h]
	\small
	\begin{center}
		\def\temptablewidth{1\textwidth}
		{\rule{\temptablewidth}{1pt}}
		\begin{tabular*}{\temptablewidth}{@{\extracolsep{\fill}}c|cc|cc|cc}
			
			mesh length & $L^1$ error & Order & $L^2$ error & Order& $L^{\infty}$ error & Order  \\
			\hline
1/5 & 2.188290e-02 & ~ & 2.402055e-02 & ~ & 3.360007e-02 & ~ \\
1/10 & 8.603723e-04 & 4.67 & 9.747310e-04 & 4.62 & 1.385732e-03 & 4.60 \\
1/20 & 2.857557e-05 & 4.91 & 3.176742e-05 & 4.94 & 4.674622e-05 & 4.89 \\
1/40 & 9.035141e-07 & 4.98 & 1.000819e-06 & 4.99 & 1.482588e-06 & 4.98 \\
1/80 & 2.826315e-08 & 5.00 & 3.132381e-08 & 5.00 & 4.649773e-08 & 4.99 \\ 					
		\end{tabular*}
		{\rule{\temptablewidth}{0.1pt}}
	\end{center}
	\vspace{-4mm} \caption{\label{accuracy-weno5-linear} Accuracy test for the 1-D sin-wave
		propagation by the conventional reconstruction procedure with the linear WENO5 reconstruction. $\Delta t = 0.2 \Delta x$.  }
\end{table}

\begin{table}[!h]
	\small
	\begin{center}
		\def\temptablewidth{1\textwidth}
		{\rule{\temptablewidth}{1pt}}
		\begin{tabular*}{\temptablewidth}{@{\extracolsep{\fill}}c|cc|cc|cc}
			
			mesh length & $L^1$ error & Order & $L^2$ error & Order& $L^{\infty}$ error & Order  \\
			\hline
1/5 & 2.188290e-02 & ~ & 2.402055e-02 & ~ & 3.360007e-02 & ~ \\
1/10 & 8.603723e-04 & 4.67 & 9.747310e-04 & 4.62 & 1.385732e-03 & 4.60 \\
1/20 & 2.857557e-05 & 4.91 & 3.176742e-05 & 4.94 & 4.674622e-05 & 4.89 \\
1/40 & 9.035141e-07 & 4.98 & 1.000819e-06 & 4.99 & 1.482588e-06 & 4.98 \\
1/80 & 2.826314e-08 & 5.00 & 3.132381e-08 & 5.00 & 4.649772e-08 & 4.99 \\ 				
		\end{tabular*}
		{\rule{\temptablewidth}{0.1pt}}
	\end{center}
	\vspace{-4mm} \caption{\label{accuracy-weno5-ao-linear} Accuracy test for the 1-D sin-wave
		propagation by the conventional reconstruction procedure with the linear WENO5-AO reconstruction. $\Delta t = 0.2 \Delta x$.  }
\end{table}

\begin{table}[htp]
	\small
	\begin{center}
		\def\temptablewidth{1\textwidth}
		{\rule{\temptablewidth}{1pt}}
		\begin{tabular*}{\temptablewidth}{@{\extracolsep{\fill}}c|cc|cc|cc}
			
			mesh length & $L^1$ error & Order & $L^2$ error & Order& $L^{\infty}$ error & Order  \\
			\hline
1/5 & 2.452730e-02 & ~ & 2.695828e-02 & ~ & 3.770457e-02 & ~ \\
1/10 & 1.118455e-03 & 4.45 & 1.231008e-03 & 4.45 & 1.759656e-03 & 4.42 \\
1/20 & 3.063525e-05 & 5.19 & 3.514416e-05 & 5.13 & 5.329365e-05 & 5.05 \\
1/40 & 9.074879e-07 & 5.08 & 1.021807e-06 & 5.10 & 1.522067e-06 & 5.13 \\
1/80 & 2.827664e-08 & 5.00 & 3.141615e-08 & 5.02 & 4.675543e-08 & 5.02 \\ 				
		\end{tabular*}
		{\rule{\temptablewidth}{0.1pt}}
	\end{center}
	\vspace{-4mm} \caption{\label{accuracy-weno5-z} Accuracy test for the 1-D sin-wave
		propagation by the conventional reconstruction procedure with the WENO5 reconstruction. $\Delta t = 0.2 \Delta x$.  }
\end{table}

\begin{table}[htp]
	\small
	\begin{center}
		\def\temptablewidth{1\textwidth}
		{\rule{\temptablewidth}{1pt}}
		\begin{tabular*}{\temptablewidth}{@{\extracolsep{\fill}}c|cc|cc|cc}
			
			mesh length & $L^1$ error & Order & $L^2$ error & Order& $L^{\infty}$ error & Order  \\
			\hline
1/5 & 2.190375e-02 & ~ & 2.404960e-02 & ~ & 3.365065e-02 & ~ \\
1/10 & 8.607461e-04 & 4.67 & 9.753283e-04 & 4.62 & 1.388495e-03 & 4.60 \\
1/20 & 2.859334e-05 & 4.91 & 3.177448e-05 & 4.94 & 4.681808e-05 & 4.89 \\
1/40 & 9.036721e-07 & 4.98 & 1.000905e-06 & 4.99 & 1.483438e-06 & 4.98 \\
1/80 & 2.826547e-08 & 5.00 & 3.132498e-08 & 5.00 & 4.650866e-08 & 5.00 \\ 				
		\end{tabular*}
		{\rule{\temptablewidth}{0.1pt}}
	\end{center}
	\vspace{-4mm} \caption{\label{accuracy-weno5-AO-weight1} Accuracy test for the 1-D sin-wave
		propagation by the new reconstruction procedure with the WENO5-AO reconstruction. $\Delta t = 0.2 \Delta x$. The linear weights are chosen as $\gamma_{Hi}=0.85,\gamma_{Lo}=0.85$.  }
\end{table}

\begin{table}[htp]
	\small
	\begin{center}
		\def\temptablewidth{1\textwidth}
		{\rule{\temptablewidth}{1pt}}
		\begin{tabular*}{\temptablewidth}{@{\extracolsep{\fill}}c|cc|cc|cc}
			
			mesh length & $L^1$ error & Order & $L^2$ error & Order& $L^{\infty}$ error & Order  \\
			\hline
1/5 & 2.397029e-02 & ~ & 2.643570e-02 & ~ & 3.664496e-02 & ~ \\
1/10 & 9.827463e-04 & 4.61 & 1.121792e-03 & 4.56 & 1.842517e-03 & 4.31 \\
1/20 & 2.927870e-05 & 5.07 & 3.225416e-05 & 5.12 & 4.527186e-05 & 5.35 \\
1/40 & 9.050562e-07 & 5.02 & 1.002233e-06 & 5.01 & 1.481292e-06 & 4.93 \\
1/80 & 2.826820e-08 & 5.00 & 3.132788e-08 & 5.00 & 4.650571e-08 & 4.99 \\ 				
		\end{tabular*}
		{\rule{\temptablewidth}{0.1pt}}
	\end{center}
	\vspace{-4mm} \caption{\label{accuracy-weno5-AO-weight2} Accuracy test for the 1-D sin-wave
		propagation  by the new reconstruction procedure with the WENO5-AO reconstruction. $\Delta t = 0.2 \Delta x$. The linear weights are chosen as $\gamma_{Hi}=0.1,\gamma_{Lo}=0.1$.  }
\end{table}

\begin{table}[htp]
	\small
	\begin{center}
		\def\temptablewidth{1\textwidth}
		{\rule{\temptablewidth}{1pt}}
		\begin{tabular*}{\temptablewidth}{@{\extracolsep{\fill}}c|cc|cc|cc}
			
			mesh length & $L^1$ error & Order & $L^2$ error & Order& $L^{\infty}$ error & Order  \\
			\hline
			1/5 & 2.779405e-02 & ~ & 3.037832e-02 & ~ & 4.278560e-02 & ~ \\
			1/10 & 2.277863e-03 & 3.61 & 2.526109e-03 & 3.59 & 3.587982e-03 & 3.58 \\
			1/20 & 2.273179e-04 & 3.32 & 2.519989e-04 & 3.33 & 3.581546e-04 & 3.32 \\
			1/40 & 2.646643e-05 & 3.10 & 2.937815e-05 & 3.10 & 4.163084e-05 & 3.10 \\
			1/80 & 3.247784e-06 & 3.03 & 3.606775e-06 & 3.03 & 5.104785e-06 & 3.03 \\  			
		\end{tabular*}
		{\rule{\temptablewidth}{0.1pt}}
	\end{center}
	\vspace{-4mm} \caption{\label{accuracy-weno5-z-all-collision} Accuracy test for the 1-D sin-wave
		propagation by the new reconstruction procedure with the WENO5 reconstruction. $\Delta t = 0.2 \Delta x$.  }
\end{table}

\subsubsection{Acoustic wave}

\begin{figure}[htbp]	
	\centering
	\subfigure[]{
	\label{acoustic-wave-1}
	\includegraphics[width=0.48\textwidth]{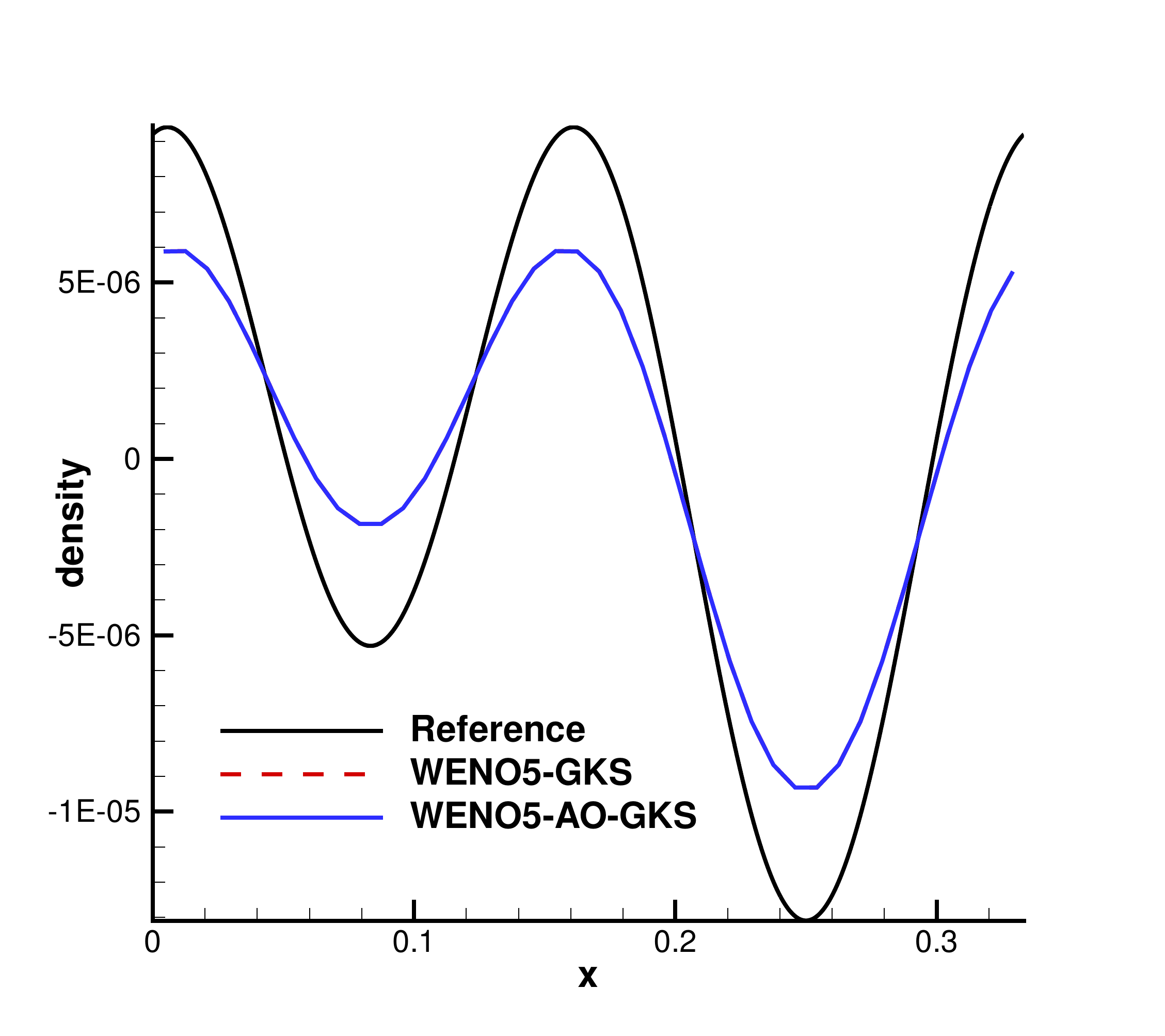}}
    \subfigure[]{
	\label{acoustic-wave-2}
	\includegraphics[width=0.48\textwidth]{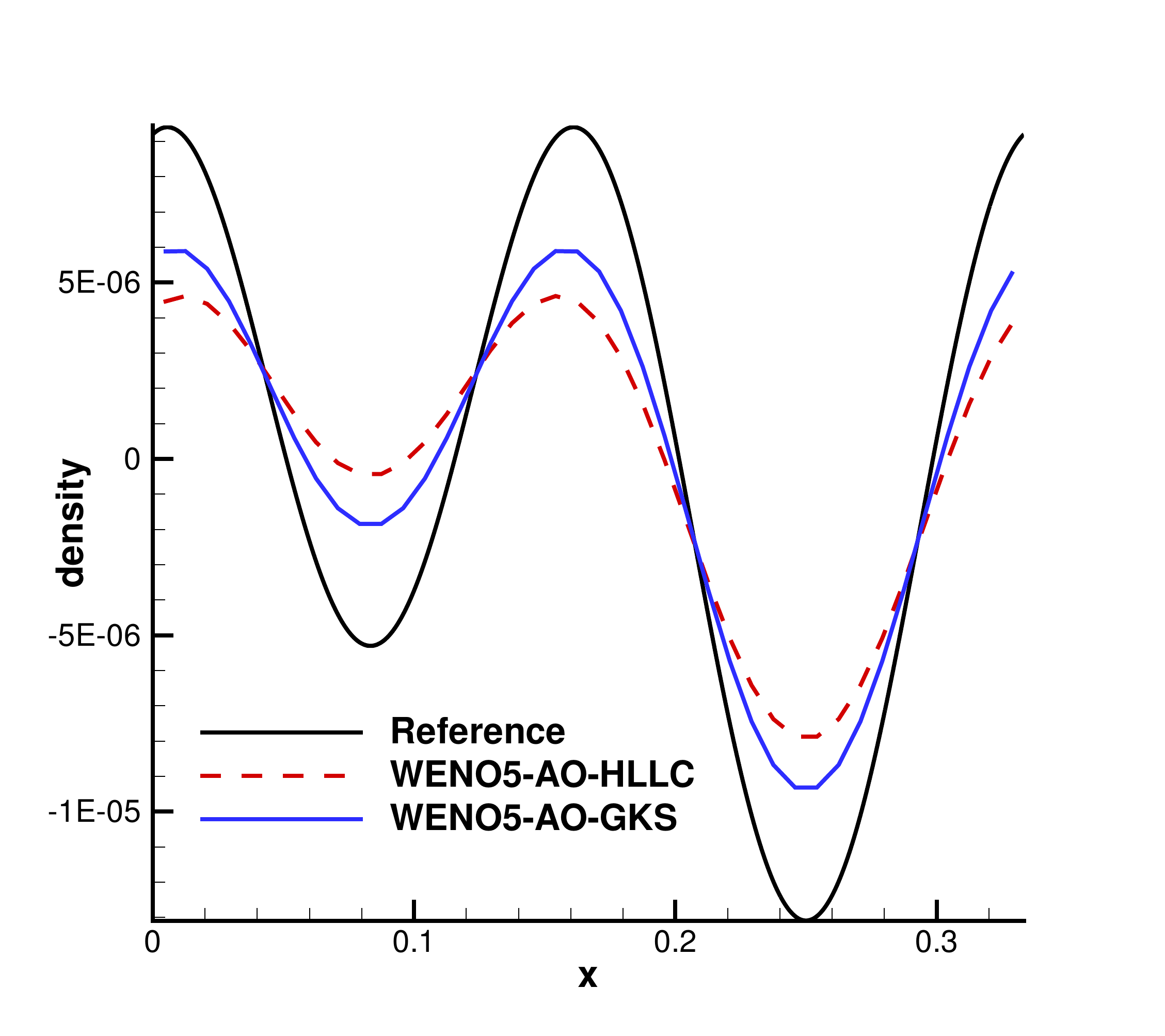}	}	
	\caption{Acoustic wave. CFL=0.5. T=1. Mesh 40. (a) comparisons between the conventional WENO-GKS and new WENO-A-GKS. (b) comparisons between different solvers.    }
	\label{acoustic-wave} 
\end{figure}

The initial conditions for a one-dimensional acoustic wave propagation  are given as follows \cite{bai2019new}
\begin{align*}
&U=U_{\infty}+\delta U, \delta U=\epsilon a_{\infty} \cos(\omega x), U_{\infty}=0\\
&\rho= \rho_{\infty}+\delta \rho,\delta \rho=\epsilon \rho_{\infty} \cos(2\omega x), \rho_{\infty}=1.1771 ,\\
&\frac{p}{p_{\infty}}=(\frac{\rho}{\rho_{\infty}})^r, p_{\infty}=101325.0 ,\\
&a_{\infty}=\sqrt{\gamma\frac{p_{\infty}}{\rho_{\infty}}},
\end{align*}
where $\epsilon=10^{-5}$ is the magnitude of initial perturbation, and $\omega=6\pi$
is the wavenumber of initial perturbations in velocity. The specific heat ratio is
$\gamma =1.4$. An analytical solution \cite{bai2019new} is given from
the approximate acoustic wave equation,
\begin{align} \label{acoustic-1d-ref}
\begin{split}
\rho (x,t)=& \rho_{\infty}+ \frac{1}{2}\epsilon \rho_{\infty} [ \cos(2\omega(x-a_{\infty}t))+ \cos(2\omega(x+a_{\infty}t)) + \\
& \cos(\omega(x-a_{\infty}t))- \cos(\omega(x+a_{\infty}t))], \\
U (x,t)=& \frac{1}{2}\epsilon a_{\infty} [ \cos(2\omega(x-a_{\infty}t))- \cos(2\omega(x+a_{\infty}t)) + \\
& \cos(\omega(x-a_{\infty}t))+ \cos(\omega(x+a_{\infty}t))], \\
p (x,t)=& p_{\infty}+ \frac{1}{2}\gamma\epsilon p_{\infty} [ \cos(2\omega(x-a_{\infty}t))+ \cos(2\omega(x+a_{\infty}t)) + \\
& \cos(\omega(x-a_{\infty}t))- \cos(\omega(x+a_{\infty}t))],
\end{split}
\end{align}
with the period $T=\frac{2\pi}{\omega a_{\infty}} \approx 9.6\times 10^{-4}$.
The computational domain is $[0,1/3]$ with periodic boundary conditions on both sides.
We follow the recipe in \cite{zhao2019compact} for the numerical initialization.
The numerical results after the wave propagates about 1,000 periods at $t=1$ are used for comparison as shown in Fig. \ref{acoustic-wave}.
From Fig.\ref{acoustic-wave-1}, the conventional fifth-order WENO-GKS and new WENO-AO-GKS have almost the same long time behavior.
The relative error from both schemes is no more than $0.001\%$ at each cell.
It demonstrates that the usage of the kinetic-style weighting to reconstruct the slopes of equilibrium state in Section \ref{new-1-d-step-2} instead of the pure linear interpolation in Section \ref{1-d-step-4} will not introduce extra numerical dissipation.
It is observed from Fig.\ref{acoustic-wave-2} that under the same WENO-AO reconstruction, the gas-kinetic solver has significant superiority over the schemes based on the Riemann solvers, such as HLLC in such a smooth case.

\subsubsection{One dimensional Riemann problems}
The reference solutions for the following 1-D Riemann problems are obtained by WENO5-GKS with 10000 uniform mesh points.

\noindent{\sl{(a) Sod problem}}

\begin{figure}[htbp]	
	\centering
	\includegraphics[width=0.46\textwidth]{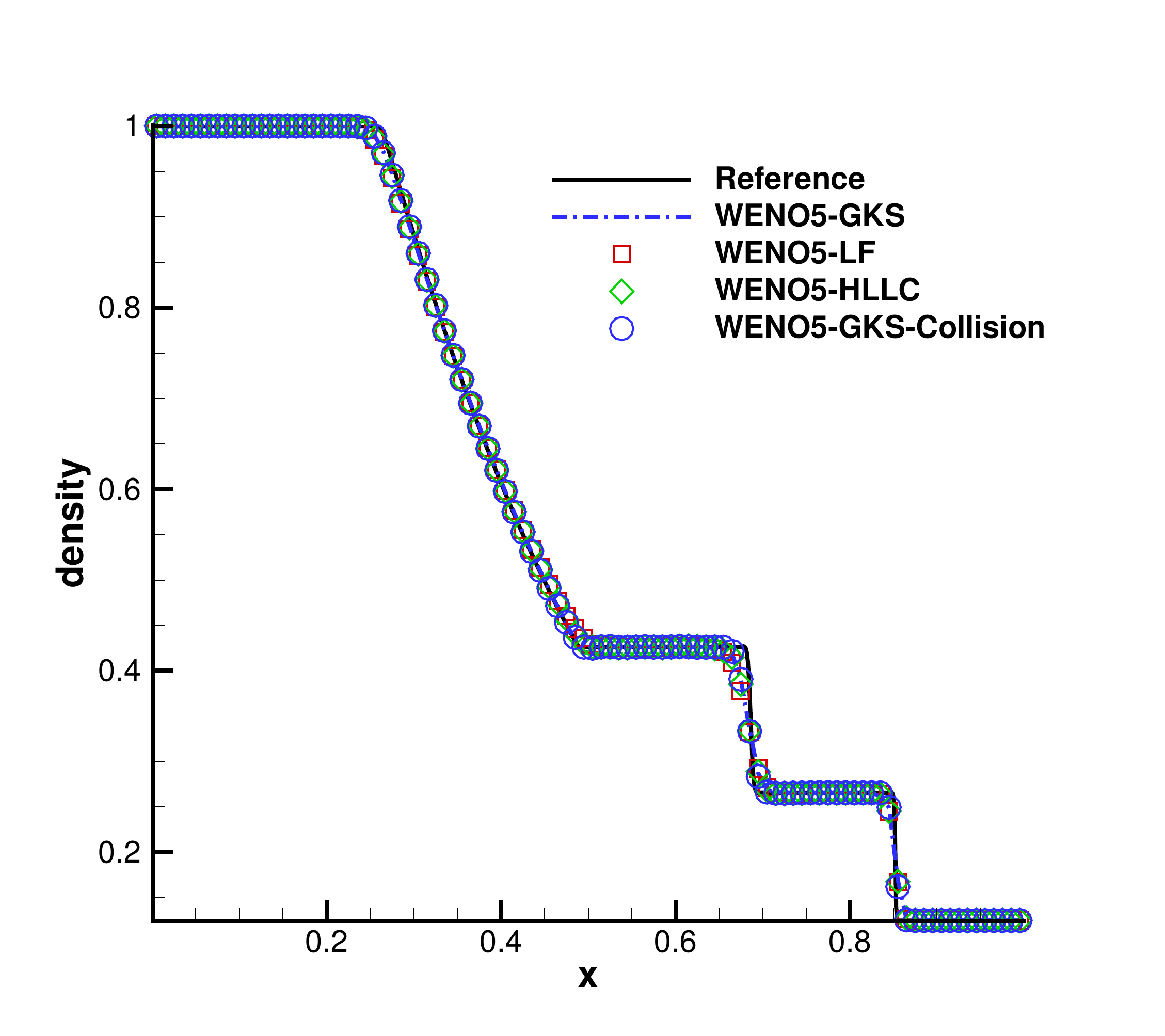}
	\includegraphics[width=0.46\textwidth]{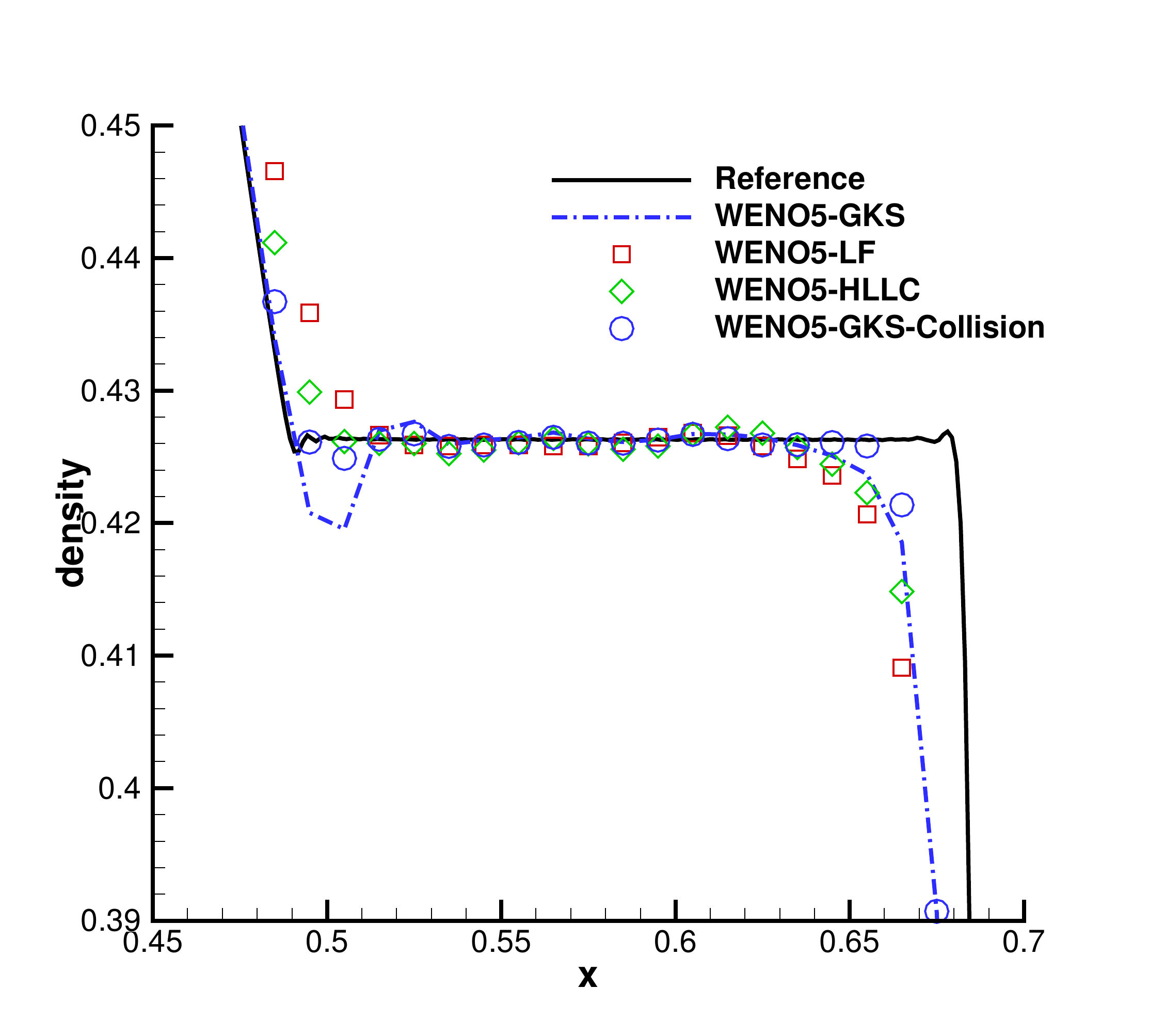}
	\caption{Sod problem: the density distributions and local enlargements with 100 cells.
		     The comparisons are conducted between
		     high-order Riemann solver based methods and GKS. CFL=0.5. T=0.2.  }
	\label{sod-rm-compare} 
\end{figure}

\begin{figure}[htbp]	
	\centering
	\includegraphics[width=0.46\textwidth]{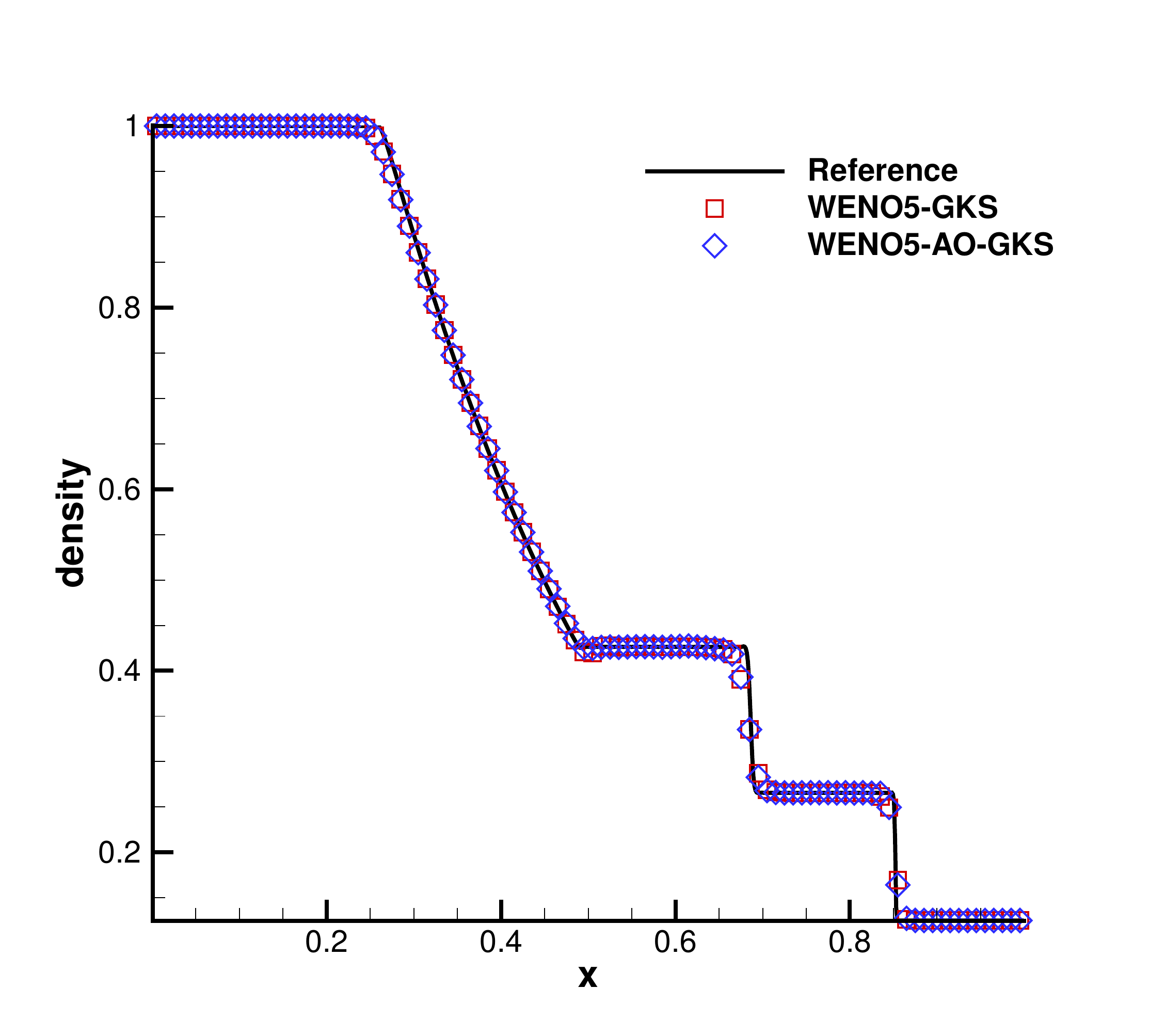}
	\includegraphics[width=0.46\textwidth]{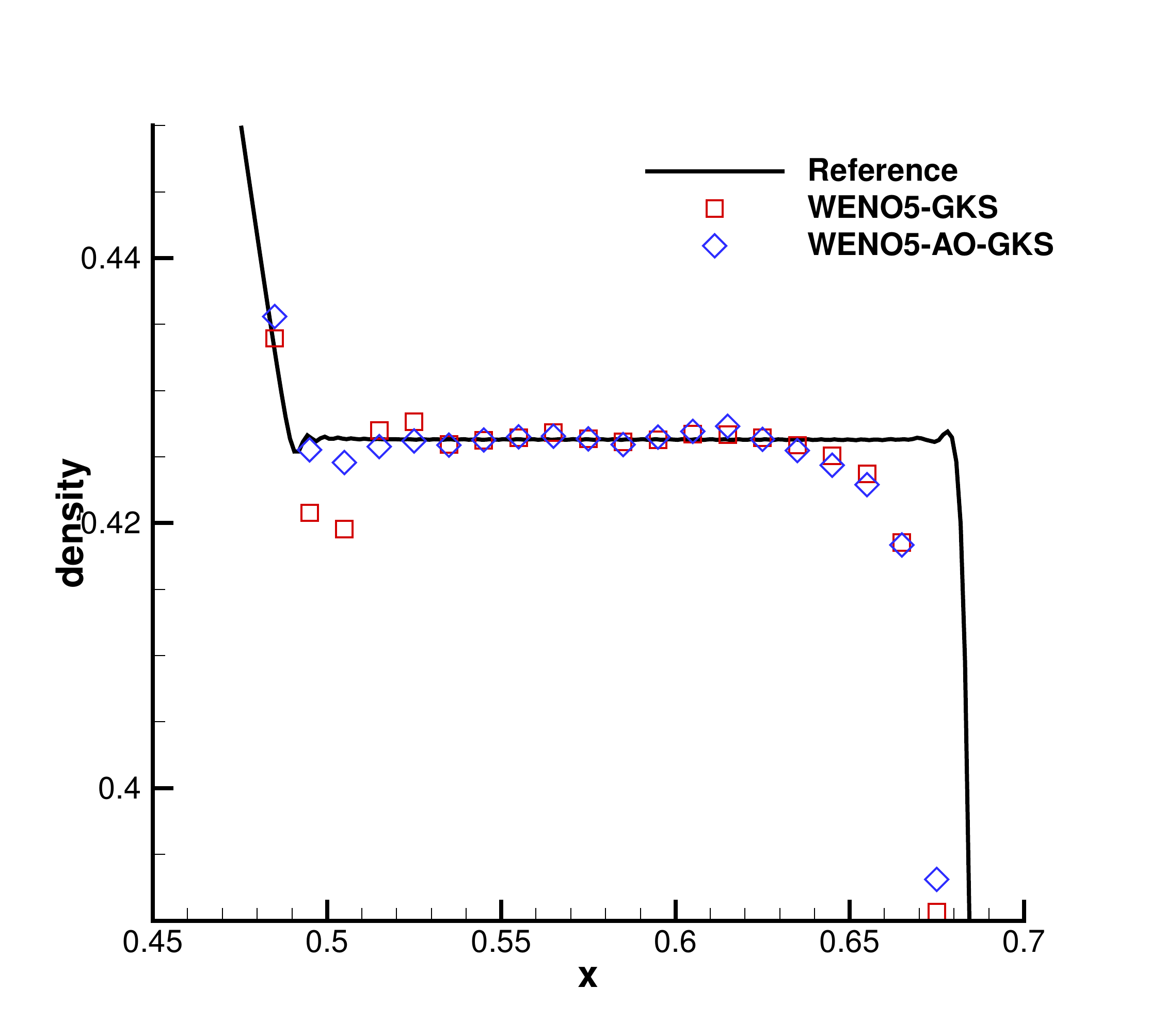}
	\caption{Sod problem: the density distributions  and local enlargements with 100 cells
		     by the conventional WENO5-GKS and new WENO5-AO-GKS. CFL=0.5. T=0.2.  }
	\label{sod-weno-compare} 
\end{figure}

The initial conditions for the Sod problem are given by
\begin{equation*}
(\rho,U,p)=\left\{\begin{aligned}
&(1, 0, 1), 0<x<0.5,\\
&(0.125,0,0.1),  0.5 \leq x<1,
\end{aligned} \right.
\end{equation*}
where $100$ uniform mesh points are used in simulation
and the solutions are presented at $t=0.2$.
We first compare the results by high-order Riemann solver-based methods with the conventional HGKS.
The same WENO5-Z reconstruction is used for all cases.
From the local enlargements in Fig. \ref{sod-rm-compare}, the solutions from the Riemann solvers have almost no undershoot around the corner of the rarefaction wave while the conventional WENO5-GKS has an observable oscillation.
Moreover, if the construction of the equilibrium state in Section \ref{1-d-step-4}  for the conventional WENO5-GKS is replaced
by the kinetic-weighting method in Section \ref{new-1-d-step-2} (named as ``WENO5-GKS-Collision"),
where the upwind mechanics is introduced in the determination of $ g^{c}_x$, the oscillation could be significantly reduced.
Similarly the undershoot is essentially eliminated for the new WENO5-AO GKS as shown in Fig. \ref{sod-weno-compare}.

\noindent{\sl{(b) Shu-Osher problem}}

The second test is the Shu-Osher problem \cite{shu1989efficient}, and
the initial conditions are
\begin{align*}
(\rho,U,p) =\begin{cases}
(3.857134, 2.629369, 10.33333), &  0<x \leq 1,\\
(1 + 0.2\sin (5x), 0, 1),  &  1 <x<10.
\end{cases}
\end{align*}

\begin{figure}[htbp]	
	\centering
	\subfigure[]{
		\label{shu-osher-1}
		\includegraphics[width=0.48\textwidth]{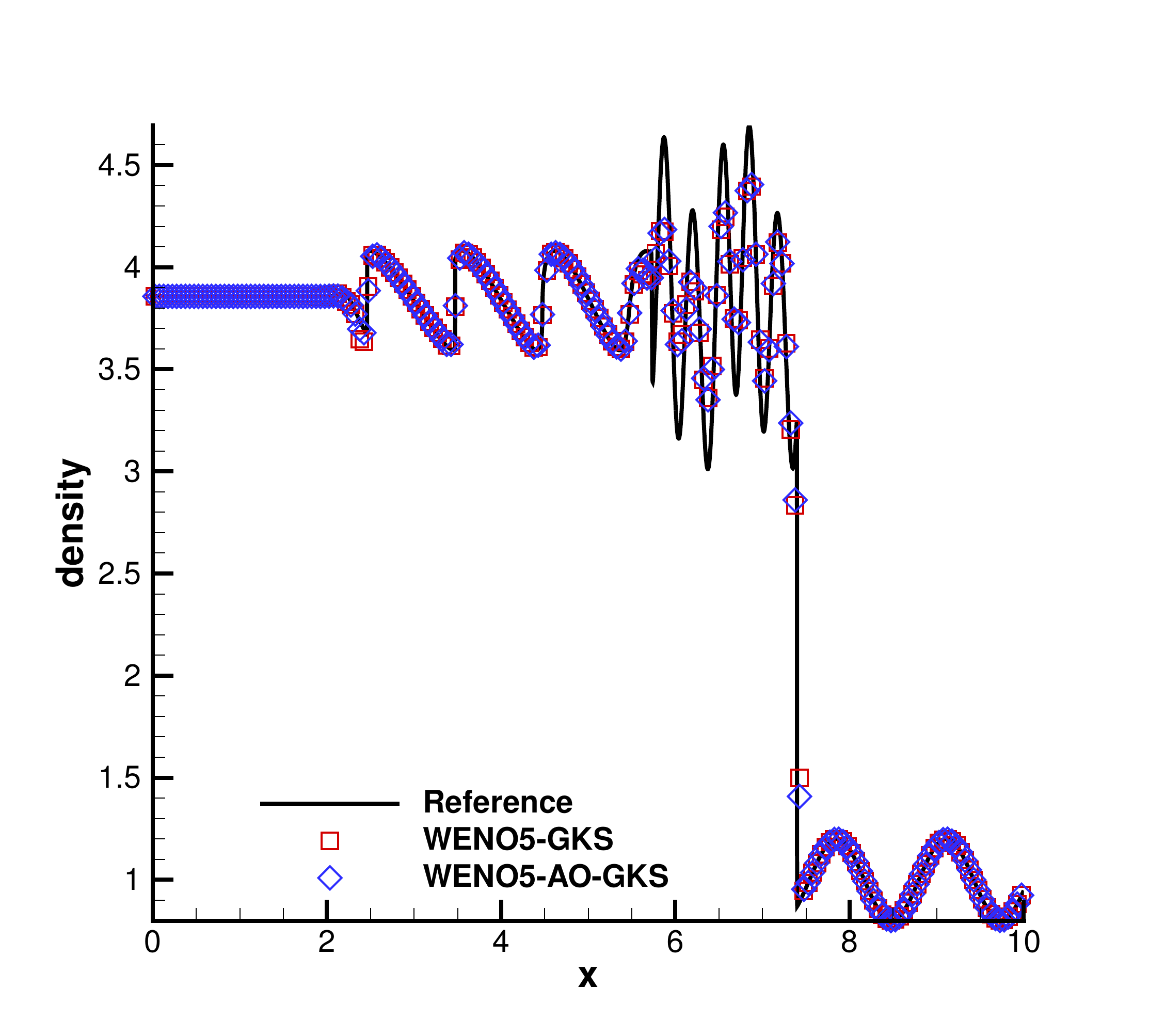}
		\includegraphics[width=0.48\textwidth]{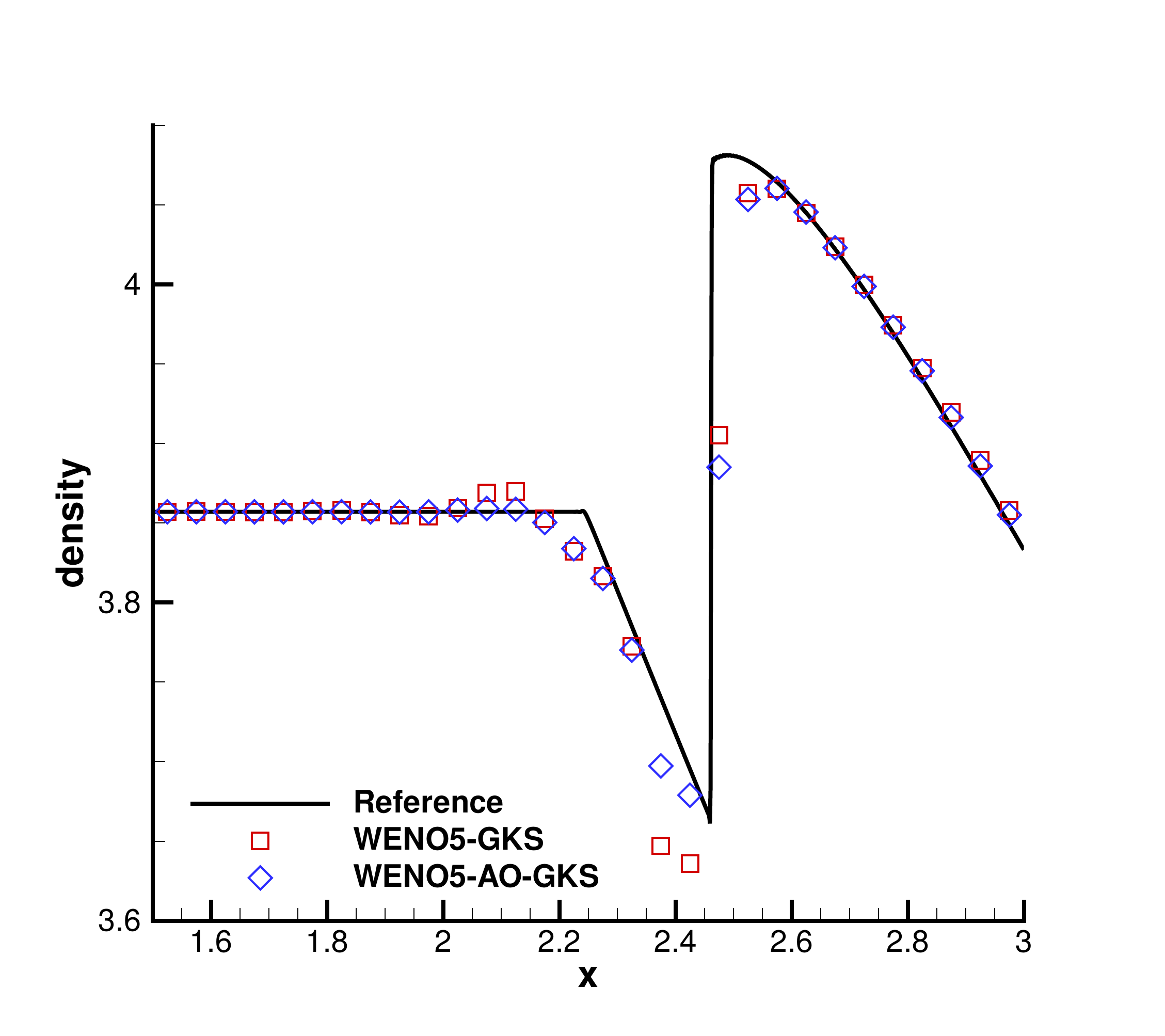}
	}
	\subfigure[]{
		\label{shu-osher-2}
		\includegraphics[width=0.48\textwidth]{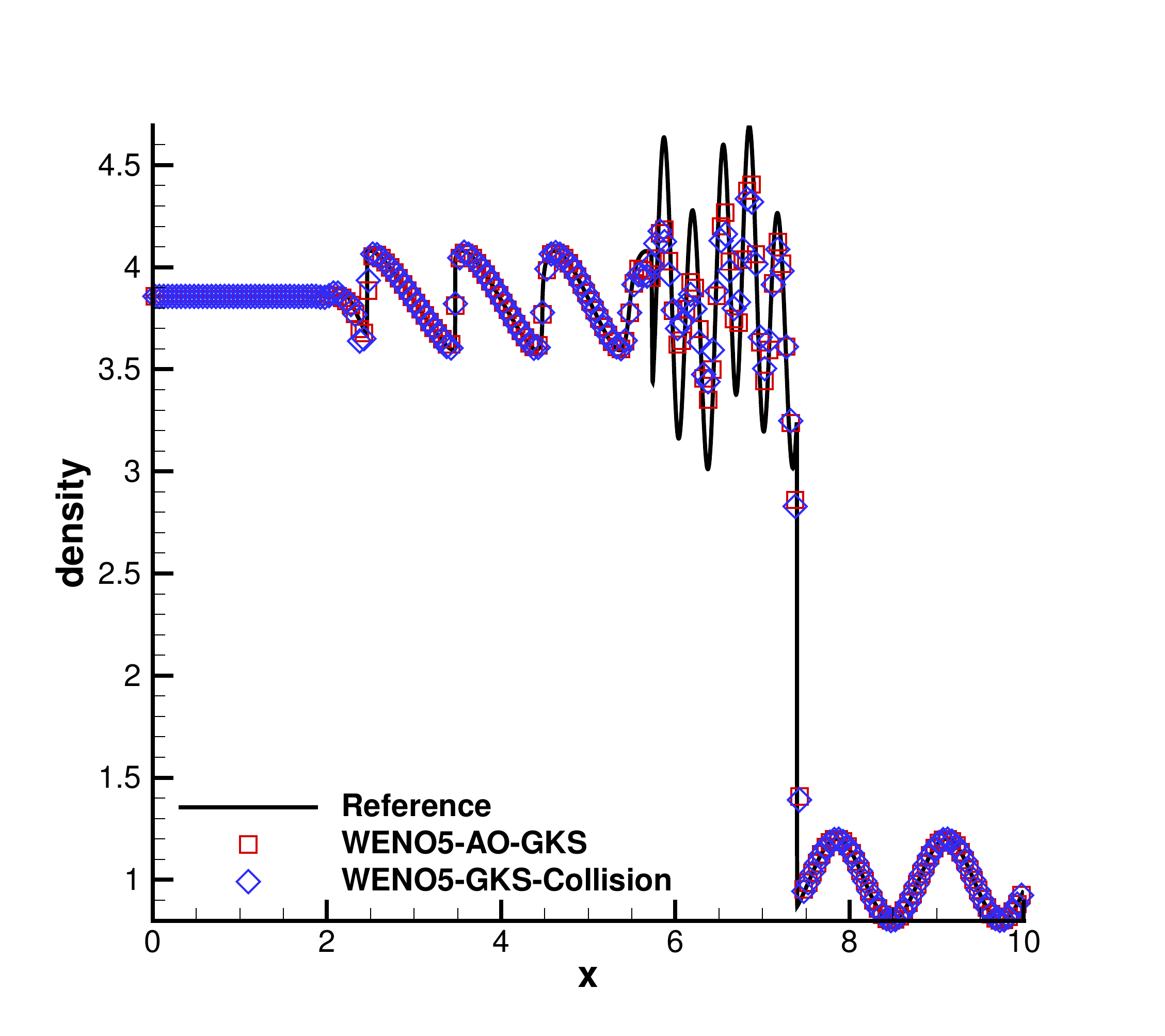}
		\includegraphics[width=0.48\textwidth]{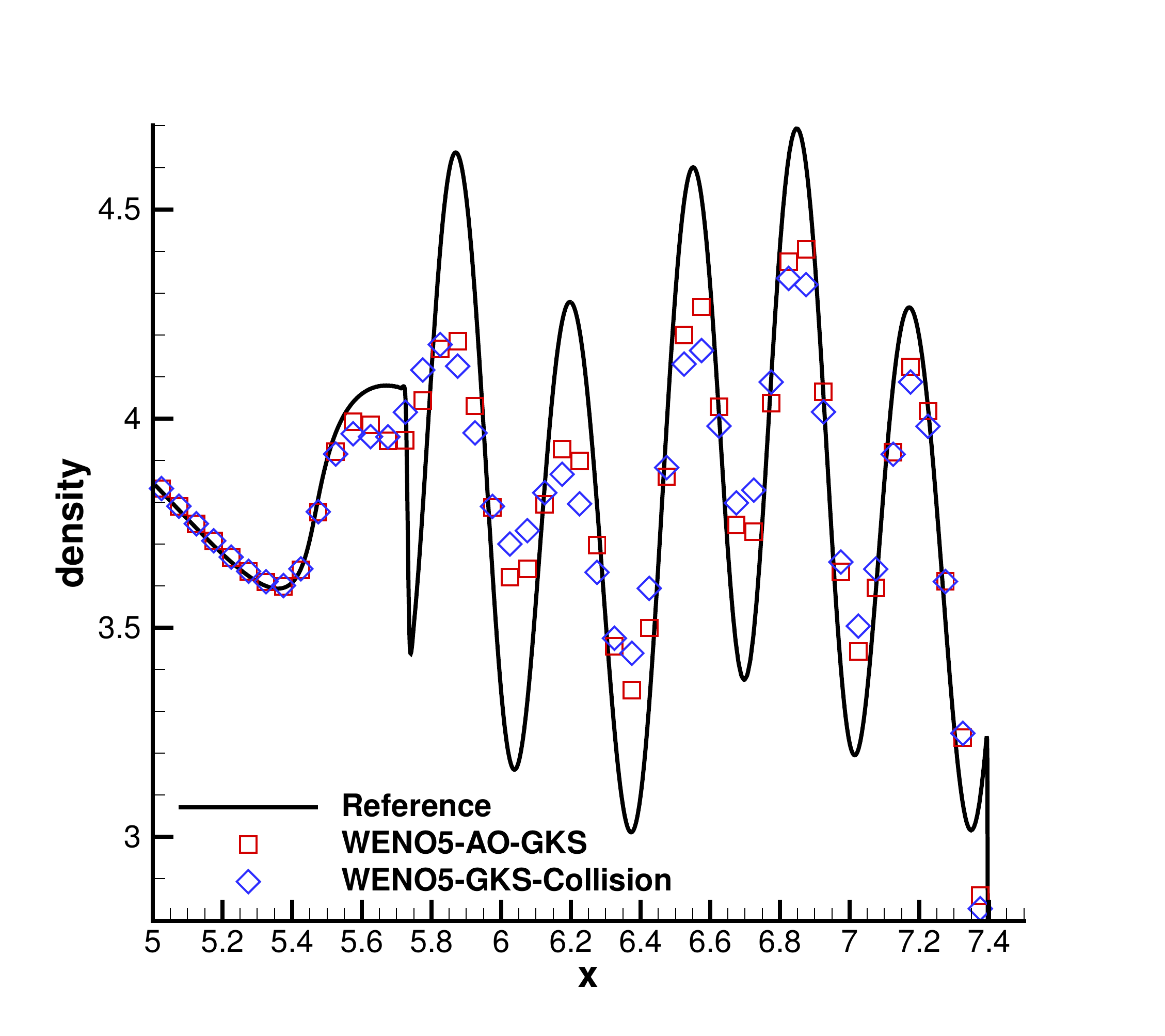}
	}	
	\caption{Shu-Osher problem: the density distributions and local enlargements with 200 cells.  CFL=0.5. T=1.8.}
\end{figure}

The computational domain is $[0, 10]$.
The non-reflecting boundary condition is given on the left, and the
fixed wave profile is extended on the right.
The computed density profiles and local enlargements for the Shu-Osher problem with $200$ mesh
points at $t = 1.8$ are shown in Fig. \ref{shu-osher-1}. The performances of the conventional HGKS with
WENO5 reconstruction and new HGKS with WENO5-AO reconstruction are almost identical in resolving the sinusoidal wave on the right.
However, the WENO5-GKS yields spurious oscillations in the local enlargements around $x=2.4$.
Moreover, if the kinetic-weighting treatment is applied for the equilibrium state, the resolutions for the smooth wave from HGKS would be significantly reduced, as shown in Fig. \ref{shu-osher-2}.
In contrast, the linear wave could be resolved nicely by the new HGKS while no overshoot/undershoot occurs in the corresponding locations.

\noindent{\sl{(c) Blast wave problem}}

\begin{figure}[htbp]	
	\centering
	\includegraphics[width=0.46\textwidth]{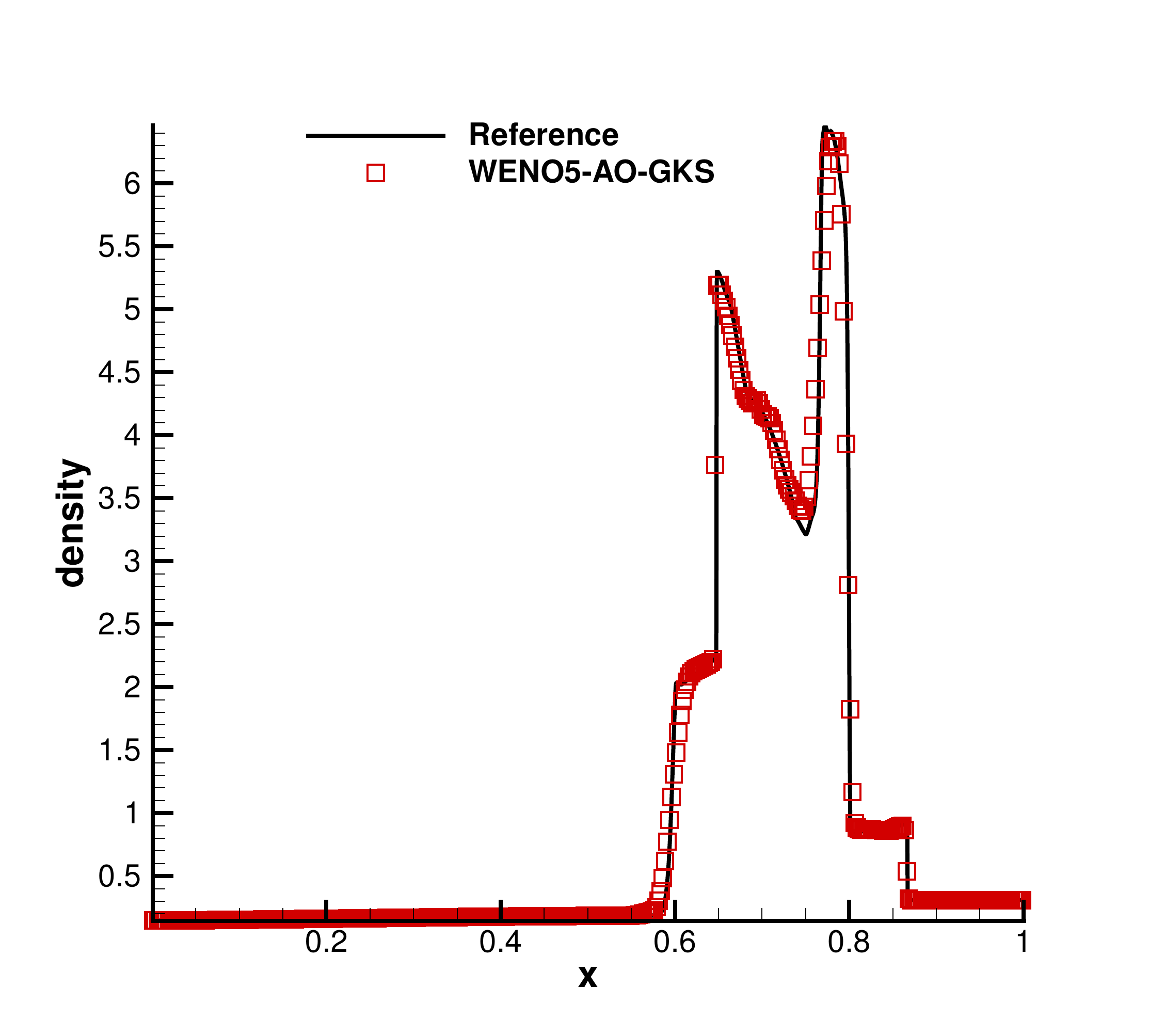}
	\includegraphics[width=0.46\textwidth]{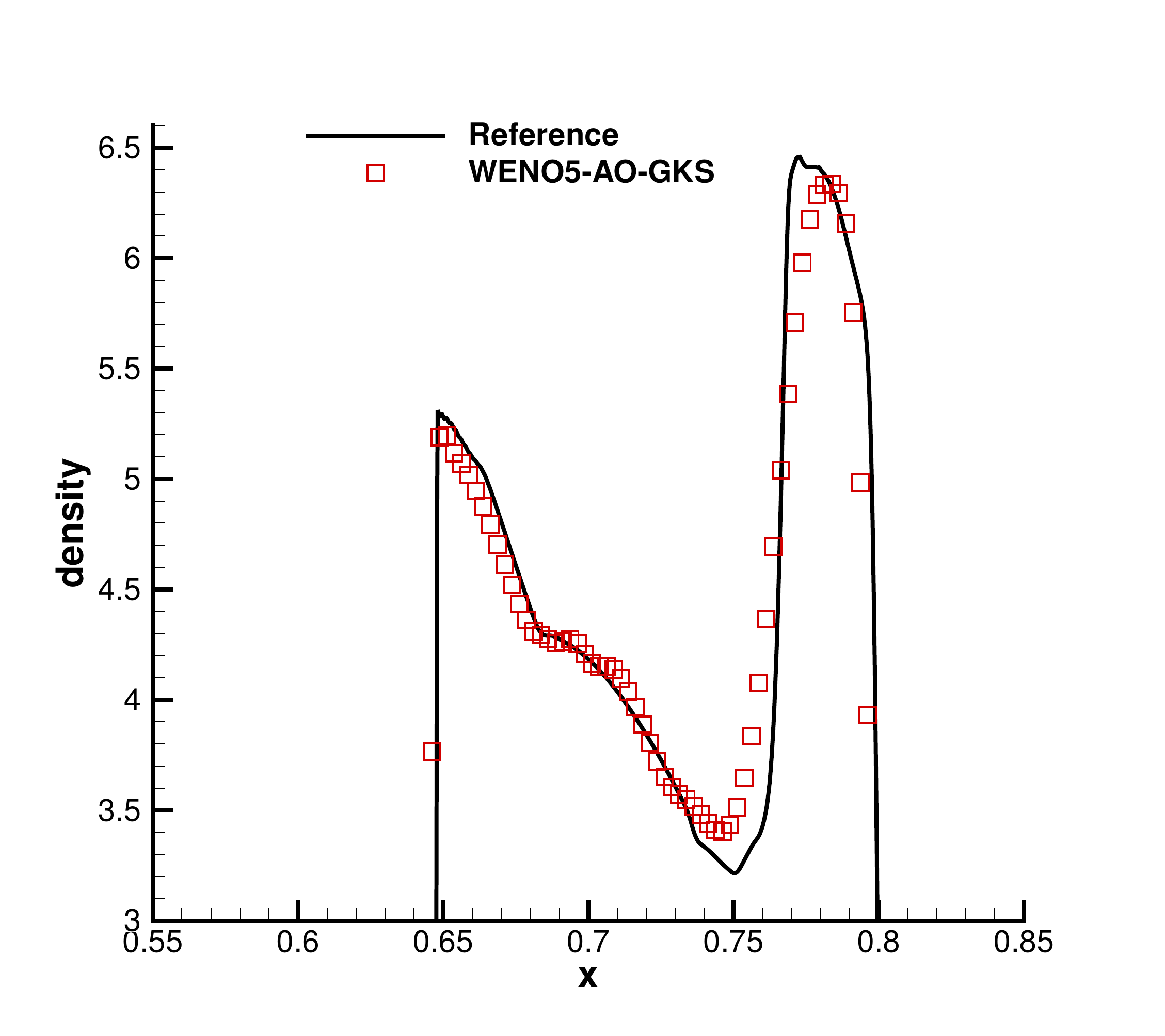}
	\caption{Blast wave problem: the density distribution and local enlargement with 400 cells. CFL=0.5. T=0.038. }
	\label{blastwave} 
\end{figure}

The initial conditions for the blast wave problem \cite{woodward1984numerical} are given as follows
\begin{equation*}
(\rho,u,p)=\left\{\begin{aligned}
&(1, 0, 1000), 0\leq x<0.1,\\
&(1, 0, 0.01), 0.1\leq x<0.9,\\
&(1, 0, 100),  0.9\leq x\leq 1.
\end{aligned} \right.
\end{equation*}
$400$ equal spaced cells are used for computation and reflection boundary conditions are applied at both sides.
The density distribution and local enlargements for the new HGKS at $t=0.038$ are presented in Fig.\ref{blastwave}.
The numerical collision time takes $c_1=0$ and $c_2=1$ and $ CFL =0.5$ as usual.
The traditional HGKS with WENO5-Z reconstruction could not pass this case with the above settings $c_1 =0$.

\subsection{2-D test cases}

\subsubsection{Accuracy test in 2-D}

Similar to 1-D case, the advection of density perturbation for the inviscid flow is chosen for accuracy test with the initial conditions

\begin{align*}
\rho(x,y)=1+0.2\sin(\pi (x+y)),\ \ \ \textbf{U}(x,y)=(1,1),  \ \ \  p(x,y)=1,
\end{align*}
within a square domain $[0, 2]\times[0, 2]$. $N \times N$ uniform mesh cells are used and the periodic boundary conditions
are applied in both directions.
The analytic solution is
\begin{align*}
\rho(x,y,t)=1+0.2\sin(\pi(x+y-t)),\ \ \  \textbf{U}(x,y,t)=(1,1),\ \ \  p(x,y,t)=1.
\end{align*}
The time steps are calculated with a $CFL=0.5$.
Both WENO5-GKS and WENO5-AO-GKS are tested with the linear weights in Table \ref{accuracy-weno5-linear-2d} and Table \ref{accuracy-weno5-ao-linear-2d}.   The results for the new method with non-linear Z-type weights are shown in Table \ref{accuracy-weno5-ao-z-2d}. The expected accuracy could be achieved for all cases.

\begin{table}[htbp]
	\small
	\begin{center}
		\def\temptablewidth{1\textwidth}
		{\rule{\temptablewidth}{1pt}}
		\begin{tabular*}{\temptablewidth}{@{\extracolsep{\fill}}c|cc|cc|cc}
			
			mesh length & $L^1$ error & Order & $L^2$ error & Order& $L^{\infty}$ error & Order  \\
			\hline
1/5 & 3.074801e-02 & ~ & 3.439108e-02 & ~ & 4.765930e-02 & ~ \\
1/10 & 1.320626e-03 & 4.54 & 1.453774e-03 & 4.56 & 2.064342e-03 & 4.53 \\
1/20 & 4.240666e-05 & 4.96 & 4.726867e-05 & 4.94 & 6.900348e-05 & 4.90 \\
1/40 & 1.377120e-06 & 4.94 & 1.529072e-06 & 4.95 & 2.235296e-06 & 4.95 \\
1/80 & 4.771096e-08 & 4.85 & 5.307786e-08 & 4.85 & 7.646075e-08 & 4.87 \\ 				
		\end{tabular*}
		{\rule{\temptablewidth}{0.1pt}}
	\end{center}
	\vspace{-4mm} \caption{\label{accuracy-weno5-linear-2d} Accuracy test for the 2-D sin-wave
		propagation by the linear WENO5 reconstruction. $CFL=0.5$.  }
\end{table}

\begin{table}[htbp]
	\small
	\begin{center}
		\def\temptablewidth{1\textwidth}
		{\rule{\temptablewidth}{1pt}}
		\begin{tabular*}{\temptablewidth}{@{\extracolsep{\fill}}c|cc|cc|cc}
			
			mesh length & $L^1$ error & Order & $L^2$ error & Order& $L^{\infty}$ error & Order  \\
			\hline
1/5 & 3.081177e-02 & ~ & 3.446190e-02 & ~ & 4.771748e-02 & ~ \\
1/10 & 1.322377e-03 & 4.54 & 1.455598e-03 & 4.57 & 2.074481e-03 & 4.52 \\
1/20 & 4.245123e-05 & 4.96 & 4.729387e-05 & 4.94 & 6.915239e-05 & 4.91 \\
1/40 & 1.377684e-06 & 4.95 & 1.529528e-06 & 4.95 & 2.237970e-06 & 4.95 \\
1/80 & 4.772228e-08 & 4.85 & 5.308750e-08 & 4.85 & 7.651685e-08 & 4.87 \\ 				
		\end{tabular*}
		{\rule{\temptablewidth}{0.1pt}}
	\end{center}
	\vspace{-4mm} \caption{\label{accuracy-weno5-ao-linear-2d} Accuracy test for the 2-D sin-wave
		propagation by the linear WENO5-AO reconstruction. $CFL=0.5$.  }
\end{table}

\begin{table}[!h]
	\small
	\begin{center}
		\def\temptablewidth{1\textwidth}
		{\rule{\temptablewidth}{1pt}}
		\begin{tabular*}{\temptablewidth}{@{\extracolsep{\fill}}c|cc|cc|cc}
			
			mesh length & $L^1$ error & Order & $L^2$ error & Order& $L^{\infty}$ error & Order  \\
			\hline
1/5 & 3.514097e-02 & ~ & 3.834933e-02 & ~ & 5.407085e-02 & ~ \\
1/10 & 1.359913e-03 & 4.69 & 1.489563e-03 & 4.69 & 2.108134e-03 & 4.68 \\
1/20 & 4.254036e-05 & 5.00 & 4.737104e-05 & 4.97 & 6.916339e-05 & 4.93 \\
1/40 & 1.377826e-06 & 4.95 & 1.529673e-06 & 4.95 & 2.238071e-06 & 4.95 \\
1/80 & 4.772252e-08 & 4.85 & 5.308776e-08 & 4.85 & 7.651741e-08 & 4.87 \\				
		\end{tabular*}
		{\rule{\temptablewidth}{0.1pt}}
	\end{center}
	\vspace{-4mm} \caption{\label{accuracy-weno5-ao-z-2d} Accuracy test for the 2-D sin-wave
		propagation by the WENO5-AO reconstruction. $CFL=0.5$.  }
\end{table}

\subsubsection{Two dimensional Riemann problems}

The two dimensional Riemann problems \cite{2dRM} are widely used
to check the performance of a scheme for high speed compressible flow. The computational domain is $[0,1]\times [0,1]$ and uniform meshes with mesh size $1/500$ are used.

\noindent{\sl{(a) Configuration 1}}

\begin{figure}[!h]	
	\centering
	\includegraphics[width=0.46\textwidth]{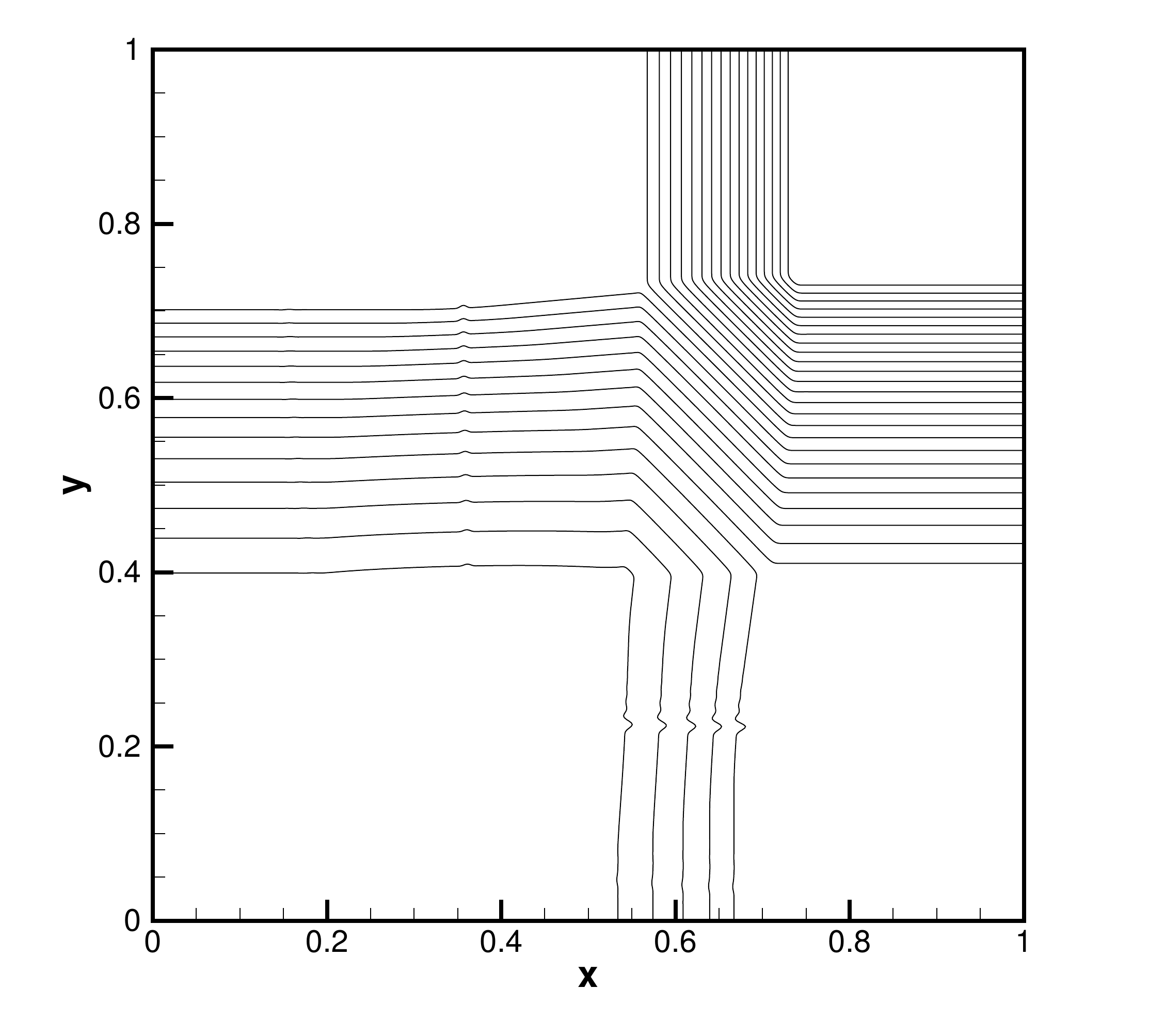}
	\includegraphics[width=0.46\textwidth]{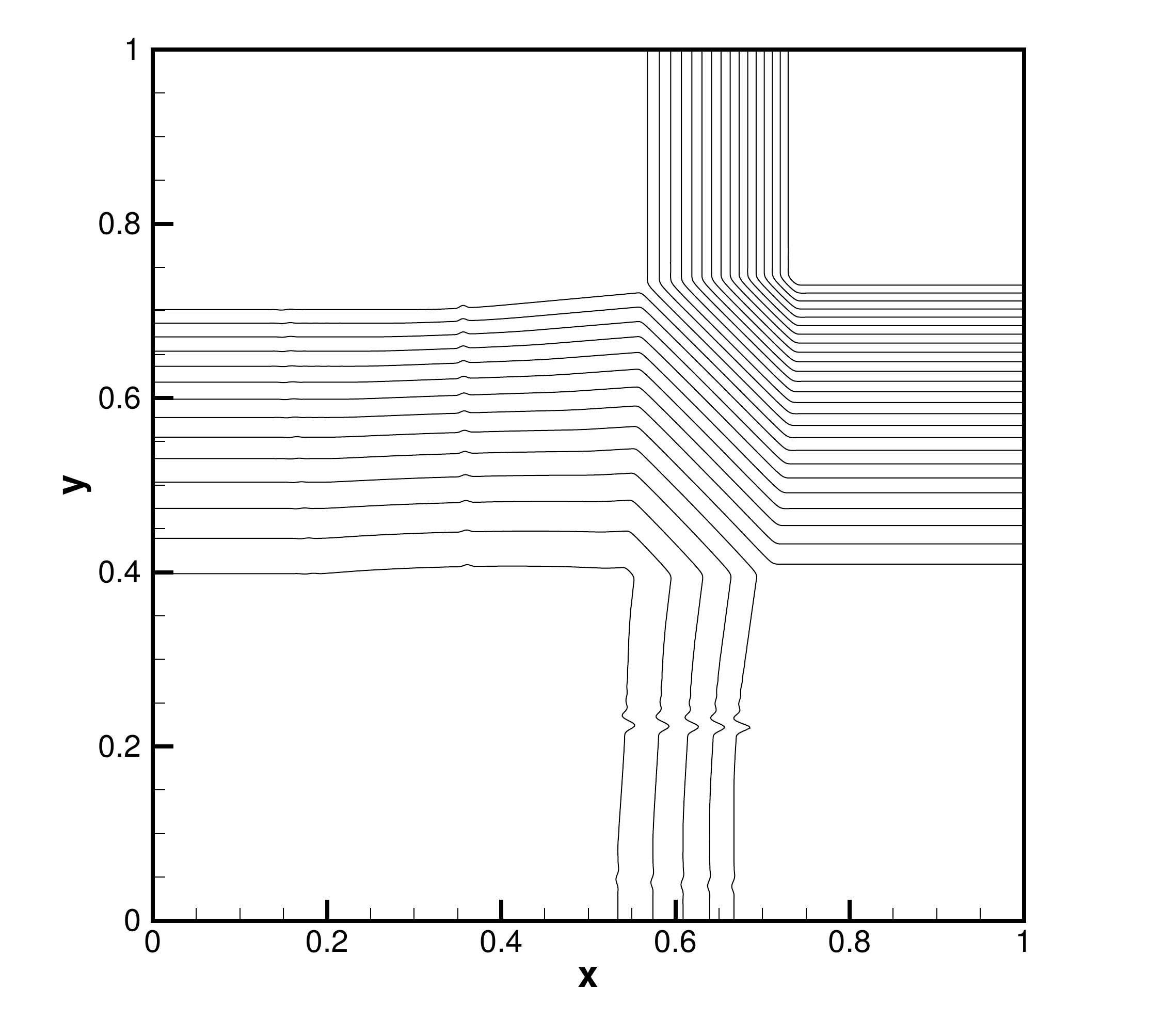}
	\caption{Two dimensional Riemann problems: the density distributions for Configuration 1. Left: the conventional WENO5-GKS. Right: the new WENO5-AO-GKS. CFL=0.5. T=0.2. Mesh: $500\times500$.}
	\label{rm-four-rarefaction}
\end{figure}

The Configuration 1 in \cite{2dRM} is tested.
Initially, there are four 1-D rarefaction waves imposed as
\begin{equation*}
(\rho,U_1,U_2,p)=\left\{\begin{aligned}
&(0.1072, -0.7259,-1.4045, 0.0439),& x<0.5,y<0.5,\\
&(0.2579, 0,-1.4045, 0.15),& x\geq 0.5,y<0.5,\\
&(1, 0,0, 1),& x\geq 0.5,y\geq 0.5,\\
&(0.5197,-0.7259,0, 0.4),& x< 0.5,y\geq 0.5.
\end{aligned} \right.
\end{equation*}
The results at t=0.2 for the original scheme and new one are given in Fig. \ref{rm-four-rarefaction}.
Although the discontinuities are weak,
the separated smooth tangential reconstructions as given in Sub-subsection \ref{2-d-reconstuction} for $ g^{c}$ may lead to a negative temperature.
As a result, a protector must be added in the original HGKS,
which replaces the reconstructed values by the first-order reconstruction if there is negative temperature detected after performing the high-order reconstruction.
By contrast, no such problem exists in the new scheme.
It partially explains why the new method is more robust than the previous one in high-dimensional case,
especially for the cases where the flow fields are extremely chaotic.
One outstanding example is the high-speed compressible isotropic turbulence as shown latter.

\noindent{\sl{(b) Configuration 6}}

\begin{figure}[!h]	
	\centering
	\includegraphics[width=0.46\textwidth]{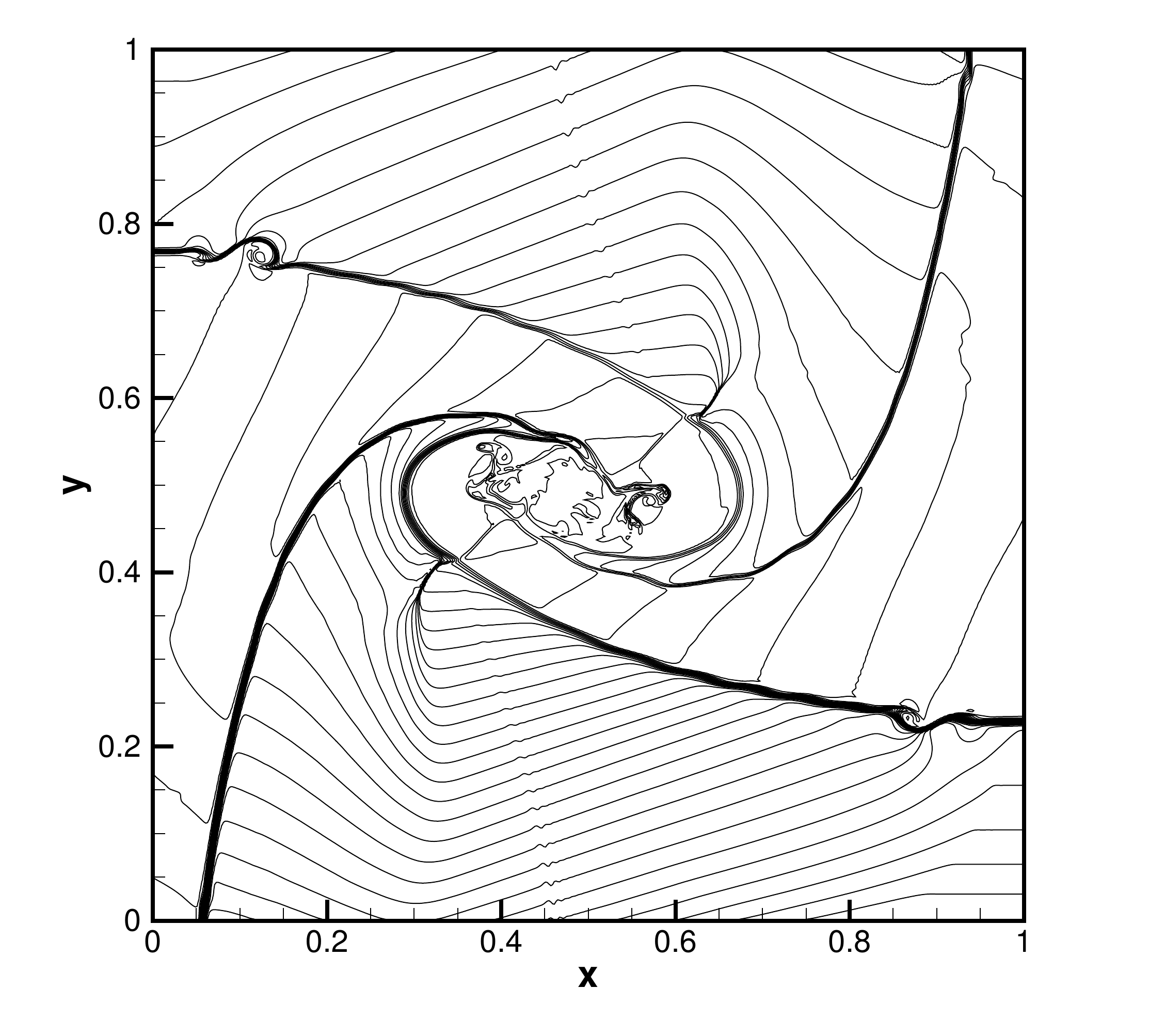}
	\includegraphics[width=0.46\textwidth]{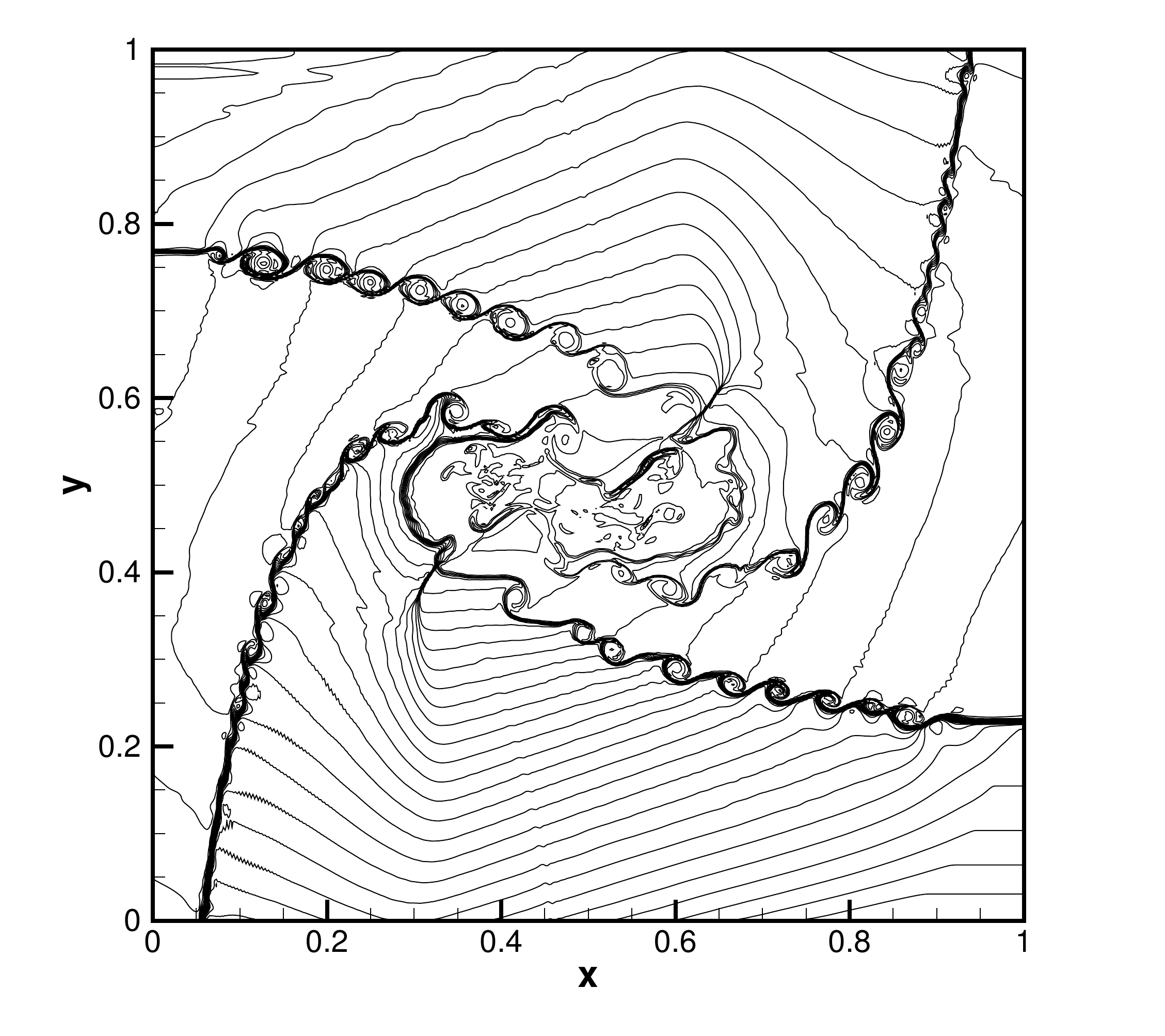}
	\caption{Two dimensional Riemann problems: the density distributions for Configuration 6. Left: the conventional WENO5-GKS. Right: the new WENO5-AO-GKS. CFL=0.95. T=0.6. Mesh: $500\times500$.  }
	\label{rm-four-jump} 
\end{figure}

The shear layer is one of the most distinguishable flow pattern for compressible flow.
Considering an ideal case, the initial conditions of Configuration 6 for four planar contact discontinuities in\cite{2dRM} are given by
\begin{equation*}
(\rho,U_1,U_2,p)=\left\{\begin{aligned}
&(1, -0.75,0.5, 1),& x<0.5,y<0.5,\\
&(3, -0.75,-0.5, 1),& x\geq 0.5,y<0.5,\\
&(1, 0.75,-0.5, 1),& x\geq 0.5,y\geq 0.5,\\
&(2,0.75,0.5, 1),& x< 0.5,y\geq 0.5,
\end{aligned} \right.
\end{equation*}
Induced by these discontinuities,
the K-H instabilities will be triggered due to the numerical viscosities.
It is commonly believed that the less numerical dissipation corresponds to larger amplitude shear instabilities  \cite{san2015rm-compare-kh-instability}.
It can be clearly observed in Fig. \ref{rm-four-jump} that
the new WENO-AO-GKS presents more vortices than the conventional WENO-GKS.
The higher order accuracy for the initial non-equilibrium states in the new reconstruction reduces the numerical dissipation.

\subsubsection{Double Mach reflection}
Designed by Woodward and Colella \cite{woodward1984numerical} the inviscid double Mach reflection problem
is widely chosen for testing the robustness of high-order methods.
The computational domain is $[0,4]\times[0,1]$.
Initially a
right-moving Mach $10$ shock with a $60^\circ$ angle against the x-axis is positioned at $(x,y)=(1/6, 0)$.
The initial pre-shock and
post-shock conditions are
\begin{align*}
(\rho,U_1,U_2, p)&=(8, 4.125\sqrt{3}, -4.125,
116.5),\\
(\rho,U_1,U_2, p)&=(1.4, 0, 0, 1).
\end{align*}
The slip boundary condition is used at the wall starting from $x =1/6$.
The post-shock condition is set for the rest of bottom boundary.
At the top boundary, the values of ghost cells follows the motion of the Mach $10$ shock.

\begin{figure}[!h]	
	\centering
	\includegraphics[height=0.24\textwidth]{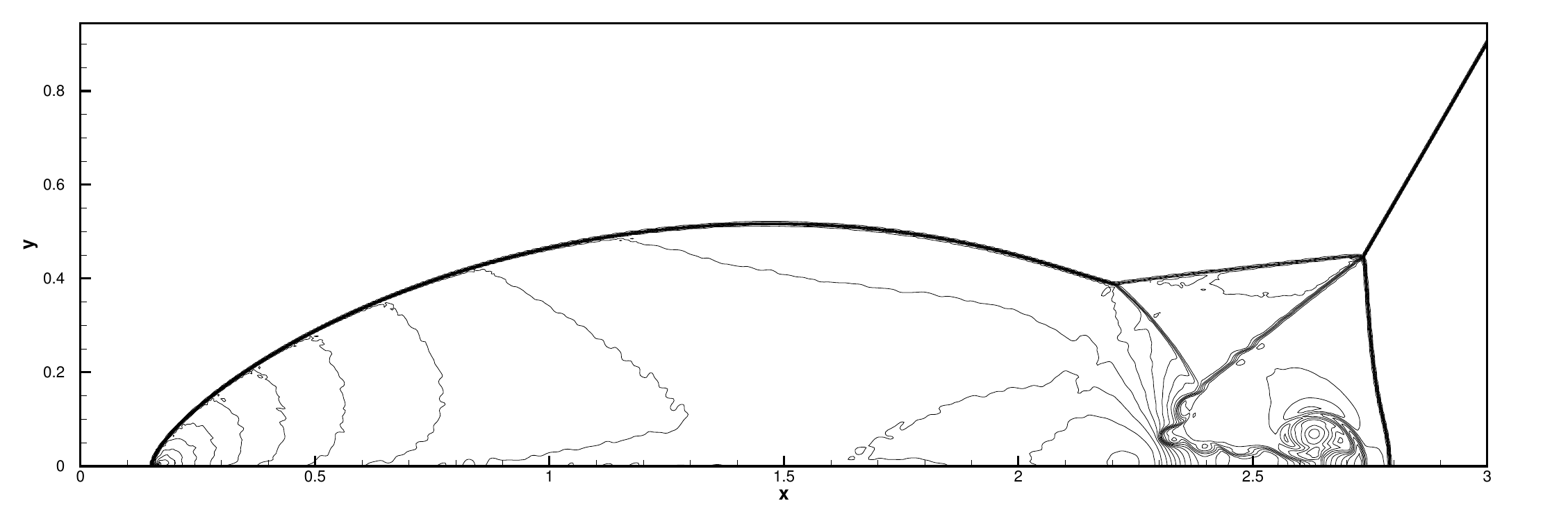}
	\includegraphics[height=0.24\textwidth]{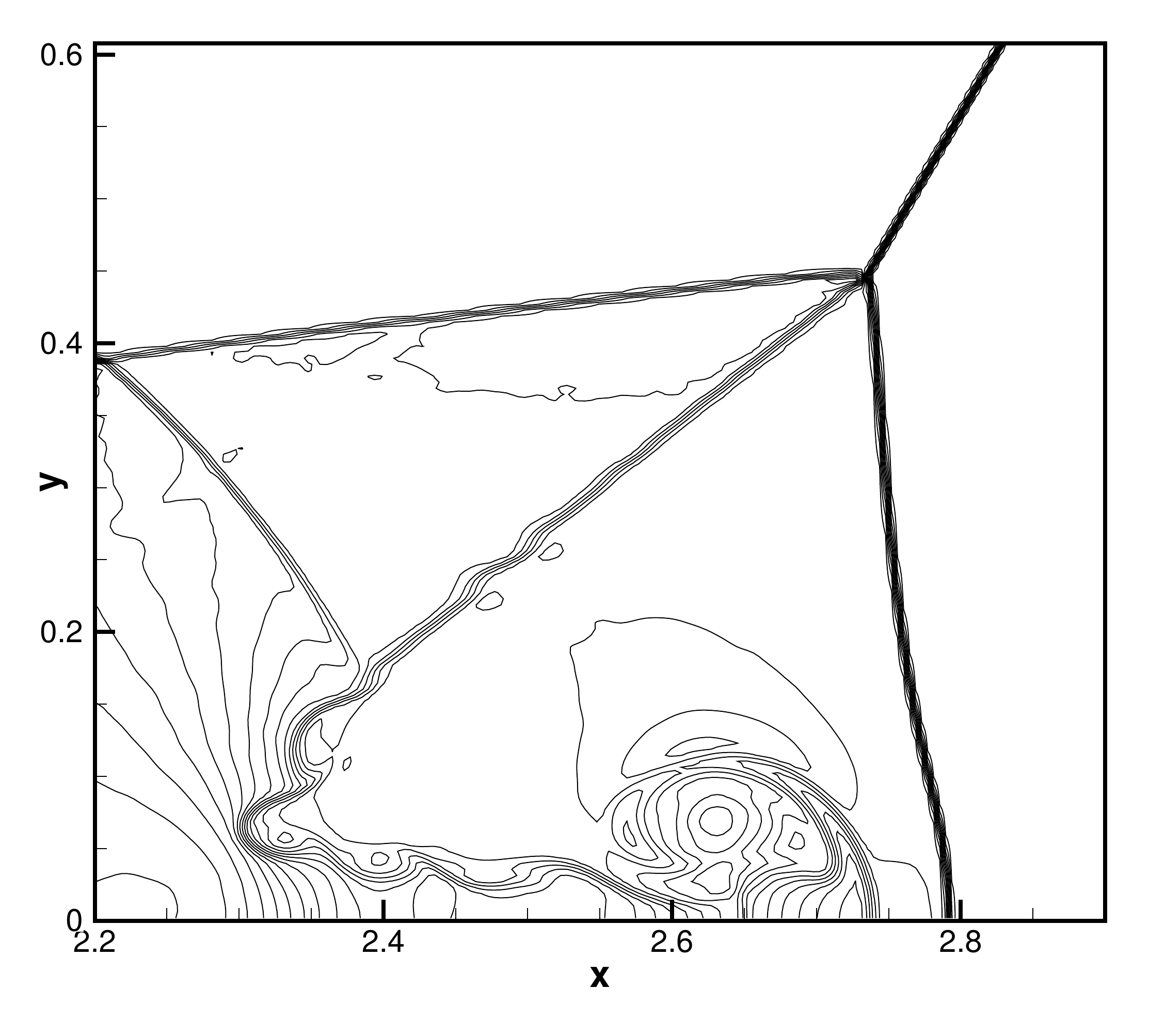}
	\caption{Double Mach. Mesh $960 \times 240$. CFL=0.5. $c_1=0.2$, $c_2=1$  }
	\label{double-mach-240-1}  
\end{figure}

\begin{figure}[!h]	
	\centering
	\includegraphics[height=0.24\textwidth]{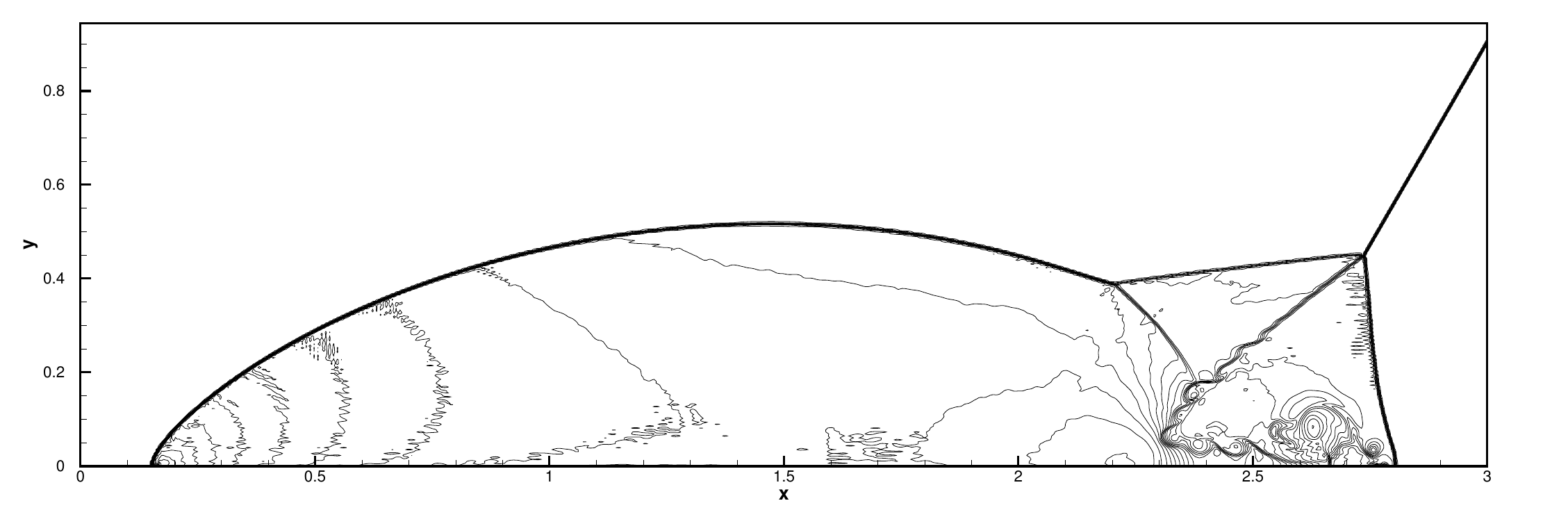}
	\includegraphics[height=0.24\textwidth]{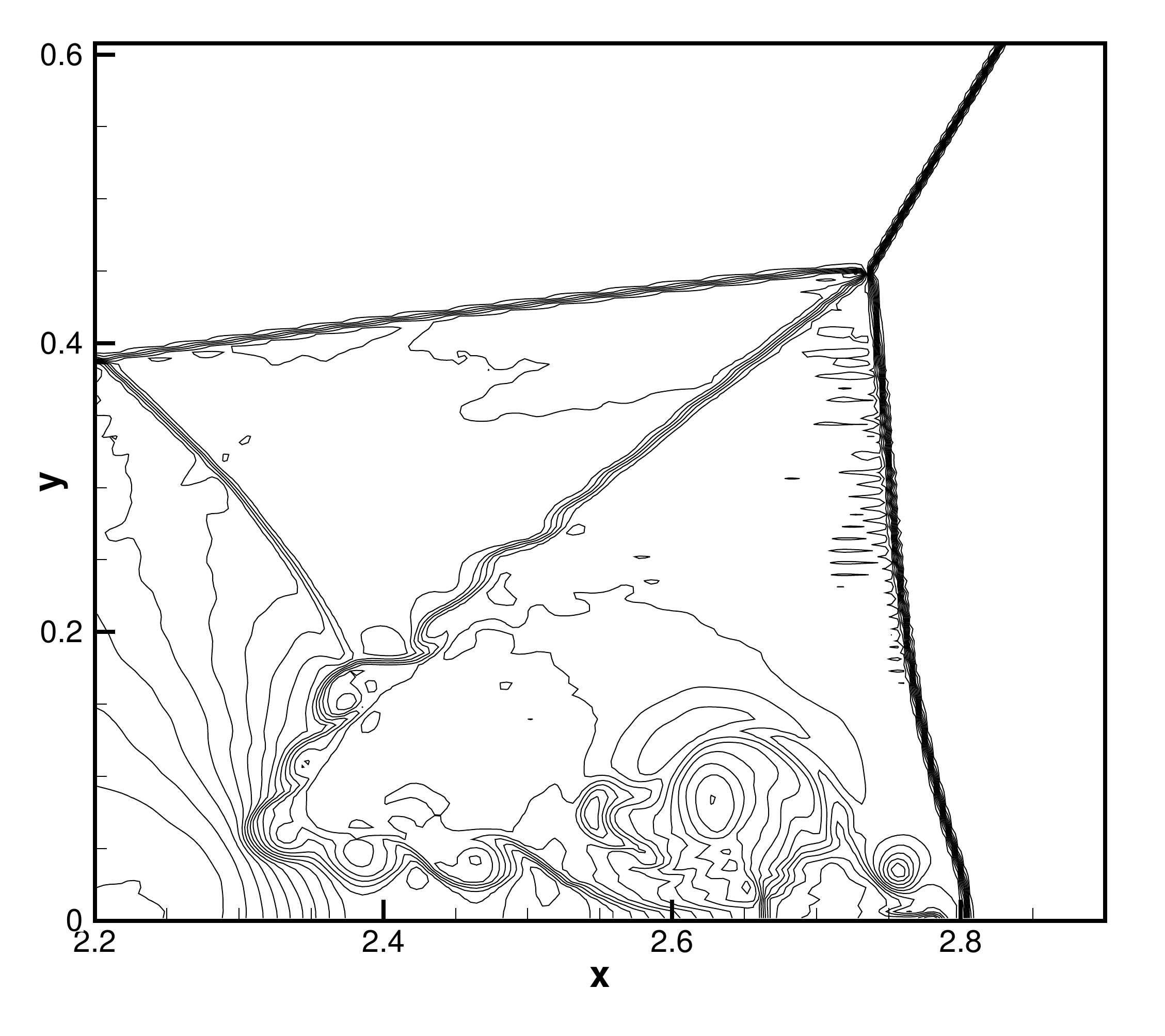}
	\caption{Double Mach. Mesh $960 \times 240$. CFL=0.8. $c_1=0$, $c_2=1$}
	\label{double-mach-240-2} 
\end{figure}

The density distributions and local enlargements with $960\times240$ uniform
mesh points at $t=0.2$ for the new method are shown in
Fig. \ref{double-mach-240-1} and Fig. \ref{double-mach-240-2}.
Suitable numerical viscosities could be added to suppress the spurious oscillations as shown in Fig. \ref{double-mach-240-1}.
The robustness of the new GKS is well validated with increasing the CFL number to 0.8,
whereas the previous WENO-Z GKS could not survive under such a large time step.

\subsubsection{Visous shock tube}
\begin{figure}[htbp]	
	\centering
		\includegraphics[width=1\textwidth]{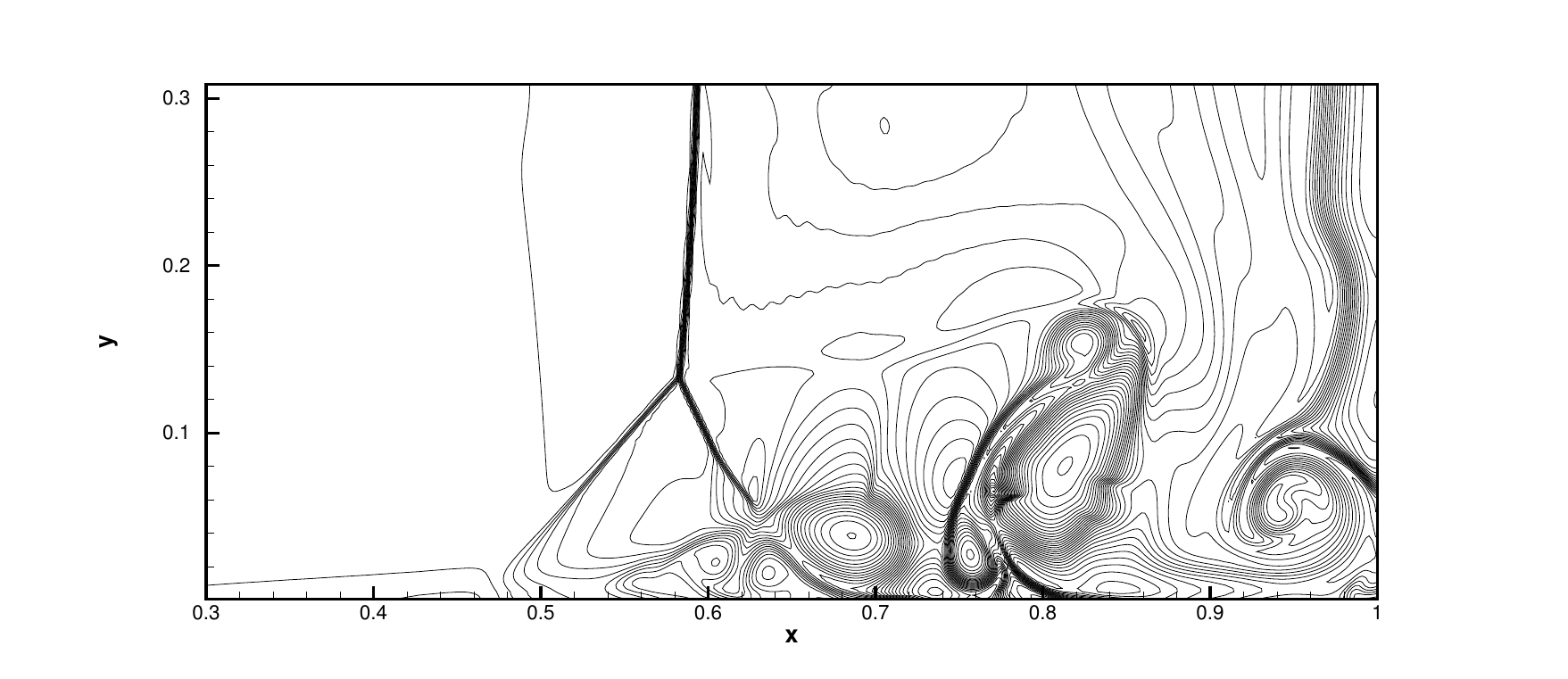}
	\caption{The density contours at t=1 for $Re=200$ viscous shock tube.  CFL=0.3. Mesh: $500 \times 250$.}
	\label{viscous-tube-re200-contour}
\end{figure}

\begin{figure}[htbp]	
	\centering
		\label{local-2}
		\includegraphics[width=1\textwidth]{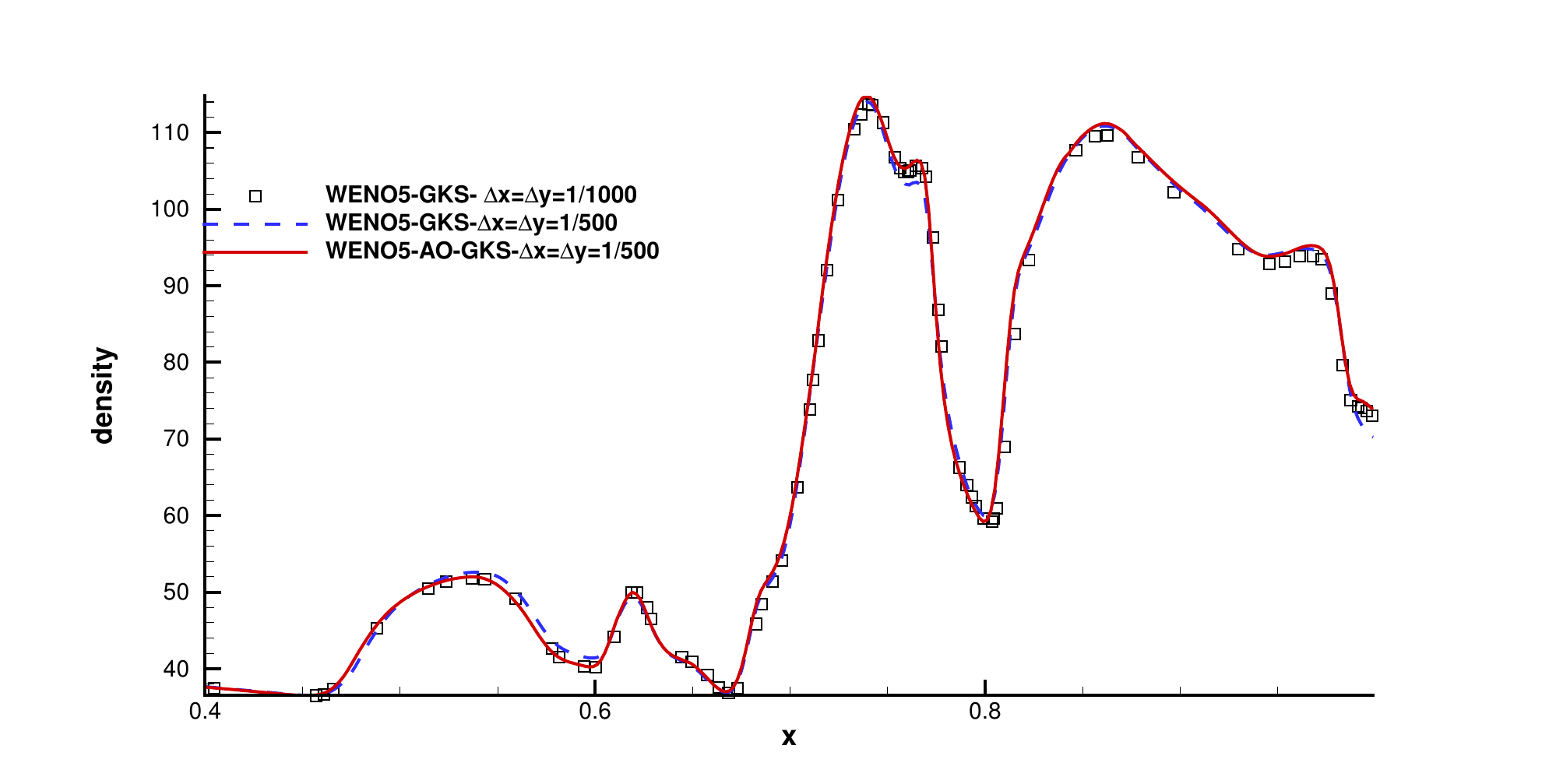}
	\caption{The density profiles along the bottom wall at t=1 for the $Re=200$ viscous shock tube. }
	\label{viscous-tube-re200-line}
\end{figure}

A viscous shock tube \cite{vistube} is used to test the performance of the new scheme for low-Reynolds number viscous flow with strong shocks.
An ideal gas is at rest in a two-dimensional unit box $[0,1]\times[0,1]$. A
membrane located at $x=0.5$ separates two different states of the
gas and the dimensionless initial states are
\begin{equation*}
(\rho,U,p)=\left\{\begin{aligned}
&(120, 0, 120/\gamma), \ \ \ &  0<x<0.5,\\
&(1.2, 0, 1.2/\gamma),  & 0.5<x<1,
\end{aligned} \right.
\end{equation*}
where $\gamma=1.4$ and Prandtl number $Pr=1$.
The Reynolds number is $Re=1/\mu=200$.
Due to symmetry, the computational domain is $[0, 1]\times[0, 0.5]$ with a symmetric boundary condition on the top boundary $x\in[0, 1], y=0.5$.
Non-slip adiabatic conditions are imposed at the other three boundaries.
The solution will develop complex two-dimensional shock/shear/boundary-layer interactions.
The dramatic changes for velocities above the bottom wall introduce strong shear stress. This is a challenging problem for high-order schemes.
The traditional HGKS with WENOZ-type weights could barely pass this case.
The JS-weights are usually used instead.
The density distributions  with
$500\times250$ uniform mesh points at $t=1.0$ from the new WENO5-AO GKS with WENOZ-type weights are shown in
Fig. \ref{viscous-tube-re200-contour}.
The density profiles along the bottom wall are also plotted and shown in Fig. \ref{viscous-tube-re200-line}.
As a comparison, the result with a fine mesh from traditional GKS in \cite{Pan2016twostage} is presented as a reference solution.

\subsection{3-D test cases}

\subsubsection{Accuracy test in 3-D}

Again, the advection of density perturbation for the inviscid flow is chosen for accuracy test with the initial conditions

\begin{align*}
\rho(x,y,z)=1+0.2\sin(\pi (x+y+z)),\ \ \ \textbf{U}(x,y,z)=(1,1,1),  \ \ \  p(x,y,z)=1,
\end{align*}
within a cubic domain $[0, 2]\times[0, 2]\times[0, 2]$. The periodic boundary conditions are adopted in all directions while
$N \times N \times N$ uniform mesh cells are used.
The analytic solution is
\begin{align*}
\rho(x,y,z,t)=1+0.2\sin(\pi(x+y+z-t)),\ \ \ \textbf{U}(x,y,z)=(1,1,1),\ \ \  p(x,y,z,t)=1.
\end{align*}

The $CFL=0.5$ is used for computation.
The both WENO5-GKS and WENO5-AO-GKS are tested with the linear weights as shown in Table \ref{accuracy-weno5-linear-3d} and Table \ref{accuracy-weno5-ao-linear-3d}. The results for the new method with non-linear Z-type weights are shown in Table \ref{accuracy-weno5-ao-z-3d}. The expected accuracy is confirmed.

\begin{table}[htbp]
	\small
	\begin{center}
		\def\temptablewidth{1\textwidth}
		{\rule{\temptablewidth}{1pt}}
		\begin{tabular*}{\temptablewidth}{@{\extracolsep{\fill}}c|cc|cc|cc}
			
			mesh length & $L^1$ error & Order & $L^2$ error & Order& $L^{\infty}$ error & Order  \\
			\hline
1/5 & 3.663706e-02 & ~ & 4.193181e-02 & ~ & 5.653477e-02 & ~ \\
1/10 & 1.703100e-03 & 4.43 & 1.864335e-03 & 4.49 & 2.716494e-03 & 4.38 \\
1/20 & 5.736655e-05 & 4.89 & 6.379782e-05 & 4.87 & 9.157851e-05 & 4.89 \\
1/40 & 2.156770e-06 & 4.73 & 2.404525e-06 & 4.73 & 3.483348e-06 & 4.72 \\
1/80 & 1.007190e-07 & 4.42 & 1.121499e-07 & 4.42 & 1.642770e-07 & 4.41 \\ 				
		\end{tabular*}
		{\rule{\temptablewidth}{0.1pt}}
	\end{center}
	\vspace{-4mm} \caption{\label{accuracy-weno5-linear-3d} Accuracy test for the 3-D sin-wave
		propagation by the linear WENO5 reconstruction. $CFL=0.5$.  }
\end{table}

\begin{table}[htbp]
	\small
	\begin{center}
		\def\temptablewidth{1\textwidth}
		{\rule{\temptablewidth}{1pt}}
		\begin{tabular*}{\temptablewidth}{@{\extracolsep{\fill}}c|cc|cc|cc}
			
			mesh length & $L^1$ error & Order & $L^2$ error & Order& $L^{\infty}$ error & Order  \\
			\hline
1/5 & 3.670079e-02 & ~ & 4.201435e-02 & ~ & 5.672087e-02 & ~ \\
1/10 & 1.705347e-03 & 4.43 & 1.866037e-03 & 4.49 & 2.699234e-03 & 4.39 \\
1/20 & 5.741795e-05 & 4.89 & 6.382396e-05 & 4.87 & 9.178298e-05 & 4.88 \\
1/40 & 2.157640e-06 & 4.73 & 2.405206e-06 & 4.73 & 3.478134e-06 & 4.72 \\
1/80 & 1.007306e-07 & 4.42 & 1.121643e-07 & 4.42 & 1.642306e-07 & 4.40 \\ 				
		\end{tabular*}
		{\rule{\temptablewidth}{0.1pt}}
	\end{center}
	\vspace{-4mm} \caption{\label{accuracy-weno5-ao-linear-3d} Accuracy test for the 3-D sin-wave
		propagation by the linear WENO5-AO reconstruction. $CFL=0.5$.  }
\end{table}

\begin{table}[htbp]
	\small
	\begin{center}
		\def\temptablewidth{1\textwidth}
		{\rule{\temptablewidth}{1pt}}
		\begin{tabular*}{\temptablewidth}{@{\extracolsep{\fill}}c|cc|cc|cc}
			
			mesh length & $L^1$ error & Order & $L^2$ error & Order& $L^{\infty}$ error & Order  \\
			\hline
1/5 & 3.844360e-02 & ~ & 4.258855e-02 & ~ & 5.786851e-02 & ~ \\
1/10 & 1.730289e-03 & 4.47 & 1.896111e-03 & 4.49 & 2.748198e-03 & 4.40 \\
1/20 & 5.749100e-05 & 4.91 & 6.389466e-05 & 4.89 & 9.180468e-05 & 4.90 \\
1/40 & 2.157708e-06 & 4.74 & 2.405289e-06 & 4.73 & 3.478366e-06 & 4.72 \\
1/80 & 1.007306e-07 & 4.42 & 1.121644e-07 & 4.42 & 1.642308e-07 & 4.40 \\ 				
		\end{tabular*}
		{\rule{\temptablewidth}{0.1pt}}
	\end{center}
	\vspace{-4mm} \caption{\label{accuracy-weno5-ao-z-3d} Accuracy test for the 3-D sin-wave
		propagation by the WENO5-AO reconstruction. $CFL=0.5$.  }
\end{table}

\subsubsection{Three dimensional Taylor-Green vortex}
The direct numerical simulation (DNS) of a three-dimensional Taylor-Green vortex \cite{debonis2013solutions} is conducted to validate the new HGKS for nearly incompressible viscous flow.
The initial flow field  is given by
\begin{align*}
&U_1=V_0\sin(\frac{x}{L})\cos(\frac{y}{L})\cos(\frac{z}{L}),\\
&U_2=-V_0\cos(\frac{x}{L})\sin(\frac{y}{L})\cos(\frac{z}{L}),\\
&U_3=0,\\
&p=p_0+\frac{\rho_0V_0^2}{16}(\cos(\frac{2x}{L})+\cos(\frac{2y}{L}))(\cos(\frac{2z}{L})+2),
\end{align*}
within a periodic cubic box $-\pi L\leq x, y, z\leq \pi L$.
The density distribution is given by retaining the constant temperature.
In the computation, $L=1, V_0=1, \rho_0=1$, and the Mach number takes $M_0=V_0/a_s=0.1$,
where $a_s$ is the sound speed. The characteristic convective time $t_c = L/V_0$.
The specific heat ratio $\gamma=1.4$ and the Prandtl number is $Pr=1$.
Numerical simulations are conducted with Reynolds number $Re=1600$.

\begin{figure}[htbp]	
	\centering
	\subfigure[Time history of kinetic energy]{
		\label{tg-re1600-ke-ao}
		\includegraphics[width=0.48\textwidth]{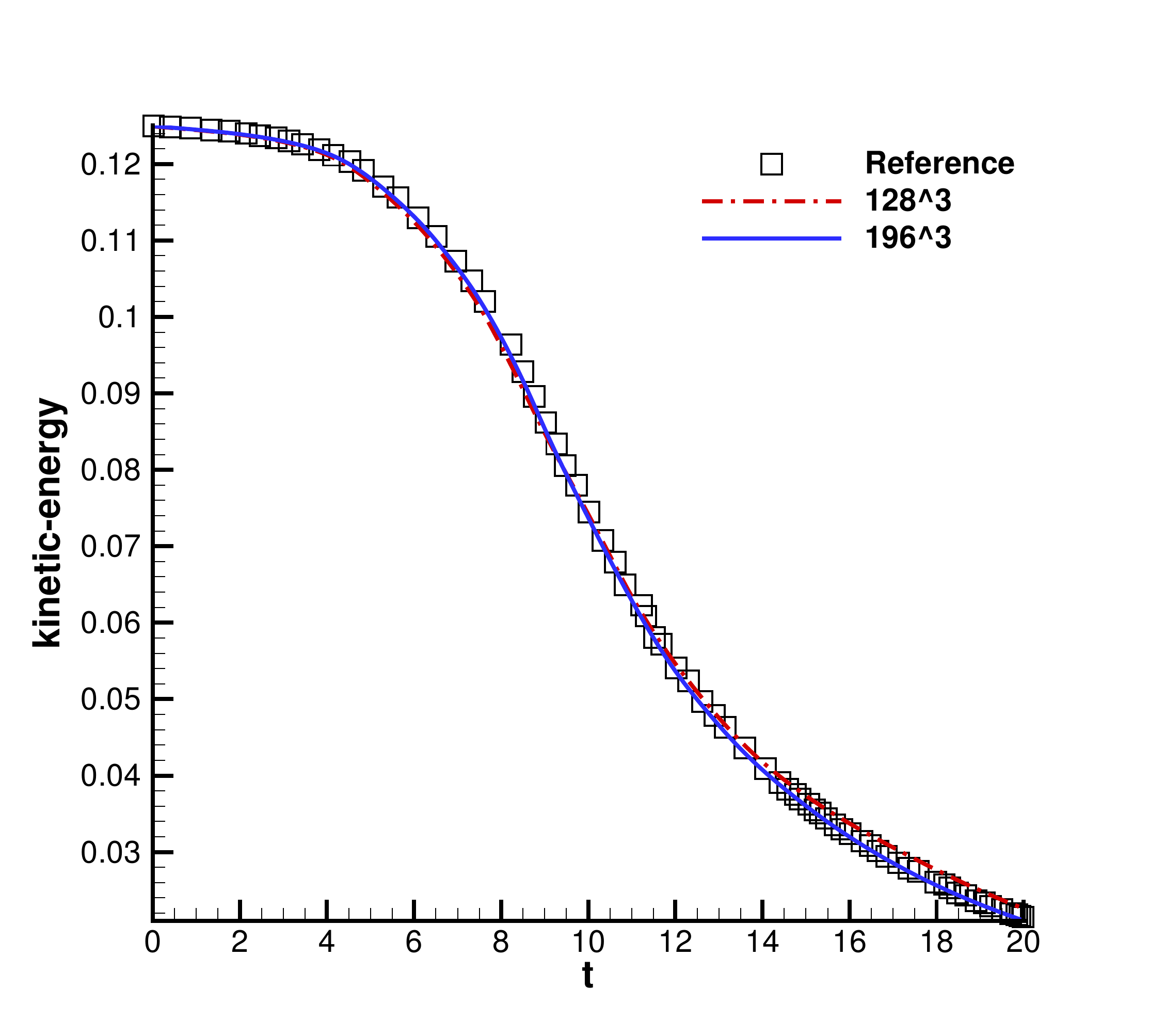}}
	\subfigure[Time history of the dissipation rate of kinetic energy]{
		\label{tg-re1600-dk-ao}
		\includegraphics[width=0.48\textwidth]{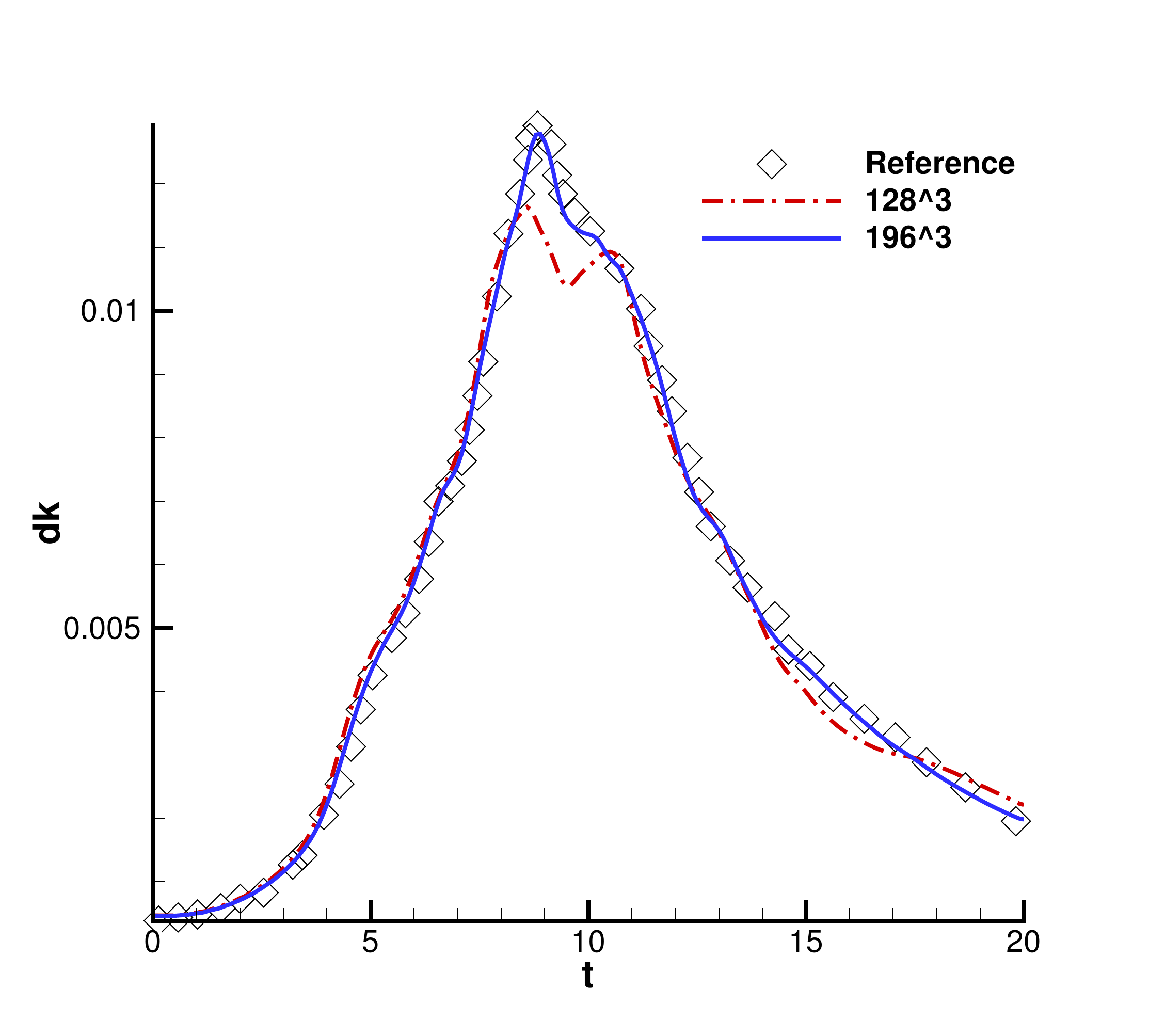}}
	\caption{Taylor-Green vortex: Re=1600. The time history of kinetic energy by the new reconstruction procedure with the linear WENO5-AO reconstruction. CFL=0.5. }
	\label{tg-re1600-ao}
\end{figure}

\begin{figure}[htbp]	
	\centering
	\subfigure[t=5]{
		\label{tg-re1600-qc-t5}
		\includegraphics[width=0.48\textwidth]{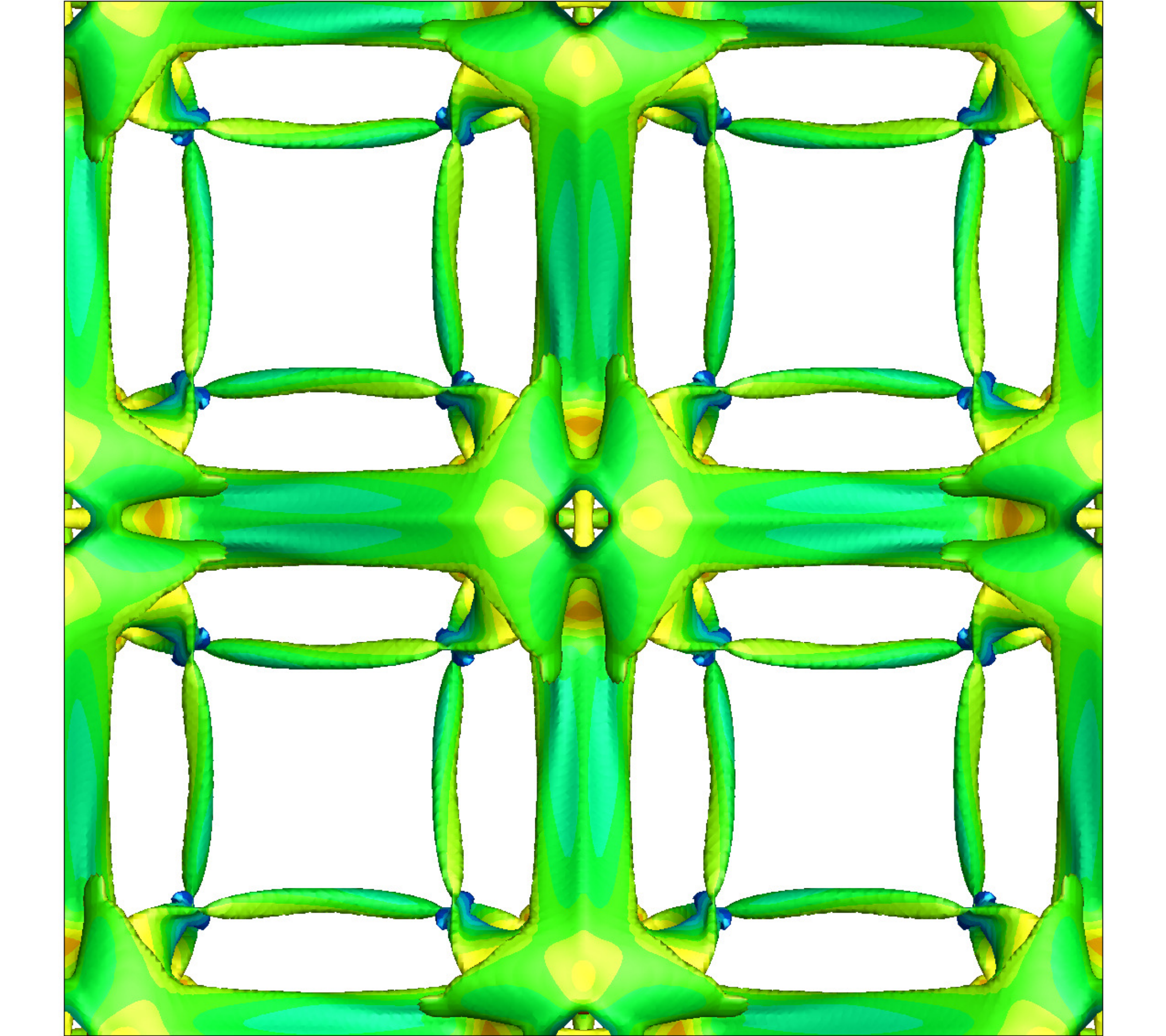}}
	\subfigure[t=10]{
		\label{tg-re1600-qc-t10}
		\includegraphics[width=0.48\textwidth]{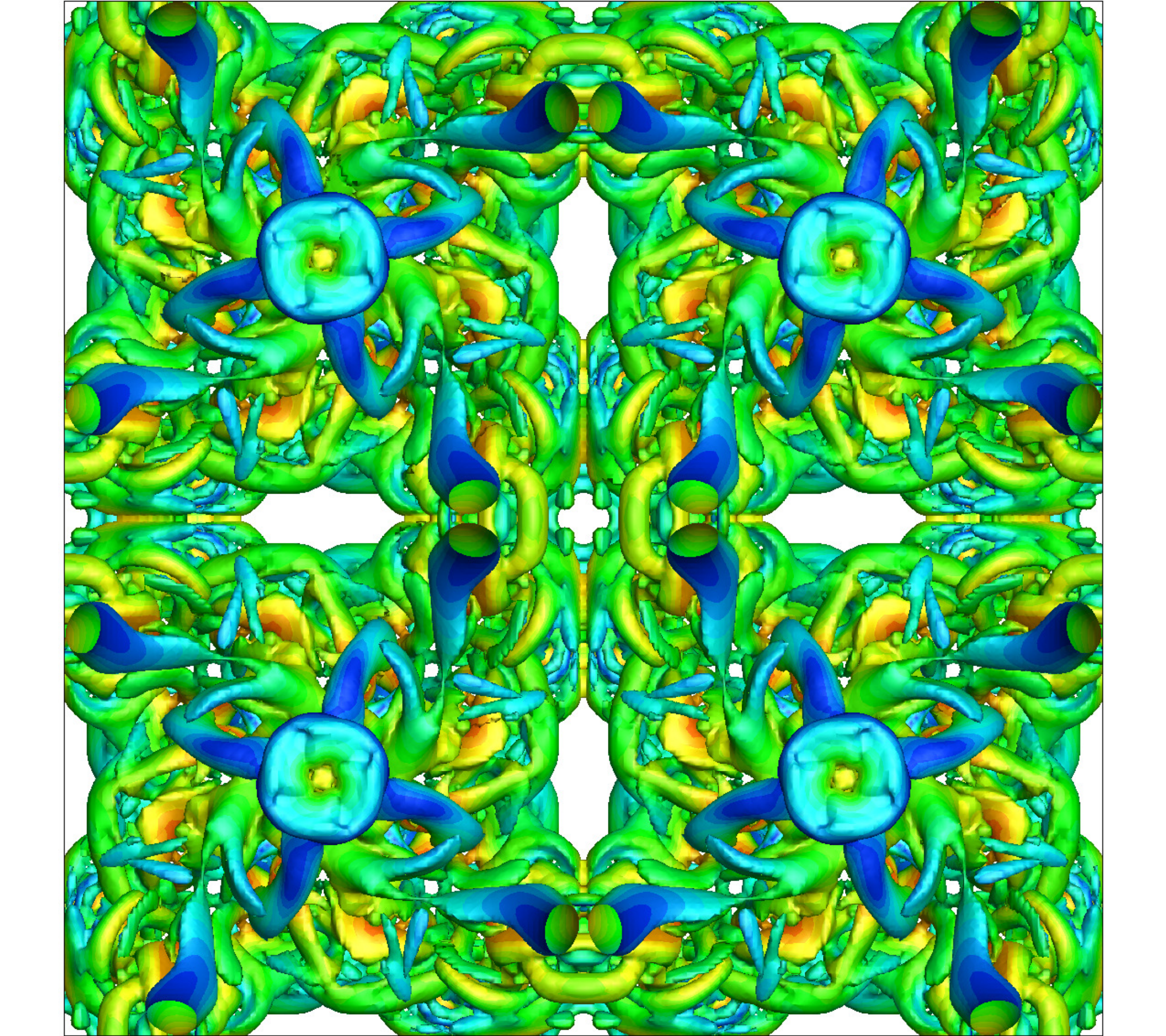}}
	\caption{Taylor-Green vortex: the iso-surfaces of Q criterion colored by ,ach number at
		time t = 5, 10 for Re = 1600. $196^3$ mesh is used. The x-y plane is shown.}
	\label{tg-contour}
\end{figure}

Two global quantities are investigated in the current study as the flow evolves in time.
The first one is the volume-averaged kinetic energy
\begin{align*}
E_k=\frac{1}{\rho_0\Omega}\int_\Omega\frac{1}{2}\rho\textbf{U}\cdot\textbf{U}d\Omega,
\end{align*}
where $\Omega$ is the volume of the computational domain.
Then the dissipation rate of the kinetic energy is given by
\begin{align*}
\varepsilon_k=-\frac{dE_k}{dt}.
\end{align*}
The linear weights of reconstruction and the smooth flux function are adopted in this case.
The numerical results of the current scheme with  $128^3$ and $196^3$ mesh points for the normalized volume-averaged kinetic energy and dissipation rate are presented in Fig. \ref{tg-re1600-ao}, which agree well with the data in \cite{debonis2013solutions}.
The iso-surfaces of $Q$ criterion colored by Mach number at $t=5$ and $10$ are shown in Fig. \ref{tg-contour}.
The vortex structures become denser and smaller with the time increment.

\subsubsection{Compressible isotropic turbulence}
\begin{figure}[htbp]	
	\centering
	\includegraphics[width=0.48\textwidth]{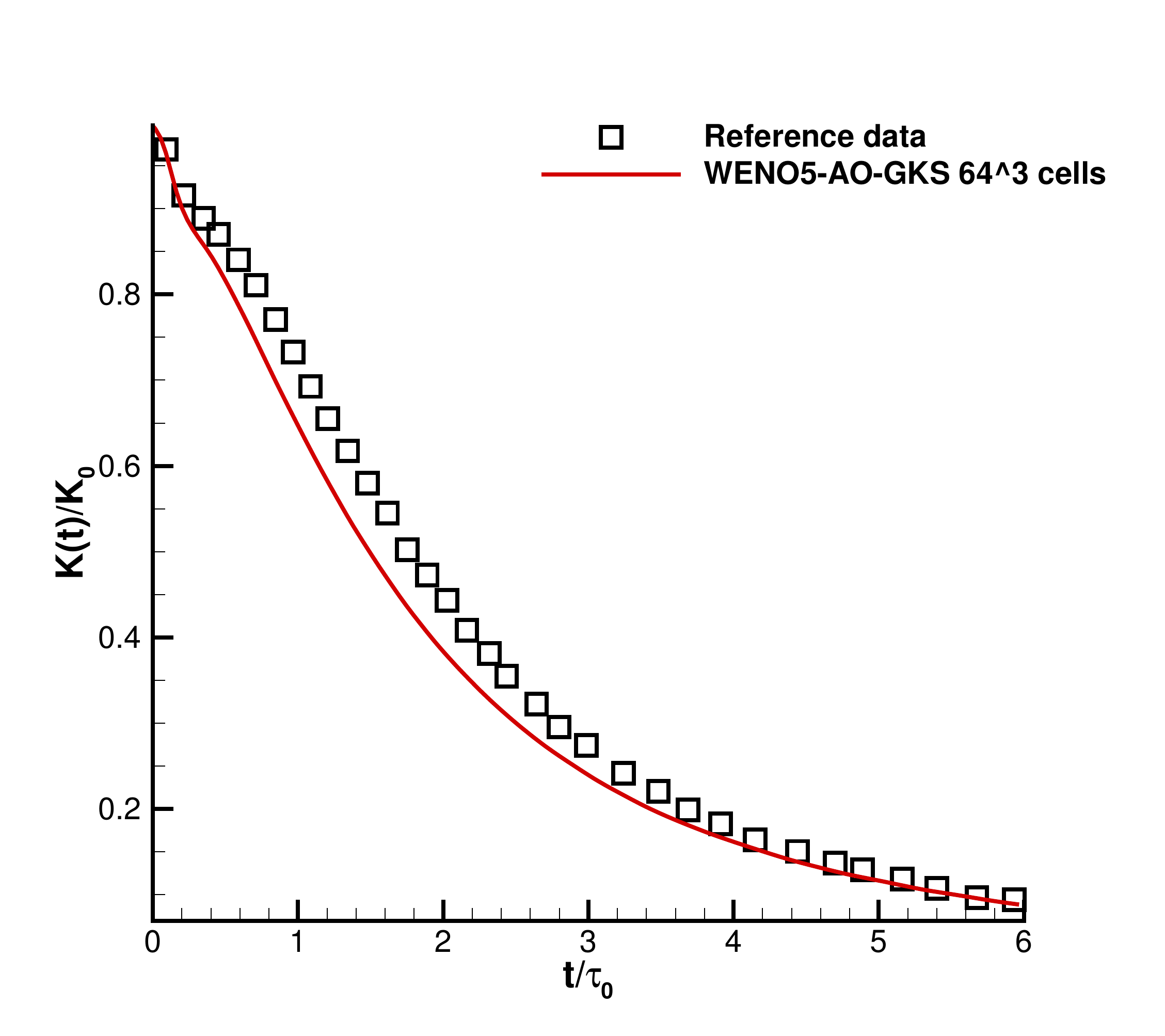}
	\includegraphics[width=0.48\textwidth]{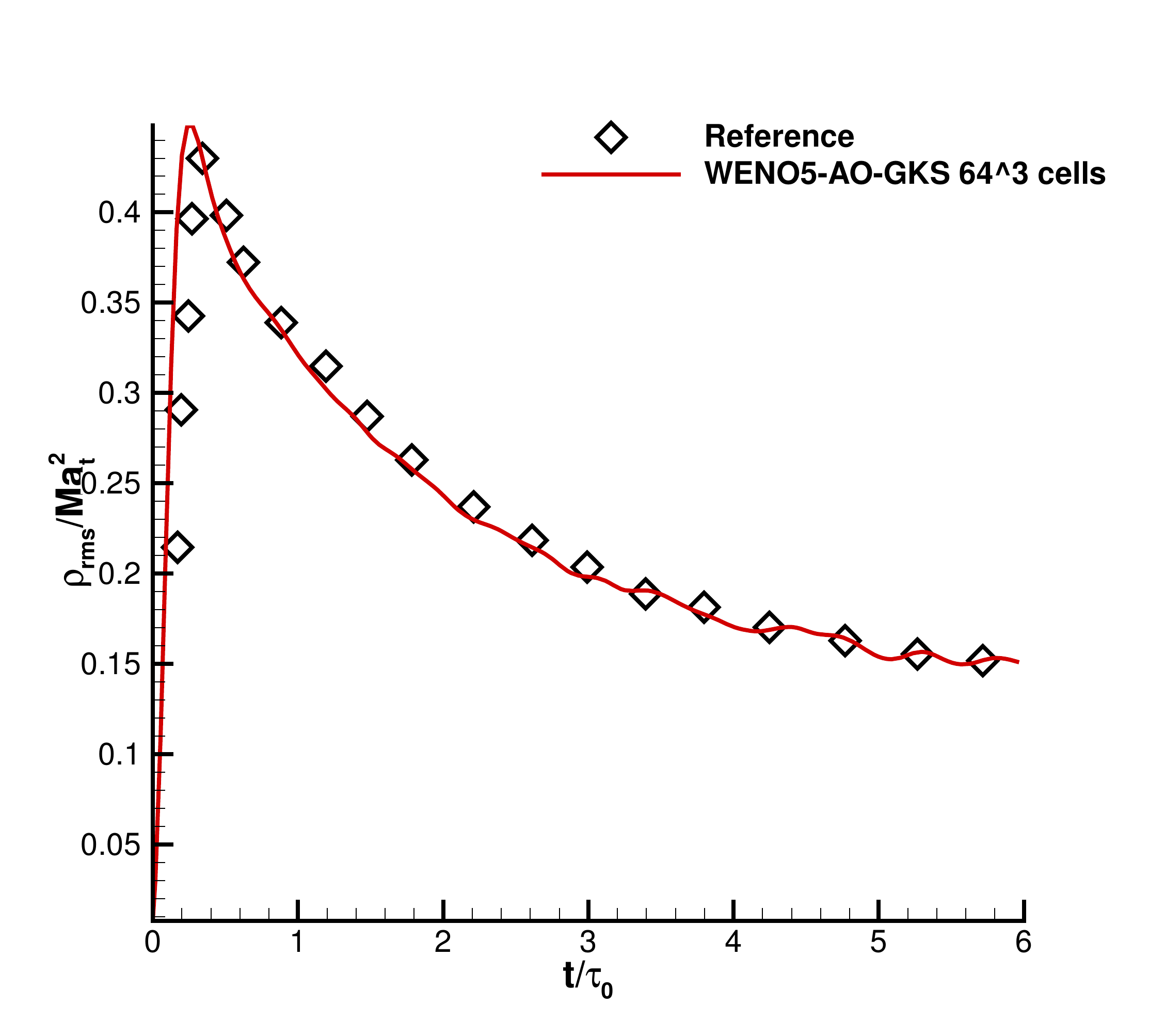}
	\caption{Compressible isotropic turbulence with $Ma_t$=0.5 Left: $K(t)/K_0$. Right: $\rho_{rms}(t)/Ma_t^2$. Mesh: $64^3$. CFL=0.5. }
	\label{cit-ma05} 
\end{figure}

A  decaying homogeneous isotropic compressible turbulence is computed within a square box defined as $-\pi \leq x, y, z\leq \pi$, and the periodic boundary conditions are used in all directions \cite{samtaney2001direct}.
Given spectrum with a specified root mean square $U'$
\begin{align*}
U'=<<\frac{\textbf{U}\cdot \textbf{U}}{3}>>^{1/2},
\end{align*}
a divergence-free random velocity field $\textbf{U}_0$ is initialized,
where $<<...>>$ is a volume average over the whole computational domain.
The specified spectrum for velocity is given by
\begin{align*}
E(k)=A_0k^4\exp(-2k^2/k_0^2),
\end{align*}
where $A_0$ is a constant to set initial kinetic energy, $k$ is the wave number, $k_0$ is the wave number at spectrum peaks.
The initial volume averaged turbulent kinetic energy $K_0$  and the initial large-eddy-turnover time $\tau_0$ is given by
\begin{align*}
K_0=\frac{3A_0}{64}\sqrt{2\pi}k_0^5,~~\tau_0=\sqrt{\frac{32}{A_0}}(2\pi)^{1/4}k_0^{-7/2}.
\end{align*}
The Taylor micro-scale and corresponding Reynolds number $Re_\lambda$ and turbulence Mach number $Ma_t$ are given as
\begin{align*}
&\lambda^2=\frac{(U')^2}{<<(\partial_1 U_1)^2>>},Ma_t=\frac{\sqrt{3}U'}{<<c_s>>}=\frac{\sqrt{3}U'}{\sqrt{\gamma T_0}},
\\
&Re_\lambda=\frac{<<\rho>>U'\lambda}{<<\mu>>}=\frac{(2\pi)^{1/4}}{4}\frac{\rho_0}{\mu_0}\sqrt{2A_0}k_0^{3/2}.
\end{align*}
The dynamic viscosity is determined by the power law
\begin{align*}
\mu=\mu_0\big(\frac{T}{T_0}\big)^{0.76},
\end{align*}
where $\mu_0$ and $T_0$ can be determined from $Re_\lambda$ and $Ma_t$ with initialized $U'$ and $\rho_0=1$. The time history of the kinetic energy, root-mean-square of density fluctuation  are defined as
\begin{align*}
K(t)=\frac{1}{2}<\rho \textbf{U}\cdot \textbf{U}>,~~\rho_{rms}(t)=\sqrt{<(\rho-\overline{\rho})^2>}.
\end{align*}
The previous direct numerical simulations by the conventional WENO-GKS \cite{pan2018two} have shown the complex structures due to the random initial flow field.
When the Mach number gets higher, the stronger shocklets generate complex shock-vortex interactions.
It becomes very challenging for high-order methods by increasing the turbulent Mach number.
Thus, a series of turbulent Mach numbers have been chosen to test the robustness of the current scheme.
A coarse mesh with $64^3$ and a fixed $Re_\lambda=72$ is used. The other parameters, i.e., $A_0=1.3\times10^{-4}, k_0=8$, are chosen according to \cite{pan2018two}.
The maximum Mach number in the flow filed is about three times of the initial turbulent Mach number.
When $Ma_t=0.5$, the pure smooth GKS solver and the WENO-AO reconstruction with linear weights could be used.
The equilibrium state  is obtained by the arithmetic average of the non-equilibrium states to further reduce the numerical dissipations.
As a result, the time history of normalized kinetic energy $K(t)/K_0$, normalized root-mean-square of density fluctuation $\rho_{rms}(t)/Ma_t^2$ agree well with the reference data under such a coarse mesh, as shown in Fig. \ref{cit-ma05}.

\begin{figure}[htbp]	
	\centering
	\subfigure[$Ma_t$=0.8]{
		\label{cit-ma08-scheme-compare}
		\includegraphics[width=0.48\textwidth]{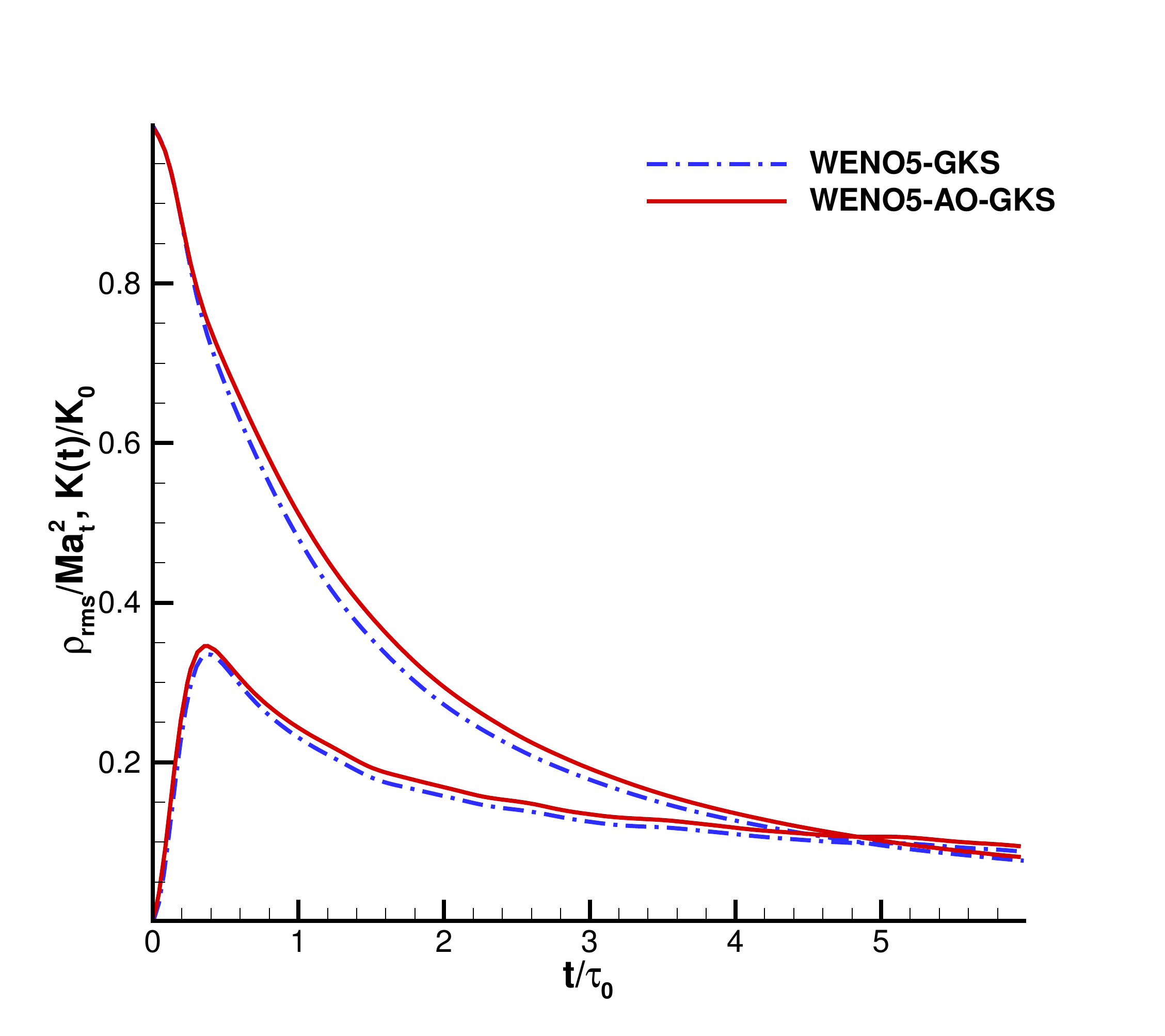}}
	\subfigure[$Ma_t$=1.0]{
		\label{cit-ma1-scheme-compare}
		\includegraphics[width=0.48\textwidth]{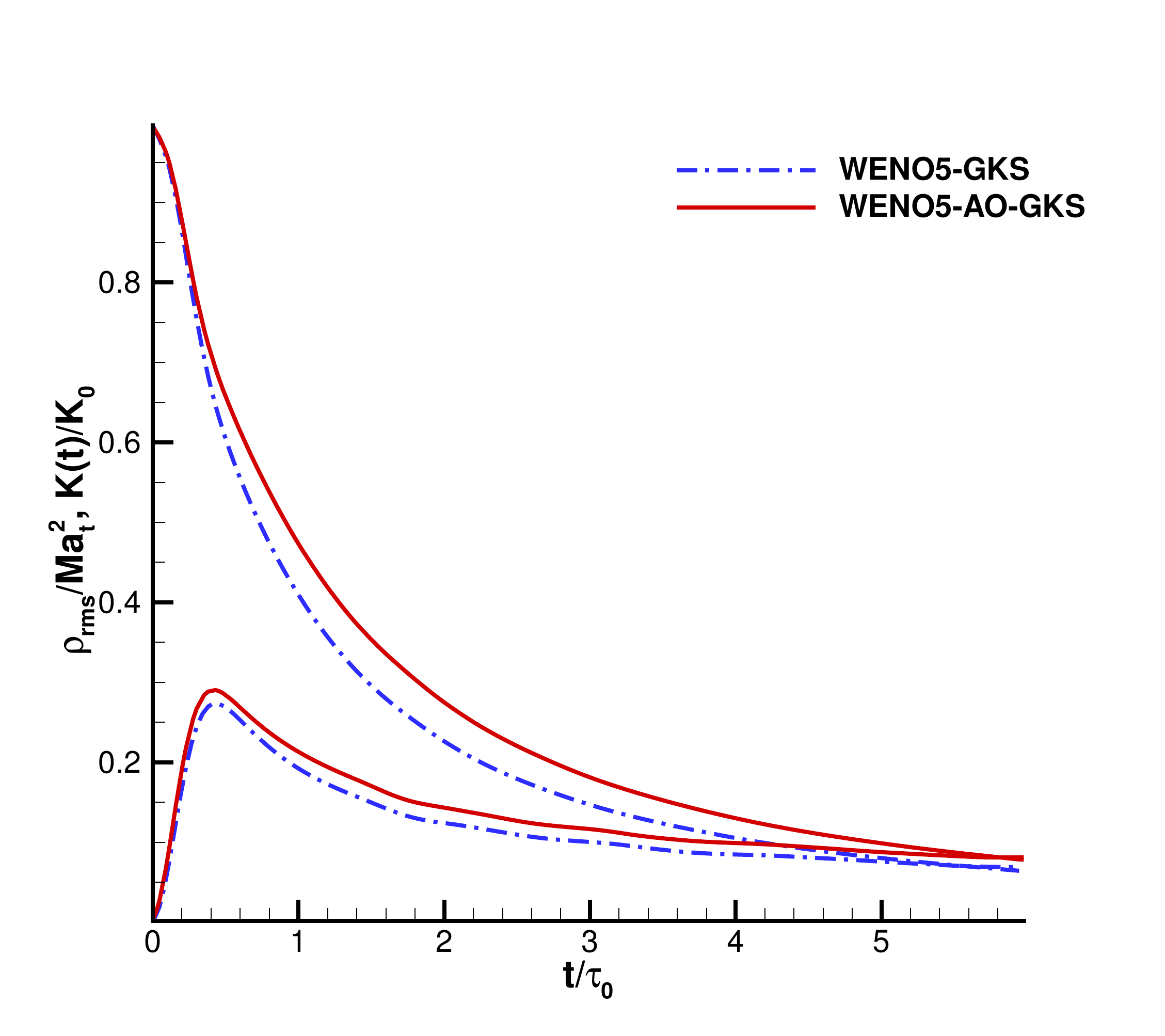}}
	\caption{Compressible isotropic turbulence: Comparison with the conventional and new HGKS. Mesh: $64^3$.}
	\label{cit-scheme-compare} 
\end{figure}

\begin{figure}[htbp]
	\centering
	\includegraphics[width=0.48\textwidth]{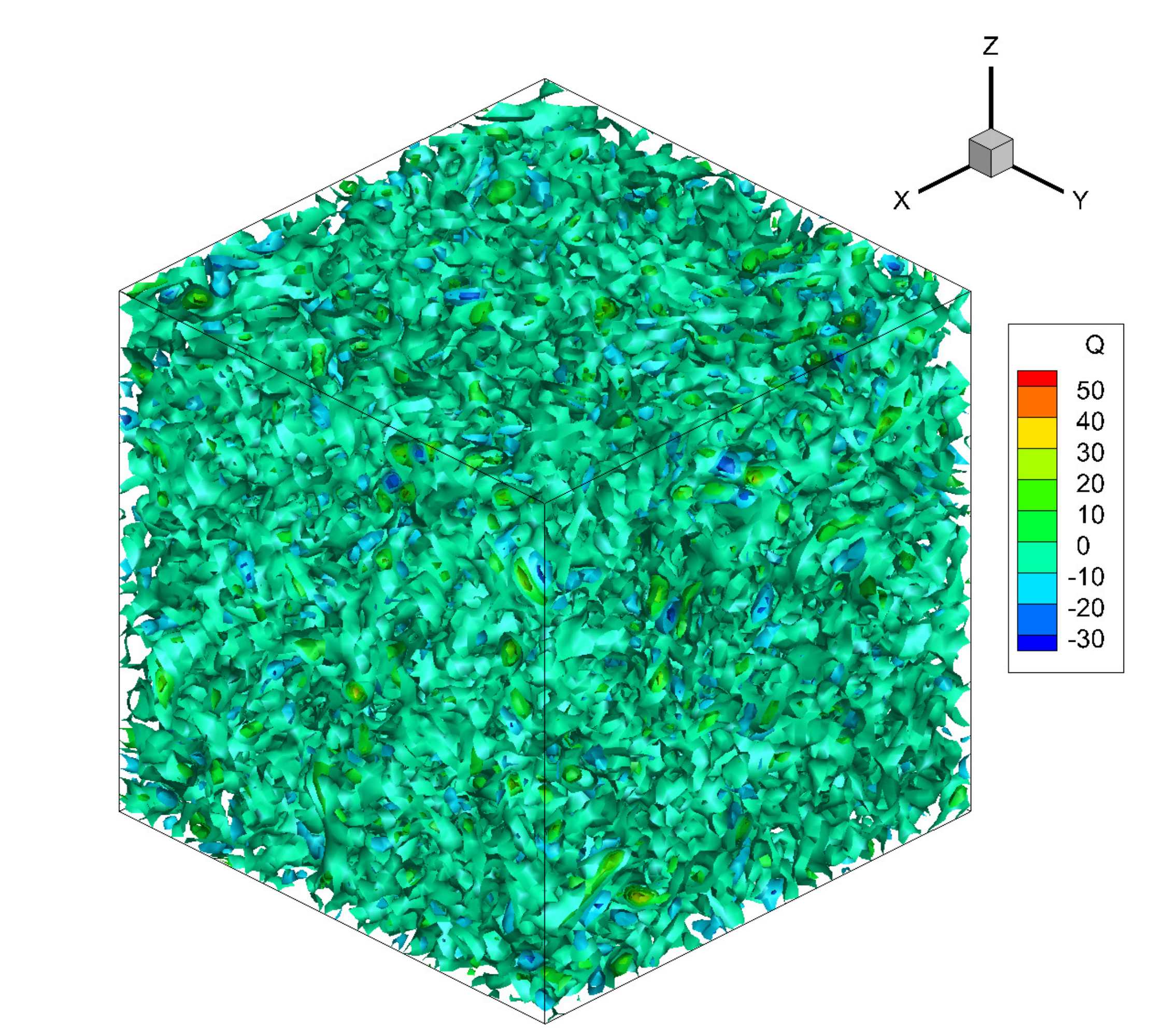}
	\includegraphics[width=0.48\textwidth]{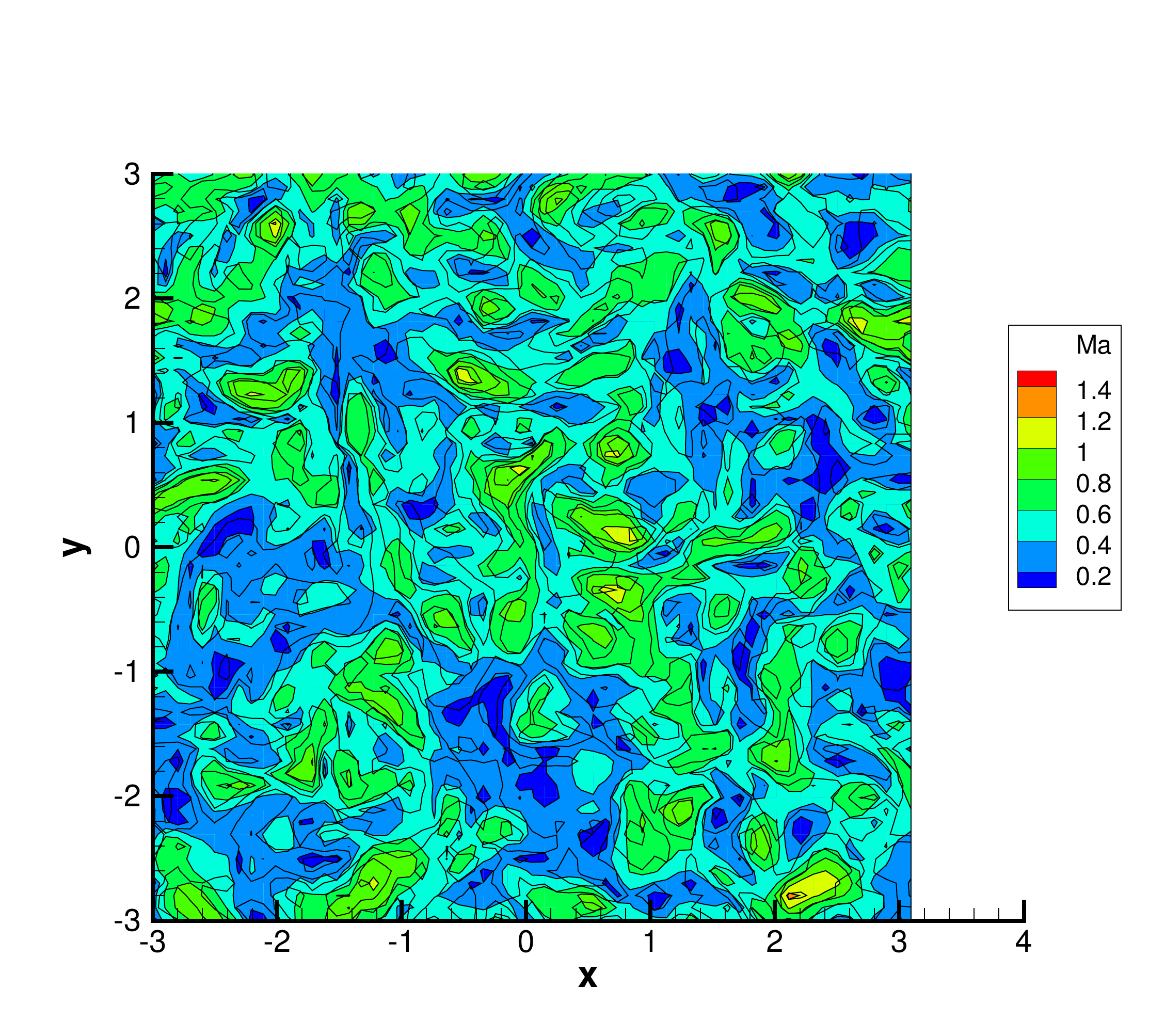}
	\caption{\label{cit-ma1-contour} Compressible homogeneous turbulence with $Ma_t=1$.
		Left: iso-surfaces of Q criterion colored by Mach number at time $t/\tau_0=1$ with $64^3$ cells; right: the Mach number distribution with $z=-\pi$ at time $t/\tau=1$.}
\end{figure}

\begin{figure}[htbp]	
	\centering
	\subfigure[The mean square root of density fluctuation]{
		\label{cit-ma-comparision-den}
		\includegraphics[width=0.48\textwidth]{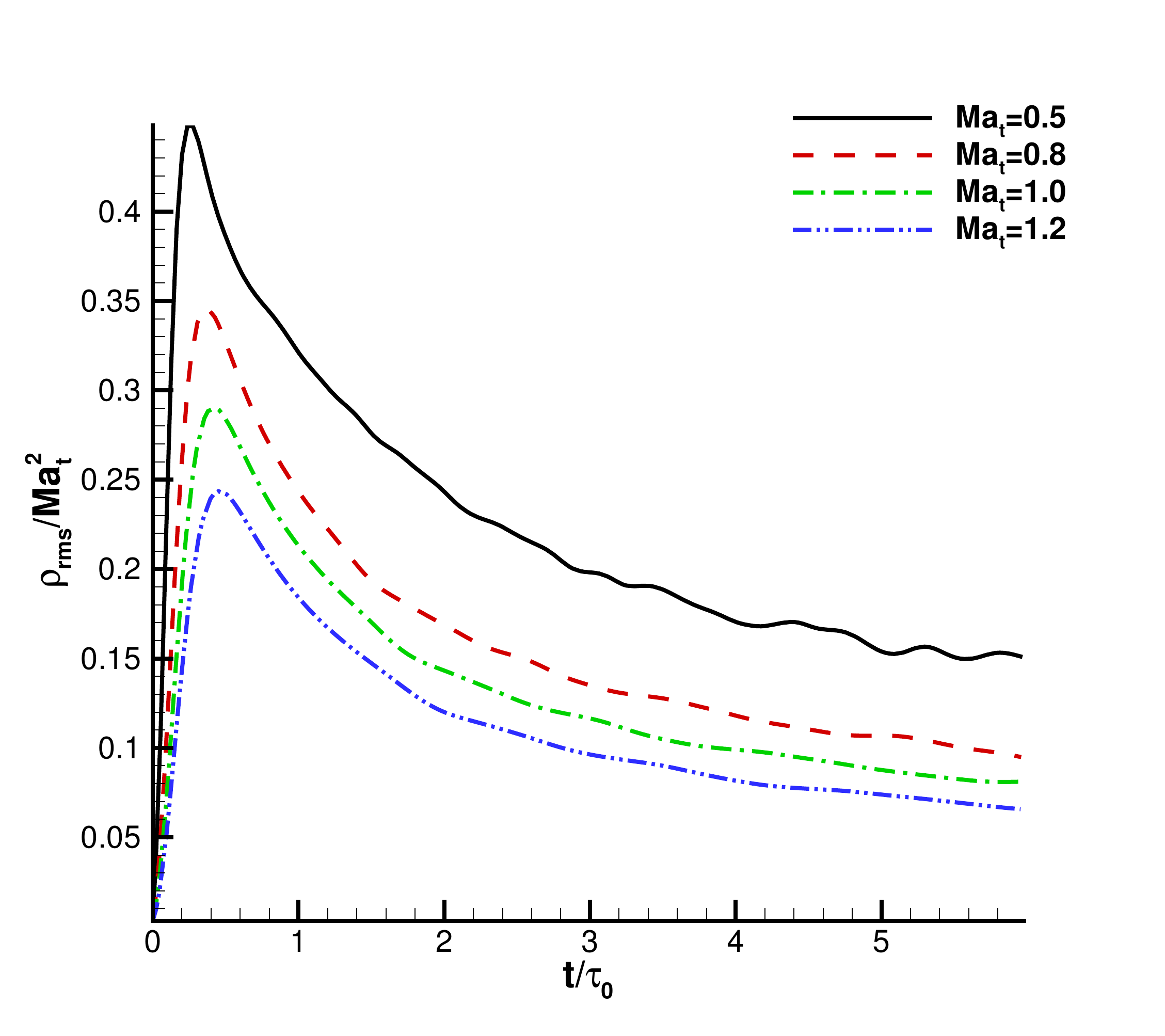}}
	\subfigure[The kinetic energy]{
		\label{cit-ma-comparision-ke}
		\includegraphics[width=0.48\textwidth]{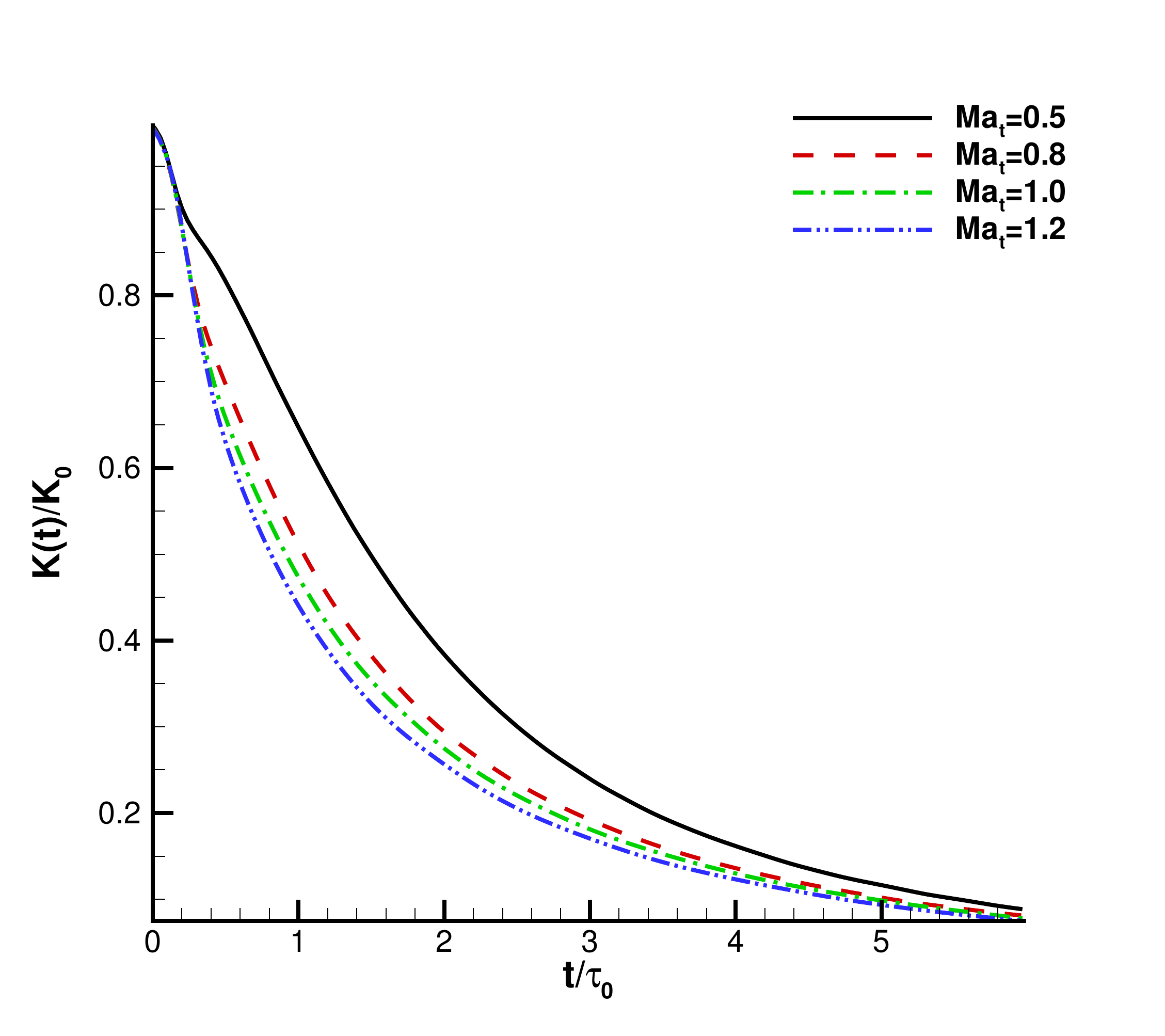}}
	\caption{Compressible isotropic turbulence: Comparison with different $Ma_t$ numbers by the new HGKS. Mesh: $64^3$. CFL=0.5.}
	\label{cit-ma-comparision} 
\end{figure}

As further increasing of $Ma_t$, the schemes could not survive with the pure smooth flux solver and linear reconstruction.
The full gas-kinetic flux solver should be used, and theoretically the conventional WENO-GKS degrades to third-order accuracy due to the low-order reconstructions for the non-equilibrium states.
For the new HGKS, it gives a strictly fifth-order spatial accuracy for both equilibrium and non-equilibrium states once the WENO5-AO-Z reconstruction is applied.
In order to improve the robustness of the scheme,the following five treatments could possibly protect the program from blowing up:
\begin{itemize}
	\item Use the full GKS solver.
	\item Change the WENOZ-type weights to WENO-JS weights.
	\item When detecting negative temperature (lambda) for face-averaged/ line-averaged/ point-wise values, the first-order reconstruction is used.
	\item Modify $\tau$ from $\tau=\frac{\mu}{p}+\delta p \Delta t$ to $\tau=\frac{\mu}{p} +\sum_{1}^{5}\delta Q \Delta t$, where $Q$ means all five primitive variables, operator $\delta Q=\frac{|Q^l-Q^r|}{|Q^l|+|Q^r|}$.	
	\item Take smaller CFL number.
\end{itemize}
A systematical comparison of the performance of different higher-order GKS with the increasing of Mach number is given in Table \ref{cit-re72-compare}.
Especially, the cases with $Ma_t=0.8$ and  $1.0$ are chosen to compare the performance of the two schemes, shown in Fig. \ref{cit-scheme-compare}.
The WENO-GKS shows more rapid dissipation rates under these cases.
The visualized results are given in Fig. \ref{cit-ma1-contour},
where the iso-surfaces of Q criterion and the selected surface slice of Mach number distribution at $z=-\pi$ are plotted.
The complex  vortexes and widespread shocklets could be clearly observed.
Lastly, the time histories of the statistical quantities with respect of different Mach numbers are shown in Fig. \ref{cit-ma-comparision}.
Generally the kinetic energy gets dissipated more rapidly with the increase of $Ma_t$.
More data have been provided in \cite{wang2017shocklet}. This case at higher Mach numbers will be further explored by HGKS.
\begin{table}[htbp]
	\small
	\begin{center}
		\def\temptablewidth{1\textwidth}
		{\rule{\temptablewidth}{0.1pt}}
		\begin{tabular*}{\temptablewidth}{@{\extracolsep{\fill}}c|c|c}		
			Mach number & Traditional WENO5-GKS & New WENO5-AO GKS\\
			\hline
			$Ma_t\le$0.5 & \tabincell{c}{Smooth reconstruction \\ Smooth GKS solver}
			& \tabincell{c}{Smooth reconstruction \\ Smooth GKS solver}  \\
			\hline
			$Ma_t$=0.8 & \tabincell{c}{ Full GKS solver \\ Limiting of negative temperature }
			& \tabincell{c}{ Full GKS solver \\ No limitation }  \\
			\hline
			$Ma_t$=1.0 & \tabincell{c}{ Full GKS solver \\ Limiting of negative temperature \\CFL=0.25}
			& \tabincell{c}{ Full GKS solver \\ Limiting of negative temperature }   \\
			\hline
			$Ma_t$=1.2 & \tabincell{c}{Only WENO-JS reconstruction \\ Full GKS solver \\ Limiting of negative temperature \\ Modification of $\tau$ \\CFL=0.25}
			& \tabincell{c}{Full GKS Solver \\ Limiting of negative temperature } \\ 	 			
		\end{tabular*}
		{\rule{\temptablewidth}{0.1pt}}
	\end{center}
	\vspace{-4mm} \caption{\label{cit-re72-compare} The validation of conventional and new HGKS under different turbulence Mach number $Ma_t$. The CFL number takes $0.5$ and the WENO-Z type reconstruction is used if the setting is not specified. Mesh size: $64^3$. }
\end{table}

\section{Conclusion}

The gas-kinetic scheme is based on a high-order gas evolution model for the flux evaluation.
The kinetic model equation is used in the construction of a time dependent gas distribution at a cell interface.
Similar to the generalized Riemann problem, the initial condition of flow variables in GKS is a piecewise continuous polynomial on both sides of a cell interface with a possible discontinuity between them.
In the previous GKS, the standard WENO-type reconstruction is used, where only point-wise values at the cell interface are reconstructed and have  the corresponding order of accuracy.
However, the GKS not only requires the point-wise values, but also the slopes of flow variables.
As a result, the use of the point-wise values from the standard WENO reconstruction and the enforcement of cell averaged flow variables
cannot get a reconstructed polynomial inside each cell with the same order of accuracy as the original order of WENO reconstruction.
At the same time, in the previous GKS both initial non-equilibrium states and the equilibrium one have to be reconstructed separately.
In order to further improve GKS, especially for the construction of higher-order GKS, the WENO-AO reconstruction has been adopted in this paper, where a whole polynomial inside each cell is obtained directly with the WENO-type reconstruction.
Even though the schemes based on the Riemann solver cannot get full benefits from WENO-AO reconstruction because only point-wise values are
required in the Riemann solution, the GKS is able to utilize the whole polynomial from WENO-AO.
Therefore, the initial non-equilibrium states inside the cell in the current HGKS can achieve the same order of accuracy of the WENO reconstruction.
Besides the improvement of order of accuracy in the initial non-equilibrium states, the equilibrium state in the HGKS is obtained through
a dynamic modeling, such as the particle collisions from the non-equilibrium distribution functions, and the separate reconstruction
for the equilibrium state in the previous GKS is avoided.
In the new HGKS, both initial non-equilibrium and equilibrium states can achieve the same order of accuracy.
Due to the absence of equilibrium reconstruction, the new scheme becomes simpler and is more flexibility in dealing with the WENO procedure at arbitrary Gaussian points than the previous scheme.
The computational efficiency has been improved as well in the current approach.
The WENO-Z-type weights can be used directly in the WENO-AO reconstruction in the current HGKS and the scheme has
the same robustness as the previous one with the WENO-JS-weights.
Another distinguishable feature of HGKS is the use of multi-stage multi-derivative approach as the time-stepping method.
More specifically, with two stages a fourth-order time accuracy has been obtained. This benefits solely from the time accurate
flux function in GKS.
Many numerical experiments are included to validate the efficiency, robustness, and accuracy of the scheme.
Accuracy tests from 1D to 3D show that the scheme meets its designed accuracy.
The scheme inherits less numerical viscosity, reduces the spurious oscillation at weak discontinuities,
and has a better shear instability resolution. In the compressible isotropic turbulence simulation, the scheme shows favorable robustness in
capturing multi-dimensional shocklet and resolve the small vortex structure accurately.
The reconstruction scheme is very important for the quality of GKS. Combining the excellent reconstruction technique and the physically reliable
evolution model, a higher-order gas-kinetic scheme with enhanced performance for the Euler and Navier-Stokes equations has been developed.
The further extension of the reconstruction procedure to high-order compact GKS on unstructured mesh will be investigated.

\section*{Acknowledgment}
The authors would like to thank Guiyu Cao, Fengxiang Zhao, Liang Pan for helpful discussion.
The current research was supported by Hong Kong Research Grant Council (16206617) and
National Natural Science Foundation of China (11772281,91852114).

\section*{Appendix A: Smooth indicators for WENO5-AO}
Following the same definition in Section \ref{new-reconstruction}, the smooth indicators for $p_k^{r3}, k=0,1,2$ are given as
\begin{align*}
\beta_k^{r3}=(p_k^{r3})_x^2+{13}/{3}(p_{k}^{r3})_{xx}^2,
\end{align*}
where
\begin{align*}
&(p_{0}^{r3})_x= -2 \bar{Q}_{-1}+ \bar{Q}_{-2}/2+3\bar{Q}_{0}/2,~~
(p_{0}^{r3})_{xx}= (\bar{Q}_{-2}-2 \bar{Q}_{-1}+\bar{Q}_{0})/2,\\
&(p_{1}^{r3})_x= (\bar{Q}_{1}-\bar{Q}_{-1})/2,~~
(p_{1}^{r3})_{xx}= (\bar{Q}_{-1}-2 \bar{Q}_{0}+\bar{Q}_{1})/2,\\
&(p_{2}^{r3})_x= 2 \bar{Q}_{1}- \bar{Q}_{2}/2-3\bar{Q}_{0}/2,~~
(p_{2}^{r3})_{xx}= (\bar{Q}_{0}-2 \bar{Q}_{1}+\bar{Q}_{2})/2.
\end{align*}
The smooth indicator for $p_3^{r5}$ is given as
\begin{align*}
\beta_3^{r5}&=((p_3^{r5})_{x} + (p_3^{r5})_{xxx}/10)^2
+ 13/3((p_3^{r5})_{xx} + 123/455 (p_3^{r5})_{xxxx})^2\\
&+781/20(p_3^{r5})_{xxx}^2+1421461/2275(p_3^{r5})_{xxxx}^2,
\end{align*}
where
\begin{align*}
&(p_{3}^{r5})_x= (-82\bar{Q}_{-1} +11\bar{Q}_{-2} + 82\bar{Q}_{1}-11\bar{Q}_{2})/120,\\
&(p_{3}^{r5})_{xx} = (40\bar{Q}_{-1} - 3\bar{Q}_{-2} -74\bar{Q}_{0} +40\bar{Q}_{1} - 3\bar{Q}_{2})/56,\\
&(p_{3}^{r5})_{xxx} = (2\bar{Q}_{-1} - \bar{Q}_{-2} - 2\bar{Q}_{1} + \bar{Q}_{2})/12,\\
&(p_{3}^{r5})_{xxxx} = (-4\bar{Q}_{-1} +\bar{Q}_{-2} + 6\bar{Q}_{0} -4\bar{Q}_{1} +\bar{Q}_{2})/24.
\end{align*}
The detailed derivations could be found in \cite{balsara2016efficient}.

\section*{Appendix B: Reconstruction at Gaussian points}
Reconstruction are needed at the Gaussian points in multi-dimensional case.
Two Gaussian points are used in the current fourth-order scheme.
Starting from the same stencils above,
the point-wise values from each sub-stencil at the Gaussian point $x_{i-1/2\sqrt{3}}$ are
\begin{align*}
	p_0^{r3}(x_{i-1/2\sqrt{3}}) &= (1-\sqrt{3}/4 ) \bar{Q}_0 + (4  \bar{Q}_{-1} - \bar{Q}_{-2}) / (4  \sqrt{3}),\\
	p_1^{r3}(x_{i-1/2\sqrt{3}}) &= \bar{Q}_0 + (\bar{Q}_{-1} - \bar{Q}_{1}) / (4  \sqrt{3}),\\
	p_2^{r3}(x_{i-1/2\sqrt{3}})&= (3  (4 + \sqrt{3})\bar{Q}_0 + \sqrt{3}  (-4  \bar{Q}_{1} + \bar{Q}_{2}))/12,\\
	p_3^{r5}(x_{i-1/2\sqrt{3}}) &= (4314  \bar{Q}_0 + (4 + 500  \sqrt{3})\bar{Q}_{-1} - (1+ 70  \sqrt{3})  \bar{Q}_{-2}, \\
	&+ (4 - 500  \sqrt{3})  \bar{Q}_{1} +(-1 + 70  \sqrt{3})  \bar{Q}_{2}) / 4320,
\end{align*}
and the first-order derivatives are
\begin{align*}
			(p_0^{r3})_x(x_{i-1/2\sqrt{3}}) &= -((-9 + \sqrt{3})\bar{Q}_0 - 2  (-6 + \sqrt{3})\bar{Q}_{-1} + (-3 + \sqrt{3})\bar{Q}_{-2}) / (6  \Delta x),\\
			(p_1^{r3})_x(x_{i-1/2\sqrt{3}}) &= -(-2  \sqrt{3}  \bar{Q}_0 + (3 + \sqrt{3})\bar{Q}_{-1} + (-3 + \sqrt{3})\bar{Q}_{1}) / (6  \Delta x),\\
			(p_2^{r3})_x(x_{i-1/2\sqrt{3}}) &= -((9 + \sqrt{3})\bar{Q}_0 - 2  (6 + \sqrt{3})\bar{Q}_{1} + (3 + \sqrt{3})\bar{Q}_{2}) / (6  \Delta x),\\
			(p_3^{r5})_x(x_{i-1/2\sqrt{3}}) &= (48  \sqrt{3}  \bar{Q}_0 - (72 + 26  \sqrt{3} ) \bar{Q}_{-1} + (9 + 	2  \sqrt{3})  \bar{Q}_{-2}, \\
				&+ (72   - 26  \sqrt{3} ) \bar{Q}_{1} - (9 -
				2  \sqrt{3})  \bar{Q}_{2}) / (108  \Delta x).
\end{align*}

The point-wise values and linear weights at
another Gaussian point $x_{i+1/2\sqrt{3}}$ are
\begin{align*}
		p_0^{r3}(x_{i+1/2\sqrt{3}}) &= (3  (4 + \sqrt{3})\bar{Q}_0 + \sqrt{3}  (-4  \bar{Q}_{-1} + \bar{Q}_{-2}))/12,\\
		p_0^{r3}(x_{i+1/2\sqrt{3}}) &= \bar{Q}_0 + (-\bar{Q}_{-1} + \bar{Q}_{1}) / (4  \sqrt{3}),\\
		p_0^{r3}(x_{i+1/2\sqrt{3}}) &= (1-\sqrt{3}/4  \bar{Q}_0) + (4  \bar{Q}_{1} - \bar{Q}_{2}) / (4  \sqrt{3}),\\
		p_3^{r5}(x_{i+1/2\sqrt{3}}) &= (4314  \bar{Q}_0 + (4 - 500  \sqrt{3})\bar{Q}_{-1} - (1- 70  \sqrt{3})\bar{Q}_{-2} ,\\
		&+ (4 + 500  \sqrt{3})  \bar{Q}_{1} - (\bar{Q}_{2} + 70  \sqrt{3})  \bar{Q}_{2}) / 4320,
\end{align*}
and the first-order derivatives are
\begin{align*}
		(p_0^{r3})_x(x_{i+1/2\sqrt{3}}) &= ((9 + \sqrt{3})\bar{Q}_0 - 2  (6 + \sqrt{3})\bar{Q}_{-1} + (3 + \sqrt{3})\bar{Q}_{-2}) /
				(6  \Delta x),\\
		(p_1^{r3})_x(x_{i+1/2\sqrt{3}}) &= (-2  \sqrt{3}  \bar{Q}_0 + (-3 + \sqrt{3})\bar{Q}_{-1} + (3 + \sqrt{3})\bar{Q}_{1}) / (6  \Delta x),\\
		(p_2^{r3})_x(x_{i+1/2\sqrt{3}}) &= ((-9 + \sqrt{3})\bar{Q}_0 - 2  (-6 + \sqrt{3})\bar{Q}_{1} + (-3 + \sqrt{3})\bar{Q}_{2}) / (6  \Delta x),\\
		(p_3^{r5})_x(x_{i+1/2\sqrt{3}}) &= -((48  \sqrt{3}  \bar{Q}_0 +( 72 - 26  \sqrt{3} ) \bar{Q}_{-1} - (9 -
				2  \sqrt{3})  \bar{Q}_{-2},\\
				& - (72  + 26  \sqrt{3} ) \bar{Q}_{1} + (9 + 2  \sqrt{3})  \bar{Q}_{2}) / (108  \Delta x)).
\end{align*}

\section*{Appendix C: Calculation of GKS flux function in 1-D}
This appendix presents some details for the
implementation of gas-kinetic flux solver \cite{GKS-2001,xu2014direct,GKS-lecture}.

For a clearer illustration, the final form of the gas kinetic distribution function along a cell interface $x_{i+1/2}$ in Eq.(\ref{flux}) is listed here again

\begin{align}\label{2nd-flux}
f(x_{i+1/2},t,u,\xi)=&(1-e^{-t/\tau}) g^{c}+((t+\tau)e^{-t/\tau}-\tau)a^{c}u g^{c}\nonumber\\
+&(t-\tau+\tau e^{-t/\tau})A^{c}  g^{c}\nonumber\\
+&e^{-t/\tau}g^l[1-(\tau+t)a^{l}u-\tau A^l)]H(u)\nonumber\\
+&e^{-t/\tau}g^r[1-(\tau+t)a^{r}u-\tau A^r)] (1-H(u)).
\end{align}

The  $f(x_{i+1/2},t,u,\xi)$ on LHS is a function of physical space $(x,t)$ and phase space $(u,\xi)$.
The interface point $x_{i+1/2} = 0$ is assumed. All coefficients on the RHS are evaluated at this point, i.e.,
$ g^{c}= g^{c}(x_{i+1/2}=0,t=0,u,\xi)$.

\subsection*{Moment calculation}
In the flux calculation according to Eq.(\ref{2nd-flux}), the moments of Maxwellian distribution functions, i.e., $ g^{c}$, $g_l$ and $g_r$,
will be evaluated. The general formulae of moment evaluations are given first.

For a one-dimensional  Maxwellian distribution
\begin{equation*}
g=\rho (\frac{\lambda}{\pi})^{\frac{K+1}{2}} e^{-\lambda ((u-U)^2+\xi ^2)},
\end{equation*}
the moments of $g$ is defined as
\begin{equation*}
\rho \displaystyle\langle| ...| \displaystyle\rangle = \int (...)g\text{d} \Xi,
\end{equation*}
the general moment formula becomes
\begin{equation*}
\displaystyle\langle| u^n\xi^{2l} |\displaystyle\rangle =
\displaystyle\langle| u^n |\displaystyle\rangle
\displaystyle\langle| \xi^{2l} |\displaystyle\rangle,
\end{equation*}
where $n$, $l$ are integers (owing to the symmetrical property
of $\xi$, the moments of $\xi$ are always even-order). With the
integral from $-\infty$ to $+\infty$, we have
\begin{align*}
\displaystyle\langle| u^0 |\displaystyle\rangle=&1,\\
\displaystyle\langle| u^1 |\displaystyle\rangle=&U,\\
...&\\
\displaystyle\langle| u^{n+2} |\displaystyle\rangle=U\displaystyle\langle| u^{n+1}|\displaystyle\rangle&+\frac{n+1}{2\lambda}\displaystyle|\langle u^{n}|\displaystyle\rangle.
\end{align*}
Due to the Heaviside function, the half integral from  $0$ to $+ \infty$ is denoted as $\displaystyle\langle| ... |\displaystyle\rangle _{>0}$,
and from $-\infty$ to $0$ as $\displaystyle\langle| ... |\displaystyle\rangle _{<0}$,
\begin{align*}
\displaystyle\langle| u^0 |\displaystyle\rangle_{>0}&=\frac{1}{2} \rm{erfc}(-\sqrt{\lambda}U),\\
\displaystyle\langle| u^1 |\displaystyle\rangle_{>0}&=U\displaystyle\langle| u^0|\displaystyle\rangle_{>0}+\frac{1}{2}\frac{e^{-\lambda U^2}}{\sqrt{\pi \lambda}},\\
&...\\
\displaystyle\langle| u^{n+2}|\displaystyle\rangle_{>0}&=
U\displaystyle\langle| u^{n+1}|\displaystyle\rangle_{>0}
+\frac{n+1}{2\lambda}\displaystyle\langle| u^{n}|\displaystyle\rangle_{>0},
\end{align*}
and
\begin{align*}
\displaystyle\langle| u^0|\displaystyle\rangle_{<0}&=\frac{1}{2}\rm{erfc}(\sqrt{\lambda}U),\\
\displaystyle\langle| u^1|\displaystyle\rangle_{<0}&=U\displaystyle\langle| u^0|\displaystyle\rangle_{<0}-\frac{1}{2}\frac{e^{-\lambda U^2}}{\sqrt{\pi \lambda}},\\
&...\\
\displaystyle\langle| u^{n+2}|\displaystyle\rangle_{<0}&=U\displaystyle\langle| u^{n+1}|\displaystyle\rangle_{<0}+\frac{n+1}{2\lambda}\displaystyle\langle| u^{n}|\displaystyle\rangle_{<0},
\end{align*}
where $\rm {erfc} $ is the standard complementary error function. The moments of $\displaystyle\langle| \xi^{2l}|\displaystyle\rangle$ from $-\infty$ to $+\infty$ are
\begin{align*}
&\displaystyle\langle| \xi^0|\displaystyle\rangle=1,\\
&\displaystyle\langle |\xi^2|\displaystyle\rangle=(\frac{K}{2\lambda}),\\
\displaystyle\langle| \xi^{2l}|\displaystyle\rangle=&\frac{K+2(l-1)}{2\lambda}\displaystyle\langle| \xi^{2(l-1)}|\displaystyle\rangle.
\end{align*}

\subsection*{Derivatives in macroscopic flow variables and microscopic distribution function}
Once the reconstruction for macroscopic flow derivatives is finished, the microscopic derivatives $a^{l,r,e},A^{l,r,e}$ in Eq.\eqref{2nd-flux} can be obtained in the following way.

From the Taylor expansion of a Maxwellian distribution, all microscopic derivatives shall have in the following form
\begin{align*}
a&=a_{1}+a_{2}u+a_{3}\frac{1}{2}(u^2+\xi^2)=a_{\beta}{\psi}_{\beta},\\
A &=A_{1}+{A}_{2}u+{A}_{3}\frac{1}{2}(u^2+\xi^2)={A}_{\beta}{\psi}_{\beta}.
\end{align*}

According to the relation between distribution function and the macroscopic variables in Eq.\eqref{f-to-convar},
we have
\begin{equation*}
\int \pmb{\psi} a g d\Xi = \frac{\partial \textbf{W}}{\partial x},
\end{equation*}
which could be expanded as
\begin{equation}\label{alpha-beta}
\left(
{\setstretch{1.3}
	\begin{array}{c}
	b_{1} \\
	b_{2} \\
	b_{3}
	\end{array}}
\right)
=
\frac{1}{\rho}\frac{\partial \textbf{W}}{\partial x}=
\frac{1}{\rho}
\left(
{\setstretch{1.3}
	\begin{array}{c}
	\frac{\partial \rho}{\partial x} \\
	\frac{\partial (\rho U)}{\partial x}\\
	\frac{ \partial (\rho E)}{\partial x}
	\end{array}}
\right)
= \displaystyle\langle| \alpha_{\beta}\psi_{\beta} \psi_{\alpha} |\displaystyle\rangle
= \displaystyle\langle| \psi_{\alpha} \psi_{\beta} | \displaystyle\rangle
\left(
{\setstretch{1.3}
	\begin{array}{c}
	a_{1} \\
	a_{2} \\
	a_{3}
	\end{array}}
\right).
\end{equation}
Denoting $\mathbf{M}=\displaystyle\langle| \psi_{\alpha} \psi_{\beta} | \displaystyle\rangle$, the above equations become a linear system
\begin{equation}\label{linear}
\mathbf{M}\textbf{a}=\textbf{b},
\end{equation}
and the coefficient matrix $\mathbf{M}$ is given by
\begin{equation*}
\mathbf{M}=
\left(
\begin{array}{ccc}
\displaystyle\langle| u^0 |\displaystyle\rangle &
\displaystyle\langle| u^1 |\displaystyle\rangle &
\displaystyle\langle| \psi _3 |\displaystyle\rangle\\
\displaystyle\langle| u^1 |\displaystyle\rangle &
\displaystyle\langle| u^2 |\displaystyle\rangle &
\displaystyle\langle| u^1 \psi _3 |\displaystyle\rangle\\
\displaystyle\langle| \psi _3 |\displaystyle\rangle &
\displaystyle\langle| u^1\psi _3 |\displaystyle\rangle &
\displaystyle\langle| \psi _3 ^2 |\displaystyle\rangle\\
\end{array}
\right)
=
\left(
\begin{array}{ccc}
1 &
U &
B_1\\
U &
U^2+1/2 \lambda &
B_2\\
B_1 &
B_2 &
B_3
\end{array}
\right) ,
\end{equation*}
where
\begin{equation*}
\begin{split}
&B_1=\frac{1}{2}(U^2+V^2+\frac{K+1}{2 \lambda}),\\
&B_2=\frac{1}{2}(U^3+\frac{(K+3)U}{2 \lambda}),\\
&B_3=\frac{1}{4}( U^4+\frac{(K+3)(U^2)}{\lambda}
+\frac{(K+1)(K+3)}{4 \lambda ^2}).\\
\end{split}
\end{equation*}
Denoting
\begin{equation*}
\begin{split}
R_3=2b_3-(U^2+\frac{K+1}{2\lambda})b_1, \quad
R_2=b_2-Ub_1,
\end{split}
\end{equation*}
the solution of Eq.\eqref{linear} can be written as
\begin{align*}
a_3&=\frac{4\lambda^2}{K+1}(R4-2UR_2),\\
a_2&=2\lambda R_2-Ua_3,\\
a_1&=b_1-Ua_2-\frac{1}{2}a_3(U^2+\frac{K+1}{2\lambda}).
\end{align*}
Thus, once the reconstructions for
macroscopic flow variables and their  derivatives are provided, the micro first-order spatial
derivatives can be calculated.

According to the compatibility condition Eq.\eqref{compatibility}, the corresponding Euler equations can be derived
\begin{equation*}
\displaystyle\langle au+A \displaystyle\rangle = 0.
\end{equation*}
Then the coefficient $A$ for the temporal evolution of a equilibrium state can be obtained by solving the following equation
\begin{equation*}
\displaystyle\langle A \displaystyle\rangle =-\displaystyle\langle au \displaystyle\rangle.
\end{equation*}

\subsection*{Time Integration for third-order flux solver}
Since the distribution function is time dependent, the total transport in one time step from $t_n$ to $t_n+\Delta t$ yields
\begin{equation*}
\begin{split}
\mathbb{F}_{i+1/2}(W^n,\delta)
&=\int_{t_n}^{t_n+\delta}F_{i+1/2}(W^n,t)\text{d}t\\
&=\int_{t_n}^{t_n+\delta}\int
u \psi f(x_{i+1/2},t,u,\xi)\text{d}\Xi\text{d}t\\
&=\int \int_{t_n}^{t_n+\delta} u \psi f(x_{i+1/2},t,u,\xi) \text{d}t \text{d}\Xi ,
\end{split}
\end{equation*}
where
\begin{equation*}
\begin{split}
\int_{t_n}^{t_n+\delta} u f(x_{i+1/2},t,u,\xi) \text{d}t
&=(\tau e^{-t/\tau} + \delta - \tau)u  g^{c}
\\&+\{\tau [\tau - e^{-t/\tau} (\delta + \tau) - \tau (e^{-t/\tau} - 1)] - \tau\delta  \}  u^2 a^{c}   g^{c}
\\&+[1/2 \delta^2 - \tau^2 (e^{-t/\tau} - 1) - \tau  \delta] u A^{c}   g^{c}
\\&+\tau (1 - e^{-t/\tau}) [H(u)u g^l + (1-H(u)) u g^r]
\\&+[\tau (e^{-t/\tau} ( \delta + \tau) - \tau) +  \tau^2 (e^{-t/\tau} - 1)] [H(u)u^2 a^l g^l + (1-H(u)) u^2 a^r g^r]
\\&+\tau^2 (e^{-t/\tau} - 1)[H(u)u A^l g^l + (1-H(u)) u A^r g^r].
\end{split}
\end{equation*}
In smooth case, it could be simplified as
\begin{equation*}
\begin{split}
\int_{t_n}^{t_n+\delta} u f(x_{i+1/2},t,u,\xi) \text{d}t
&= \delta u  g^{c} -\tau  \delta u^2 a^{c}   g^{c}
+[1/2 \delta^2 - \tau  \delta] u A^{c}   g^{c}.
\end{split}
\end{equation*}

\bibliographystyle{plain}%
\bibliography{jixingbib}

\end{document}